\documentclass{aa}
\usepackage{graphicx}
\usepackage{txfonts}

\newcommand{\onvire}[1]{}

\newcommand{\beq}{\begin{equation}}
\newcommand{\eeq}{\end{equation}}
\newcommand{\excs}{\extracolsep{\fill}}

\usepackage{color}
\usepackage{epsfig}
\usepackage{amssymb}
\usepackage{graphicx,natbib}
\usepackage{pdflscape}
\usepackage{longtable}
\usepackage[figuresright]{rotating} 
\begin{document}
\title{Transient dust in warm debris disks}
\subtitle{Detection of Fe-rich olivine grains}
   \author{J. Olofsson
          \inst{1}
          \and 
          A. Juh\'asz
          \inst{2}
          \and 
          Th. Henning
          \inst{1}
	   \and 
          H. Mutschke
          \inst{3}
	   \and 
          A. Tamanai
          \inst{4}
	   \and 
          A. Mo\'or
          \inst{5}
	   \and 
          P. \'Abrah\'am
          \inst{5}
          }

   \offprints{olofsson@mpia.de}
   
    \institute{Max Planck Institut f\"ur Astronomie,
     K\"onigstuhl 17, 69117 Heidelberg, Germany \\
     \email{olofsson@mpia.de} \and
    Leiden Observatory, Leiden University, PO Box 9513, NL-2300 RA Leiden, The Netherlands \and
    Astrophysikalisches Institut und Universit\"ats-Sternwarte (AIU), Schillerg\"a§chen 2-3, 07745 Jena, Germany \and
    University Heidelberg, Kirchhoff-Institut f\"ur Physik, 69120 Heidelberg, Germany \and
    Konkoly Observatory, Research Centre for Astronomy and Earth Sciences, Hungarian Academy of Sciences, P.O. Box 67, H-1525 Budapest, Hungary 
   }

   \date{Received \today; accepted } 
   \abstract
   {Debris disks trace remnant reservoirs of leftover planetesimals in planetary systems. A handful of ``warm" debris disks have been discovered in the last years, where emission in excess starts in the mid-infrared. An interesting subset within these warm debris disks are those where emission features are detected in mid-infrared spectra, which points towards the presence of $\mu$m-sized dust grains, with temperatures above hundreds K. Given the ages of the host stars, the presence of these small grains is puzzling, and questions their origin and survival in time.}
   {This study focuses on determining the mineralogy of the dust around 7 debris disks with evidence for warm dust, based on Spitzer/\textsc{Irs} spectroscopic data, in order to provide new insights into the origin of the dust grains. }
   {We develop and present a new radiative transfer code (\textsc{Debra}) dedicated to SED modeling of optically thin disks. The \textsc{Debra} code is designed such as it can determine dust composition and disk properties simultaneously. We make use of this code on the SEDs of seven warm debris disks, in combination with recent laboratory experiments on dust optical properties.}
   {We find that most, if not all, debris disks in our sample are experiencing a transient phase, suggesting a production of small dust grains on relatively short timescales. Dust replenishment should be efficient on timescales of months for at least three sources. From a mineralogical point of view, we find that crystalline pyroxene grains (enstatite) have small abundances compared to crystalline olivine grains. The main result of our study is that we find evidences for Fe-rich crystalline olivine grains (Fe / [Mg + Fe] $\sim$ 0.2) for several debris disks. This finding contrasts with studies of gas-rich protoplanetary disks, where Fe-bearing crystalline grains are usually not observed.}
   {The presence of Fe-rich olivine grains, and the overall differences between the mineralogy of dust in Class\,II disks compared to debris disks suggest that the transient crystalline dust in warm debris disk is of a new generation. We discuss possible crystallization routes to explain our results, and also comment on the mechanisms that may be responsible for the production of small dust grains.}
   \keywords{Stars: individual: HD\,113766\,A, HD\,69830, BD+20\,307, HD\,15407\,A, HD\,169666, HD\,98800\,B, [GBR2007]\,ID\,8 -- planetary systems: Zodiacal dust -- circumstellar matter -- Infrared: stars --
   Techniques: spectroscopic}
\authorrunning{Olofsson et al.}
\titlerunning{Transient dust in warm debris disks}

   \maketitle
%

\section{Introduction}

Several hundred stars are known to harbor debris disks. The first one was discovered around Vega (\citealp{Aumann1984}), using the Infrared Astronomical Satellite (IRAS). As detailed in the review by \citet{Wyatt2008}, debris disks are characterized by photospheric emission down to mid-infrared (IR) wavelengths. Emission is then seen in excess at longer wavelengths, caused by thermal emission arising from dust, located in circumstellar belts, heated by the stellar radiation. One key characteristic of debris disks is that they are optically thin at all wavelengths. A comparison with Solar System's Kuiper belt or asteroid belt is often made in order to explain such systems. Since planetesimals and km-sized bodies do not contribute to the thermal emission, smaller dust grains that radiate more efficiently are therefore required to reproduce the observed emission in excess. However, the smallest dust grains are also subjected to various effects, such as Poynting-Robertson drag and radiation pressure, that can evacuate them from the system on short timescales. This suggests that the population of small dust grains is continuously replenished. Such replenishment can for example happen via destructive collisions of km-sized bodies and will eventually result in a collisional cascade that will produce the observed excess emission, and may converge to a steady-state evolution of the debris disk.

Most of the known debris disks are Kuiper belt-type disks (\citealp{Carpenter2009}, \citealp{Liseau2010}, \citealp{Lohne2012}), showing no evidences for warm dust grains in their inner regions. However, thanks to mid-IR observations, several unusual objects emerged during the last years: warm debris disks. For these systems, the emission in excess starts in the mid-IR ($\sim$\,5--10\,$\mu$m), suggesting the presence a material much closer to the central star. Our study will focus on a subset of this group: warm debris disks with emission features. These very rare objects ($\sim$\,2\% of all debris disks, \citealp{Bryden2006}, \citealp{Chen2006}) are F-, G-, K-stars, whose mid-IR spectra (5-35\,$\mu$m) show strong emission features. Such emission features are usually seen in primordial disks around young Class\,II objects, and are associated with \,$\mu$m-sized silicate dust grains. In addition to the relatively small sizes of such dust grains, the presence of a strong excess at short wavelengths ($\sim$\,5\,$\mu$m) is indicative of a high temperature for the dust grains, suggesting the dust is located close to the star. The presence of these small dust grains close to the star is intriguing since the solution of a steady-state evolution may not be suitable anymore. At the ages of the systems, the smallest grains should have been depleted, either by radiation pressure or Poynting-Robertson drag. This means the disks could be experiencing a transient phase (e.g., \citealp{Wyatt2007}). In that case, two scenarios are possible, either ({\it i}) a recent collision between two or more bodies (planetesimals, comets, asteroids), or ({\it ii}) an outer belt feeds the inner regions, in a very similar scenario to the Late Heavy Bombardment (LHB), that happened in the Solar System. To have a better understanding of the origin of this dust, and assess whether it is transient or not, one way is to study the Spectral Energy Distribution (SED) of these rare debris disks, and determine the temperature of the dust responsible for the emission, and thus its distance from the star.

A powerful diagnostic to complement such study is to make use of the emission features that arise from (sub-)\,$\mu$m-sized grains. The width, shape and peak positions of the emission features provide information on the composition of the dust grains (see \citealp{Henning2010} for a recent review). The most common dust species are from the silicate class, which have been detected in various environments, e.g., in the interstellar medium, winds of AGB stars, and objects from the Solar System. The main building blocks of the silicates are Si, O, Mg and Fe (then followed by Ca and Al). When modeling the dust mineralogy via spectral decomposition, one should therefore focus primarily on dust species that contained these four main ingredients. Silicates are an association of [SiO$_4$]$^{4-}$ tetrahedra, with inclusions of Mg or Fe (following cosmic abundances). Not only the chemical composition of the grains has an influence on the emission features; the internal periodic arrangement of the SiO$_4$ tetrahedra will have a significant impact on the emission features. Grains with long-range order are in their crystalline form, as opposed to the amorphous form. As explained in detail in \citet{Henning2010}, optical properties of crystalline and amorphous grains are significantly different, with multiple sharp emission features for the former one, while emission features are much broader for amorphous grains. It is therefore possible to distinguish between amorphous and crystalline grains, as well as different dust species. Modeling the dust mineralogy can therefore provide useful insights into dust composition and crystallization processes at stake in the debris disks, which in turn provide information on the origin of the dust. To assess if the grains trace a second generation or are the remnants of the primordial, gas-rich disks one has to search for possible differences between the mineralogy of dust in Class\,II disks compared to the dust we observe in debris disks. For several stars, \citet{Juh'asz2010} found pyroxene crystalline grains with a Fe fraction of about $\sim$\,10\%, however, a result that appears to be quite general for the petrology of crystalline olivine grains in protoplanetary disks is that they are Mg-rich (and hence Fe-poor, see \citealp{Bouwman2008}, \citealp{Olofsson2010}). The overall lack of Fe-bearing crystalline grains has been studied by several authors and can be explained by crystallization processes that take place in massive disks (gas-phase condensation, \citealp{Gail2010} or thermal annealing, \citealp{Nuth2006}). One may wonder if we observe a similar result for the dust in warm debris disks. If not, we may be witnessing the production of a second generation of dust, where crystalline grains are produced by different mechanisms. 

In the following, we present in Section\,\ref{sec:obs} the sample of warm debris disks and briefly summarize the Spitzer/\textsc{Irs} data reduction. The new radiative transfer code we use is described in details in Sect.\,\ref{sec:debra}. We also justify which dust species we choose to model the mid-IR spectra. Section\,\ref{sec:early_res} presents several ``early" results that we find relevant to SED modeling of debris disks, and we present our best fit models for individual sources in Sect.\,\ref{sec:indiv}. The discussion of our results is then separated in three parts. In Section\,\ref{sec:survival} we investigate the survival and transient nature of the dust grains as well as the time variability for some of the sources in our sample, while in Sect.\,\ref{sec:origin} we emphasize on the origin of the crystalline grains that we detect in the dust belts. Finally, we briefly compare our results to what is known for objects in our Solar System in Sect.\,\ref{sec:solar}, and we summarize our findings and possible further improvements in Sect.\,\ref{sec:conclusion}.

\section{Spitzer/\textsc{Irs} data and stellar parameters\label{sec:obs}}

\begin{table*}
  \begin{center}
    \caption{\label{tbl:stars}Stellar sample}
    \begin{tabular}{lccccccccc}
      \hline \hline
      Starname & Alias & SpT & $T_{\star}$ & $L_{\star}$ & $d_{\star}$ & Age & AOR & Epoch & Ref \\
     & & & [K] & [$L_{\odot}$] & [pc] & [Myr] & & [d-m-y] & \\
      \hline
      \object{HD\,113766\,A} & & F4V & 6800 & 4.4 & 123 & 10-16 & 3579904 & 01-03-2004 & 1, 2, 3\\
      \object{HD\,69830} &  & K0V & 5650 & 0.67 & 12.6 & 4000 & 28830720 & 14-01-2009 & 4, 5, 6 \\
      \object{BD+20\,307} & HIP\,8920 & G0 & 5750 & 1.8 & 92$\pm$11 & $\sim$1000 & 14416384 & 15-01-2006 & 7, 8, 9 \\
      \object{HD\,15407\,A} &  & F5V & 6500 & 4.5 & 54.7 & 80 & 26122496 & 09-10-2008 & 10 \\ 
      \object{HD\,169666} & & F5 & 6750 & 5.16 & 53.2 & 2100 & 15016960 & 02-07-2005 & 11 \\
      \object{HD\,98800\,B} & TWA4 & K5 & 4600 & 0.62 & 44.9 & 8-10 & 3571968 & 25-06-2004 & 12, 13, 14, 15 \\
      \object{$[$GBR2007$]$\,ID\,8} & 2MASS\,J08090250-4858172 & G & 5750 & 0.8 & 361 & 38$^{+3.5}_{-6.5}$ & 21755136 & 16-06-2007 & 16, 17 \\
     	\hline
   \end{tabular}
  \end{center}

   {\bf References:} (1) \citet{Chen2005}, (2) \citet{Chen2006}, (3) \citet{Lisse2008}, (4) \citet{Beichman2005}, (5) \cite{Lisse2007}, (6) \citet{Beichman2011}, (7) \cite{Weinberger2008}, (8) \citet{Zuckerman2008a}, (9) \citet{Weinberger2011}, (10) \citet{Melis2010}, (11) \citet{Mo'or2009}, (12) \cite{Koerner2000}, (13) \cite{Prato2001}, (14) \citet{Furlan2007}, (15) \citet{Verrier2008}, (16) \citet{Gorlova2007}, (17) \citet{Naylor2006}
\end{table*}

The stellar sample (Table\,\ref{tbl:stars}) was gathered from the literature and investigating the Spitzer archive, looking for warm debris disks around solar analogs (F-, G-, K-type stars). The Spitzer/\textsc{Irs} spectra were retrieved from the Spitzer archive and reduced using the FEPS pipeline (see \citealp{Bouwman2008}).

All stellar parameters, namely $T_{\star}$, $L_{\star}$ and $d_{\star}$ (effective temperature, luminosity and distance, respectively), were taken from literature. To reproduce the photospheric emission we use Kurucz model that best matches the photometric near-IR data and will minimize the excess emission in the 5--8\,$\mu$m range of the \textsc{Irs} data. The spectral decomposition procedure is extremely sensitive to the photospheric emission model used, especially for objects that display small mid-IR excess (e.g., HD\,69830, see Fig.\,\ref{fig:SED_1}). 

As several Spitzer/\textsc{Irs} AOR (Astronomical Observation Request) contained both high and low spectral resolution observations at different wavelengths (e.g. Short-Low and Long-High), we rebin the spectra to the minimal spectral resolution in order to obtain an uniform spectral resolution over the entire wavelength range. Doing so, the fitting process does not favor spectral regions with higher spectral resolution.

Four objects in our sample are known multiple systems: HD\,113766, HD\,98800, BD+20\,307 and HD\,15407. For the first two objects, given the small separation of the different components and the slit size of the \textsc{Irs} instrument, the photospheric emission of the dust-free component has to be removed in order to consistently determine the dust location around the star. For HD\,113766 we use the luminosities as reported in \citet{Lisse2008} for HD\,113766\,A (around which the dust is located). For HD\,98800 we use the Keck diffraction-limited observations of the quadruple system by \citet{Prato2001} to model photospheric emission of the two components. \citet{Prato2001} indeed provide the individual fluxes for the two spectroscopic binaries A \& B (the dust being located around the spectroscopic binary HD\,98800\,B). For these two objects we will now refer to HD\,113766\,A and HD\,98800\,B. BD+20\,307 is a spectroscopic binary (\citealp[3.5 days period, ][]{Weinberger2008}) of two, almost identical, late F-type stars and the disk is circumbinary, meaning no correction has to be applied to the \textsc{Irs} spectrum. HD\,15407 is a binary system with a 21.21" separation. The dust is located around the primary HD\,15407\,A (F5) and the secondary HD\,15407\,B (K2, \citealp{Melis2010}) does not appear in the \textsc{Irs} slit, therefore no correction to the spectrum has to be performed for this object.

\section{Modeling the optically thin emission\label{sec:debra}}

To model the SEDs including the \textsc{Irs} spectra, we use the ``DEBris disk RAdiative transfer" code (\textsc{Debra} hereafter), a radiative transfer code dedicated to optically thin disks. Originally adapted from the ``Debris disk radiative transfer simulation tool" (\citealp{Wolf2005}), the code has been improved in order to determine the dust mineralogy as probed by the \textsc{Irs} spectra.

\subsection{The disk setup}

\begin{figure}
\begin{center}
\hspace*{-0.5cm}\includegraphics[angle=0,width=\columnwidth,origin=bl]{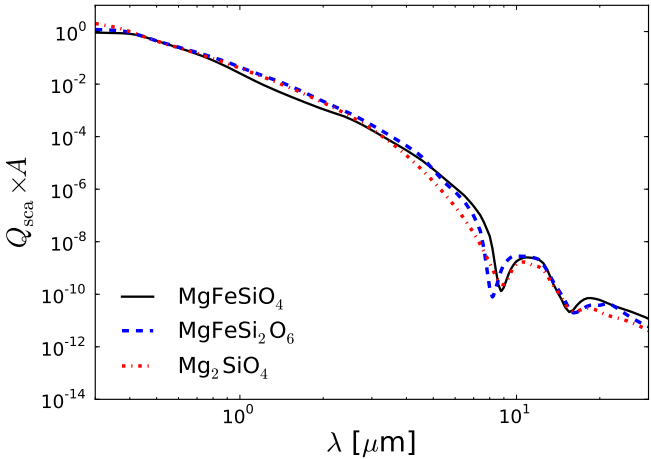}
\caption{\label{fig:qsca}$Q_{\mathrm{sca}} \times A$ for 0.1 \,$\mu$m-sized grains as a function of wavelength for three dust species. $Q_{\mathrm{sca}}$ being the scattering efficiency and $A$ the albedo of the grain (see Eqn\,\ref{eqn:albedo}).}
\end{center}
\end{figure}

The \textsc{Debra} code is designed to compute the SED of a ``star + disk" system in the optically thin regime. The star is defined by its Kurucz model and distance $d_{\star}$, and the radiative transfer code can then compute the reprocessed (thermal and scattered) emission arising from the circumstellar dust. The disk is defined by the following parameters: inner and outer radii $r_{\mathrm{in}}$ and $r_{\mathrm{out}}$, mass of dust $M_{\mathrm{dust}}$ and a volume density power exponent $\alpha$ ($\leq 0$). The density profile is assumed to be a power-law ($\sum (r) = \sum (r/r_0) ^{\alpha}$) between $r_{\mathrm{in}}$ and $r_{\mathrm{out}}$ (sampled on $nr = 80$ radial points). For a given set of parameters, one can then compute the total emission arising from the disk over a large wavelength range (for more details, see \citealp{Wolf2005}). For all the sources presented in this study, we choose $\lambda_{\mathrm{min}} = 0.3$\,$\mu$m and $\lambda_{\mathrm{max}} = 100$\,mm. The broad wavelength range is required to sample the heating (stellar) and cooling (dust) radiation field to calculate the dust temperature correctly. 

We can question whether scattered light emission should be considered when trying to model mid-IR (5--35\,$\mu$m) spectra. If the albedo $A$ of a single particle of size $s$ is defined as follows:
\begin{eqnarray}\label{eqn:albedo}
A = \frac{Q_{\mathrm{sca}}(\lambda,s)}{Q_{\mathrm{abs}}(\lambda,s) + Q_{\mathrm{sca}}(\lambda,s)},
\end{eqnarray}
where $Q_{\mathrm{abs}}(\lambda,s)$ and $Q_{\mathrm{sca}}(\lambda,s)$ are the absorption and scattering efficiencies (see Sect.\,\ref{sec:dust} for more details), then the scattered light emission of this single dust grain is proportional to $Q_{\mathrm{sca}}(\lambda,s) \times A$ (\citealp{Wolf2005}). Figure\,\ref{fig:qsca} shows how this quantity decreases dramatically with increasing wavelengths ($\sim$\,4 orders of magnitude smaller at 5\,$\mu$m than at 1\,$\mu$m), for three different amorphous dust species, for a grain size $s = $\,0.1\,$\mu$m. Consequently, we choose not include the contribution of scattered light in the modeling process.

The last free parameter of our model is the grain size distribution, which we assume to be a differential power-law of index $p$ (d$n(s) \propto s^{p}$d$s$, $p \leq 0$) between $s_{\mathrm{min}}$ and $s_{\mathrm{max}}$. One given model is therefore the sum of the stellar flux and thermal emission from grains of different sizes, for several dust species, and is defined by four parameters $r_{\mathrm{in}}$, $r_{\mathrm{out}}$, $\alpha$ and $p$. The mass of dust $M_{\mathrm{dust}}$ within the disk is determined when fitting the dust composition over the \textsc{Irs} spectral range. The fitting is performed via the Levenberg-Marquardt algorithm on a linear combination of all the thermal emission contributions from the considered dust species. For a single dust grain of size $s$, the thermal emission flux $F_{\mathrm{emis}}$, received by an observer at distance $d_{\star}$, is computed as follows:
\begin{eqnarray}
F_{\mathrm{emis}}(r,\lambda,s) = \left( \frac{s}{d_{\star}} \right) ^2 \times Q_{\mathrm{abs}}(\lambda,s) \times \pi \times B_{\lambda} (\lambda, T_{d}),
\end{eqnarray}
where $B_{\lambda} (\lambda, T_{d})$ is the Planck function at temperature $T_d$, which depends on grain sizes, compositions and distance from the star $r$ (see Sect.\,\ref{sec:temperature}). Finally, to determine the best fit over the entire SED (i.e., the disk geometry as well as the dust composition) we use a genetic algorithm (\texttt{pikaia}, \citealp{Charbonneau1995}) over the four free parameters mentioned above. The code iterates over 100 generations, with 50 individuals per generation, and fits the dust composition at every iteration. Each time, the goodness of the fit is estimated with a reduced $\chi^2_{\mathrm{r}}$ and the best fit is the one with the lowest $\chi^2_{\mathrm{r}}$.

\subsection{The dust setup\label{sec:dust}}

\begin{table*}
  \begin{center}
    \caption{\label{tbl:dust}Dust species used. State refers to the amorphous (``A") or crystalline (``C") form.}
    \begin{tabular}{lcccccc}
      \hline \hline
      Species & State & Chemical & Density & Mg / [Mg + Fe] & Shape & Ref \\
                   &          & formula   & [g.cm$^{-3}$] & [\%]      & & \\
      \hline
	Amorphous silicate & A & MgFeSiO$_4$ & 3.71 & 50 & DHS &  1 \\
	(olivine stoichiometry) & & & & & &  \\
	Amorphous silicate & A & Mg$_2$SiO$_4$ & 3.88 & 100 & DHS &  2 \\
	(olivine stoichiometry) & &  & & & & \\
	Amorphous silicate & A & MgFeSi$_2$O$_6$ & 3.20 & 50 & DHS &  1 \\
	(pyroxene stoichiometry) & & & & & & \\
	Amorphous silicate & A & MgSiO$_3$ & 2.71 & 100 & DHS & 2 \\
	(pyroxene stoichiometry) & & & & & & \\
	Mg-rich olivine & C & Mg$_{1.85}$Fe$_{0.15}$SiO$_4$ & 3.33 & 92.5 & Aerosol &  3 \\
	(San Carlos sample) & & & & & &  \\
	Fe-rich olivine & C & Mg$_{1.6}$Fe$_{0.4}$SiO$_4$ & 3.33 & 80 & Aerosol & 3 \\
	(Sri Lanka sample) & & & & & & \\
	Ortho-enstatite & C & MgSiO$_3$ & 2.78 & 100 & DHS & 4 \\
	Silica ($\beta$-cristobalite) & C & SiO$_2$ & 2.27 & - & Aerosol &  5 \\
	Carbon & A & C & 1.67 & - & DHS & 6 \\ \hline
   \end{tabular}
  \end{center}
	
{\bf References:} (1) \citet{Dorschner1995}, (2) \citet{Jager2003},  (3) \citet{Tamanai2010}, (4) \citet{Jaeger1998}, (5) \citet{Tamanai2010a}, (6) \citet[][``cel600", ``cel800", ``cel1000": cellulose pyrolized at 600, 800 and 1000\,$^{\circ}$C, respectively]{Jager1998}
\end{table*}

The \textsc{Irs} spectra of the objects in our sample show prominent emission features that are associated with warm (sub-)\,$\mu$m silicate dust grains. In order to reproduce the data we use several dust species, from amorphous (with olivine and pyroxene stoichiometries) and crystalline silicates (forsterite and enstatite) to crystalline $\beta$-cristobalite silica and amorphous carbon. These amorphous or crystalline dust grains, or their associated emission features, have been observed numerous times in disks around young Class\,II objects (e.g., \citealp{Bouwman2001,Bouwman2008}, \citealp{Juh'asz2010}, \citealp{Oliveira2011}, \citealp{Olofsson2009,Olofsson2010}, \citealp{Sargent2009}), and disks around evolved Class\,III objects (e.g., \citealp{Beichman2011}, \citealp{Lisse2007,Lisse2008,Lisse2009}, \citealp{Weinberger2011}). Amorphous grains of olivine (Mg$_{2x}$Fe$_{2(1-x)}$SiO$_4$) and pyroxene (Mg$_{2x}$Fe$_{2(1-x)}$Si$_2$O$_6$) stoichiometries are also the dominant species in the interstellar medium ($\sim$98\%, \citealp{Kemper2005}, \citealp{Min2007}). They are the major building blocks for the dust content in circumstellar environments. And yet, many uncertainties remain on the optical properties of these grains. For instance \cite{Tamanai2006} show the differences in band positions and band widths of the emissivity profiles for several species (crystalline forsterite and enstatite or amorphous Mg$_2$SiO$_4$) when using two different laboratory techniques (KBr pellets versus aerosol measurements). These measured emissivity profiles are also different from absorption efficiencies calculated using optical constant measurements. One must be aware of these unfortunate limitations, and interpret carefully the results from spectral decomposition when using several dust constituents. This is the main reason why we focus our study on silicate dust grains and choose not to include additional, more complex dust species, to limit degeneracies in the fitting process. 
Other dust components would have been included only if significant signatures were detected in the residuals. All the dust components used in this study are listed in Table\,\ref{tbl:dust}.

\subsubsection{Absorption efficiencies\label{sec:mac}}

There are several ways to obtain absorption efficiencies $Q_{\mathrm{abs}}$, such as optical constant combined with a scattering theory or using laboratory extinction measurements (\citealp{Henning2010a}). When possible (and relevant, see Sect.\,\ref{sec:forst}) we opt for optical constant as it enables us to separate the contributions from different grain sizes, while aerosol measurements do not disentangle such contributions as the sample is a mixture of grains with different sizes. For the amorphous silicate grains we use the Distribution of Hollow Spheres (DHS) scattering theory, with a filling factor $f_{\mathrm{max}}$ of 0.7. \citet{Min2007} have shown these absorption efficiencies best match the extinction profile toward the galactic center at 10 and 20\,$\mu$m. For the crystalline species we also use the DHS scattering theory, but with a filling factor $f_{\mathrm{max}}$ of 1.0, for which the absorption efficiencies usually best match the observed band positions and compare well with other laboratory measurements (see \citealp{Juh'asz2010}). For the amorphous carbon grains we use the DHS theory with a filling factor $f_{\mathrm{max}}$ of 1. 

We compute absorption efficiencies for ten discrete grain sizes, logarithmically spaced between $s_{\mathrm{min}} = 0.1$\,$\mu$m and $s_{\mathrm{max}}$. For the crystalline silicate and silica grains, we limit $s_{\mathrm{max}}$ to 1\,$\mu$m. According to crystallization models, one does not expect to produce pure crystals of several $\mu$m via thermal annealing. Grains are more likely to grow via collisional aggregation of both small crystalline and amorphous material (e.g. \citealp{Min2008}). Additionally, optical properties of large crystalline grains can become degenerate with optical properties of other dust species, since the recognizable emission features of small crystals fade away when increasing $s$. Limiting $s_{\mathrm{max}}$ to 1\,$\mu$m is in line with previous studies of dust in protoplanetary disks (e.g., \citealp{Bouwman2008}, \citealp{Juh'asz2010}, \citealp{Olofsson2010}) where crystalline grains are found to be much smaller than amorphous grains. Concerning the maximum grain size for amorphous grains, we tested several values for $s_{\mathrm{max}}$ and the outcomes of this exercise are discussed in Sect.\,\ref{sec:smax}.

\subsubsection{Temperature determination\label{sec:temperature}}

Young massive Class\,II circumstellar disks contain large amount of dust, but also contains gas. When modeling \textsc{Irs} spectra for these objects, one can therefore assume all the dust grains, located in the optically thin regions of the disks, are in thermal contact and have the same temperature at a given distance $r$. However debris disks are devoid of gas and the previous assumption cannot hold anymore as no medium can efficiently support thermal exchange between dust grains. Consequently, we have to compute the temperature of the grains individually. The temperature $T_d$ of a given dust grain does not only depend on its distance to the star, but also on how efficiently it can absorb stellar light. The absorption efficiency being traced by $Q_{\mathrm{abs}}(\lambda,s)$, the relation between $T_d$ and $r$ can be expressed as follows:
\begin{eqnarray}
\label{eqn:radius}
r(s,T_d) = \frac{R_{\star}}{2} \times \sqrt{ \frac{ \int_{\lambda} F_{\star}(\lambda, T_{\star}) \times Q_{\mathrm{abs}}(\lambda,s) ~ \mathrm{d} \lambda}{\int_{\lambda} \pi \times B_{\lambda}(\lambda,T_d) \times Q_{\mathrm{abs}}(\lambda,s)~  \mathrm{d} \lambda}},
\end{eqnarray}
where $F_{\star}(\lambda, T_{\star})$ is the stellar surface flux at temperature $T_{\star}$, and $R_{\star}$ the stellar radius. Equation\,\ref{eqn:radius} shows that $r$, and hence $T_d$, depends on the absorption efficiencies used, especially in the near-IR where $F_{\star} (\lambda)$ peaks. For the F-, G-, K-type stars of our sample $F_{\star}$ peaks around $\sim$\,0.5--0.6\,$\mu$m, a wavelength included in our spectral range ($\lambda_{\mathrm{min}} = 0.3$\,$\mu$m). Inverting numerically Eqn.\,\ref{eqn:radius}, for each dust species and for each grain sizes we are able to consistently compute their temperature depending on their distance to the star.

\subsubsection{Forsterite and crystalline olivine grains\label{sec:forst}}

\begin{figure}
\begin{center}
\hspace*{-0.5cm}\includegraphics[angle=0,width=\columnwidth,origin=bl]{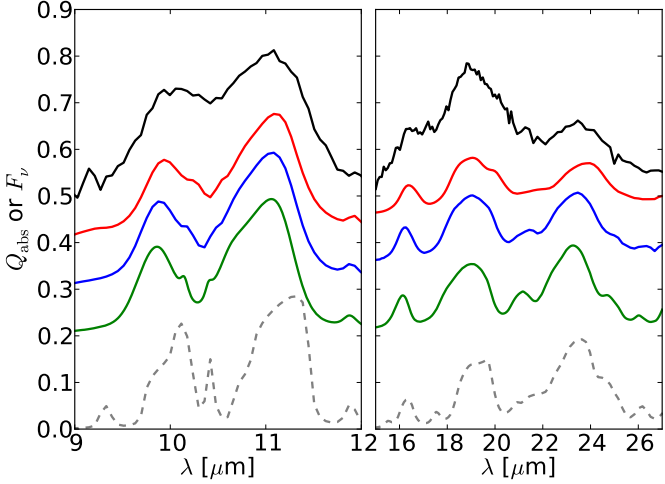}
\caption{\label{fig:qSoTa}From bottom to top: $Q_{\mathrm{abs}}$ values for forsterite (dashed grey, \citealp{Sogawa2006}), aerosol measurements of forsterite (green, \citealp{Tamanai2006}), Mg-rich (blue), Fe-rich olivine grains (red, \citealp{Tamanai2010}) and stellar subtracted spectrum of HD\,69830 in arbitrary units (black). To ease readability, $Q_{\mathrm{abs}}$ values and the spectrum are scaled and offset}
\end{center}
\end{figure}

\begin{figure}
\begin{center}
\hspace*{-0.5cm}\includegraphics[angle=0,width=\columnwidth,origin=bl]{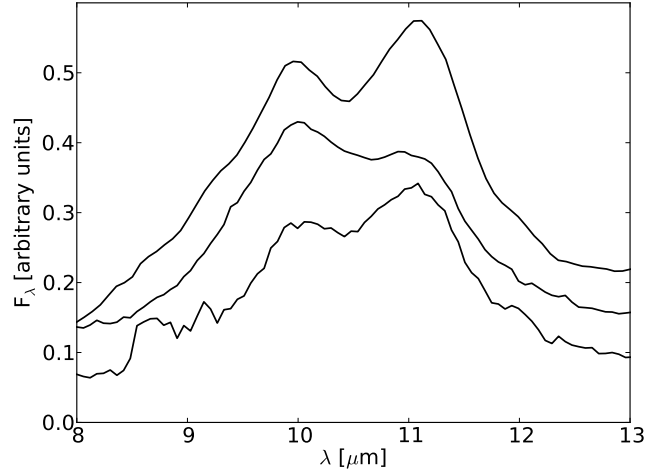}
\caption{\label{fig:11mu}10\,$\mu$m features for photosphere subtracted spectra showing the most prominent crystalline olivine features (from bottom to top: HD\,69830, [GBR2007]\,ID\,8 and HD\,113766\,A). Spectra have been scaled and offset}
\end{center}
\end{figure}

Absorption efficiencies are the tools used to interpret observations, and one should always make sure to use the most relevant ones. As mentioned previously, these efficiencies suffer from uncertainties at different levels. Laboratory measurements are unfortunately subjected to various environmental effects such as electromagnetic polarization of the embedding medium (KBr, CsI or PE) when using the pellet technique (see \citealp{Tamanai2006,Tamanai2009}). Additionally when using optical constant, scattering theories have there own caveats, as discussed for example in \citet{Min2007} for both Mie and DHS theories. By inspecting the emission features that are usually associated with crystalline olivine grains, we can determine which laboratory data are best suited to reproduce the observations. Figure\,\ref{fig:qSoTa} shows several laboratory measurements for two wavelength ranges. From bottom to top, the dashed grey line shows absorption efficiencies calculated with the DHS theory (with $f_{\mathrm{max}} = 1$) from optical constant of \citet{Sogawa2006}, green, blue and red lines represent aerosol measurements for synthesized forsterite (Mg$_2$SiO$_4$, \citealp{Tamanai2006}), Mg-rich and Fe-rich olivine grains (\citealp{Tamanai2010}), with Mg fractions (=\,Mg/[Mg + Fe]) of 100, 92.5 and 80\,\%, respectively. The black line is the photosphere subtracted spectrum of HD\,69830. To ease readability, laboratory data and spectrum have been scaled and offset. An immediate conclusion from Figure\,\ref{fig:qSoTa} is that the calculated $Q_{\mathrm{abs}}$ values of forsterite (using optical constant from \citealp{Sogawa2006}) do not match the observed spectrum, neither around 10\,$\mu$m nor at 19 and 24\,$\mu$m. According to \citet{Mutschke2009} some of these differences, especially around 10\,$\mu$m, arise from a simplified treatment of the crystal anisotropy in scattering theories. At this point, it is important to note that HD\,69830 is not the only object for which these differences will be a problem. Figure\,\ref{fig:11mu} shows the photosphere subtracted spectra (with different scalings and offsets) for three objects that show strong emission features associated with crystalline olivine grains (from bottom to top: HD\,69830, [GBR2007]\,ID\,8 and HD\,113766\,A). Band positions and shapes are overall quite similar for the three sources. For these objects as well, the $Q_{\mathrm{abs}}$ values computed with DHS will not match the shapes and peak positions of the observed emission features. Instead, the -scaled- aerosol measurements shown in Fig.\,\ref{fig:qSoTa} provide a much better match. Consequently, we choose not to use the absorption efficiencies computed with the optical constant from \cite{Sogawa2006}, and prefer to make use of the aerosol measurements, to represent for crystalline olivine grains in our modeling. The great advantage of the aerosol technique is that it minimizes not only the contamination by environmental effects, but also eliminate possible computational artifacts inherited from the scattering theory used.

\begin{figure*}
\begin{center}
\hspace*{-0.5cm}\includegraphics[angle=0,width=2\columnwidth,origin=bl]{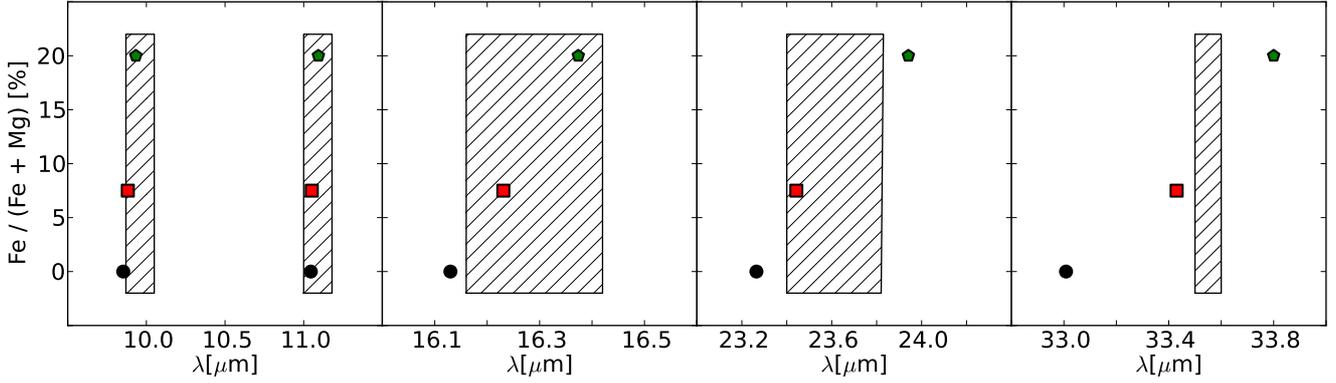}
\caption{\label{fig:peak}Peak positions of main emission features associated with crystalline olivine grains (9.8, 11, 16.2, 23--24 and 33\,$\mu$m), as a function of Fe fraction. Black circles, red squares and green pentagons correspond to aerosol measurements of pure forsterite, Mg- and Fe-rich samples, respectively (\citealp{Tamanai2006,Tamanai2010}). Hatched regions correspond to the range of peak positions in the observed spectra of HD\,69830, HD\,113766\,A, ID\,8 and HD\,169666.}
\end{center}
\end{figure*}

The only difference between the three aerosol measurements shown in Figure\,\ref{fig:qSoTa} is the iron content. The Mg fractions are 100, 92.5 and 80\% for the different samples (the subscripts of Fe and Mg fractions of the two natural samples are determined with energy dispersive X-ray detection). As shown in Figure\,\ref{fig:qSoTa}, the iron content mostly makes a difference at wavelengths longer than 20\,$\mu$m, while it has very little influence around 10\,$\mu$m. For the Fe-rich olivine grains, the 19\,$\mu$m feature is slightly broader and the feature at 23--24\,$\mu$m is shifted toward longer wavelengths, a result consistent with the measurements from \citet{Koike2003}. To investigate this effect deeper, Figure\,\ref{fig:peak} shows the peak positions for aerosol measurements, for several emission features associated with crystalline olivine grains, as a function of the Fe content. Peak positions for aerosol measurements for synthesized forsterite (\citealp{Tamanai2006}) are represented by black circles, while peak positions for Mg- and Fe-rich olivine grains (\citealp{Tamanai2010}) are represented by red squares and green pentagons, respectively. The hatched areas on Figure\,\ref{fig:peak} show the range of peak positions for these specific emission features in the observed spectra of HD\,69830, HD\,113766\,A, $[$GBR2007$]$\,ID\,8 and HD\,169666. From Figure\,\ref{fig:peak}, one can immediately see that the observed peak positions are better reproduced by Mg- or Fe-rich olivine grains, while synthesized forsterite does not seem to provide a good match (within a few tenths of $\mu$m). Since these three measurements are otherwise pretty similar, we decide to keep only the two most relevant samples with respect to our observations (Mg- and Fe-rich samples), in order to limit the number of degenerate parameters in the modeling approach. The relevance of including crystalline olivine grains with a Fe fraction up to 20\% is further discussed in Section\,\ref{sec:ferich}. 

The disadvantages of using the aerosol measurements instead of optical constant are: {\it (i)} the overall level of measurements is difficult to constrain. The experiment is non-quantitative, in the sense that it is not possible to determine how many particles have entered the aerosol chamber, and thus the total dust mass is uncertain. And {\it (ii)}, the information about the contribution of different grain sizes is lost. It is unfortunately not possible to select one finite grain size while injecting them in the aerosol apparatus. However, it is still possible to select a maximum grain size, which is $\sim$1\,$\mu$m in the case of the two samples discussed here. One should finally be aware that extinction spectra obtained by aerosol spectroscopy contain a scattering contribution, which is however negligible since the typical particle sizes are much smaller compared to the wavelength of observation. Therefore, the aerosol extinction spectra are equivalent to absorption in the frame of our work.

\begin{figure}
\begin{center}
\hspace*{-0.5cm}\includegraphics[angle=0,width=\columnwidth,origin=bl]{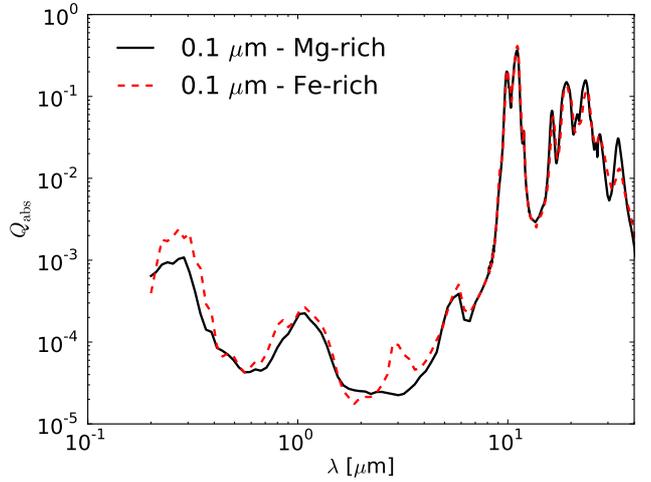}
\caption{\label{fig:qfo}$Q_{\mathrm{abs}}$ values for 0.1\,$\mu$m-sized Mg- and Fe-rich olivine grains (solid black line and red dashed line, respectively), using the near-IR measurements from \citet{Zeidler2011} combined with aerosol measurements from \citet{Tamanai2010}}
\end{center}
\end{figure}

In a recent study, \citet{Zeidler2011} determined the optical constant for several dust species in the near-IR, including the Mg- and Fe-rich olivines discussed in this Section. The increased Fe content (7.5 to 20\%) does not dramatically change the absorption efficiencies in the near-IR. This is illustrated in Figure\,\ref{fig:qfo} where we show the $Q_{\mathrm{abs}}$ values for Mg- and Fe-rich olivine 0.1\,$\mu$m-sized grains (solid black and dashed red lines, respectively). To compute these $Q_{\mathrm{abs}}$ values for different grain sizes, we first use the DHS theory (with $f_{\mathrm{max}} = 1$) for different grain sizes (between 0.1 and 1.0\,$\mu$m) on the combination of both near-IR optical constant from \citet{Zeidler2011} and \citet{Sogawa2006} (at longer wavelengths). In the range $7 \leq \lambda \leq 40$\,$\mu$m (data from \citealp{Zeidler2011} stop at 7\,$\mu$m), we then replace the resulting absorption efficiencies with the aerosol measurements from \citet{Tamanai2010} that are representative of grains smaller than 1\,$\mu$m (the size distribution in the aerosol apparatus being unknown). Since aerosol data are non-quantitative, we additionally normalize their peak value to the peak value of the previously calculated $Q_{\mathrm{abs}}$ values at 11\,$\mu$m, in order to have a more accurate overall level. The motivation behind this work being that $Q_{\mathrm{abs}}$ values are required in the near-IR regime, in order to compute grain temperatures consistently (Eqn.\,\ref{eqn:radius}).

\subsubsection{Silica\label{sec:sio2}}

\begin{figure}
\begin{center}
\hspace*{-0.5cm}\includegraphics[angle=0,width=\columnwidth,origin=bl]{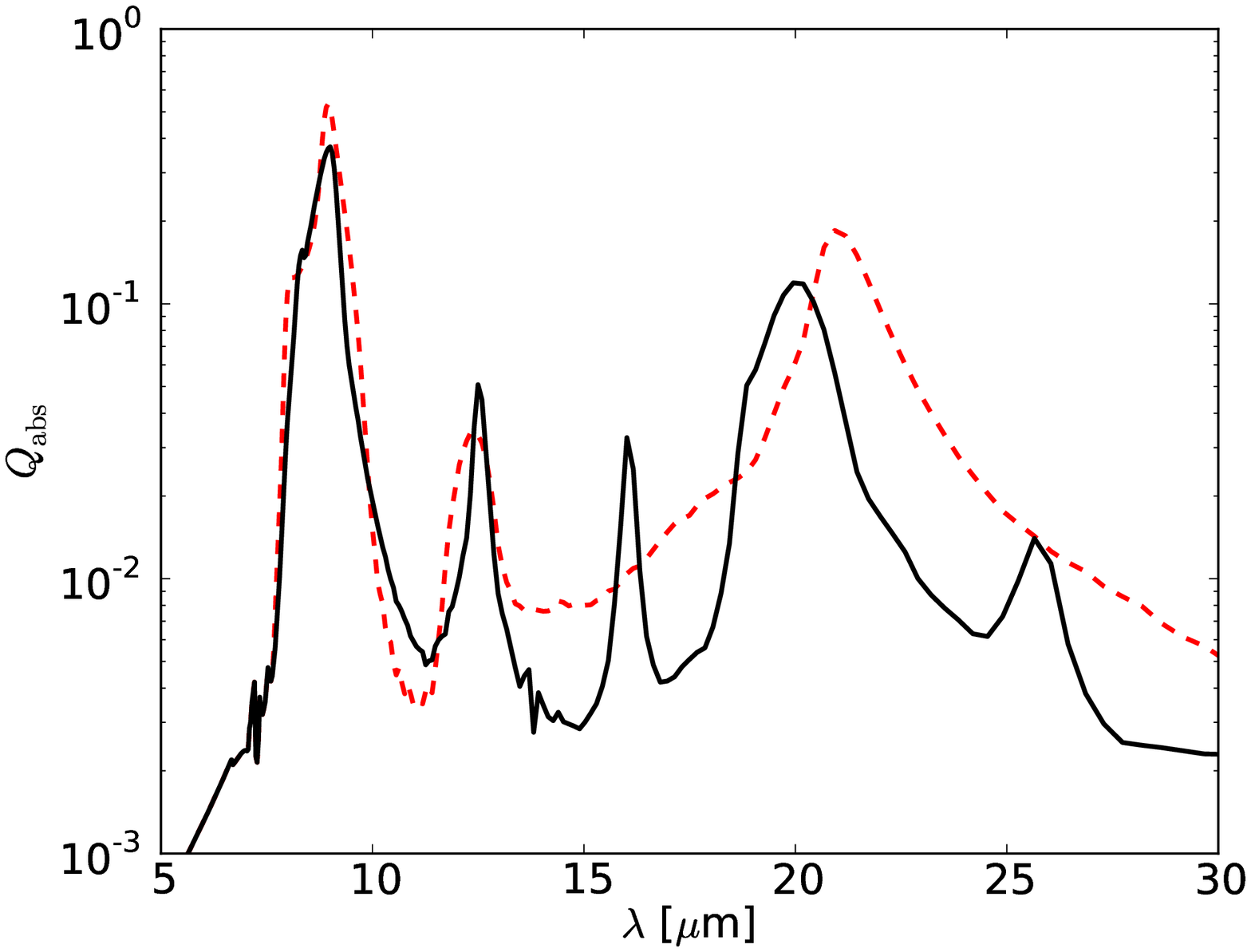}
\caption{\label{fig:qsio}$Q_{\mathrm{abs}}$ values for $\beta$-cristobalite silica grains from aerosol measurements (solid black line, \citealp{Tamanai2010a}), and 0.1\,$\mu$m-sized grains computed with optical constant from \citet[][dashed red line]{Henning1997}.}
\end{center}
\end{figure}

Silica grains have recognizable emission features around 9, $\sim$12.5, 16, and $\sim$20\,$\mu$m. Several polymorphs of silica have been studied in detail by \citet{Sargent2009a}. According to the authors, spectroscopic observations of protoplanetary disks are best reproduced by annealed silica, which contains $\beta$-cristobalite. They used the KBr pellet transmission measurements from \citet{Fabian2000}, in which $\beta$-cristobalite was formed by thermal annealing of SiO$_2$ during 5\,h at a temperature of 1220\,K. This motivates our choice for using this particular polymorph to represent for silica grains. However, even if \citet{Sargent2009a} discussed the possible influence of the KBr medium before concluding it should not have altered too much the band positions, we prefer to use aerosol measurements from \citet{Tamanai2010a} in order to avoid any possible contamination by the pellet medium. The maximum grain size inside the aerosol chamber is again 1\,$\mu$m. Figure\,\ref{fig:qsio} shows the results of aerosol measurements for $\beta$-cristobalite silica (black line). The major emission features peak at 9.0, 12.5, 16.0 and 20\,$\mu$m, which compares well with measurements from \citet[][9.1, 12.6, 16.1 and 20\,$\mu$m]{Fabian2000}. 

To obtain $Q_{\mathrm{abs}}$ values, we could not follow the same procedure as for Mg- and Fe-rich olivine grains (Sect.\,\ref{sec:forst}) since no data are available for wavelengths shorter than $\sim$\,7\,$\mu$m for $\beta$-cristobalite grains. Therefore, we use optical constant from \citet[][SiO$_2$ at 300\,K]{Henning1997}, and combine them with optical constant from iron-free MgSiO$_3$ in the spectral range 0.3--7\,$\mu$m. With the DHS theory ($f_{\mathrm{max}} = 1$) we then compute $Q_{\mathrm{abs}}$ values for 10 grain sizes between 0.1 and 1\,$\mu$m (dashed red line for 0.1\,$\mu$m in Fig.\,\ref{fig:qsio}). We then replace these absorption efficiencies with the aerosol measurements from \citet{Tamanai2010a} in the range 7--40\,$\mu$m. The choice of using MgSiO$_3$ data has in the end no measurable influence on the results. \cite{Kitamura2007} reviewed and studied optical properties of non-crystalline silica glass in different wavelength ranges, including the range 0.1--7\,$\mu$m. However, in this particular spectral range, the imaginary part of the complex refractive index are extremely low ($k_{\lambda} \sim 10^{-5}$). As further discussed in Sect.\,\ref{sec:dirty}, such low $k_{\lambda}$ values render the grains almost transparent to stellar light. Consequently, even at $r \leq$\,1\,AU from the star, such grains will have extremely low temperatures (as derived from Eqn.\ref{eqn:radius}) and will not contribute to the total emission. Therefore, silica grains need to be mixed with impurities, which will dominate the near-IR absorption efficiencies.

\subsection{Determining uncertainties}

To assess the significance of our results, we compute uncertainties on relative abundances of the different dust species, as well as disk parameters. This is achieved using the \texttt{emcee} package (\citealp{Foreman-Mackey2012}), which makes use of a Markov chain Monte Carlo (MCMC) to sample the parameter space and help providing uncertainties via Bayesian inference. For each of the 5000 runs of the MCMC procedure, $\chi^2_{\mathrm{r}}$ values and their associated probabilities ($e^{-\chi^2_r /2}$) are stored, as well as the 4 free parameters for the disk geometry plus the corresponding abundances for each dust species. Probability distributions are then generated by projecting the sum of probabilities onto the considered dimension (e.g., $r_{\mathrm{in}}$ values). A gaussian fit to the distribution provides its width $\sigma$ that we assume to be the uncertainty. An example of such a distribution is shown in Section\,\ref{sec:ferich}.

\section{Early results on SED modeling\label{sec:early_res}}

In the course of this study, we encounter a couple of issues when modeling \textsc{Irs} spectra (e.g., poorly fitted emission features, low temperatures for grains with small $Q_{\mathrm{abs}}$ values in the near-IR, or maximum grain size $s_{\mathrm{max}}$ used for the amorphous dust grains). We detail in this Section why these issues are considered as problematic, how we correct for them and what are the implications on the dust content in debris disks. In parallel, we emphasize the limits and possible biases of SED modeling and spectral decomposition for debris disks.

\subsection{``Dirty" silicates\label{sec:dirty}}

\begin{figure*}
\begin{center}
\includegraphics[angle=0,width=2\columnwidth,origin=bl]{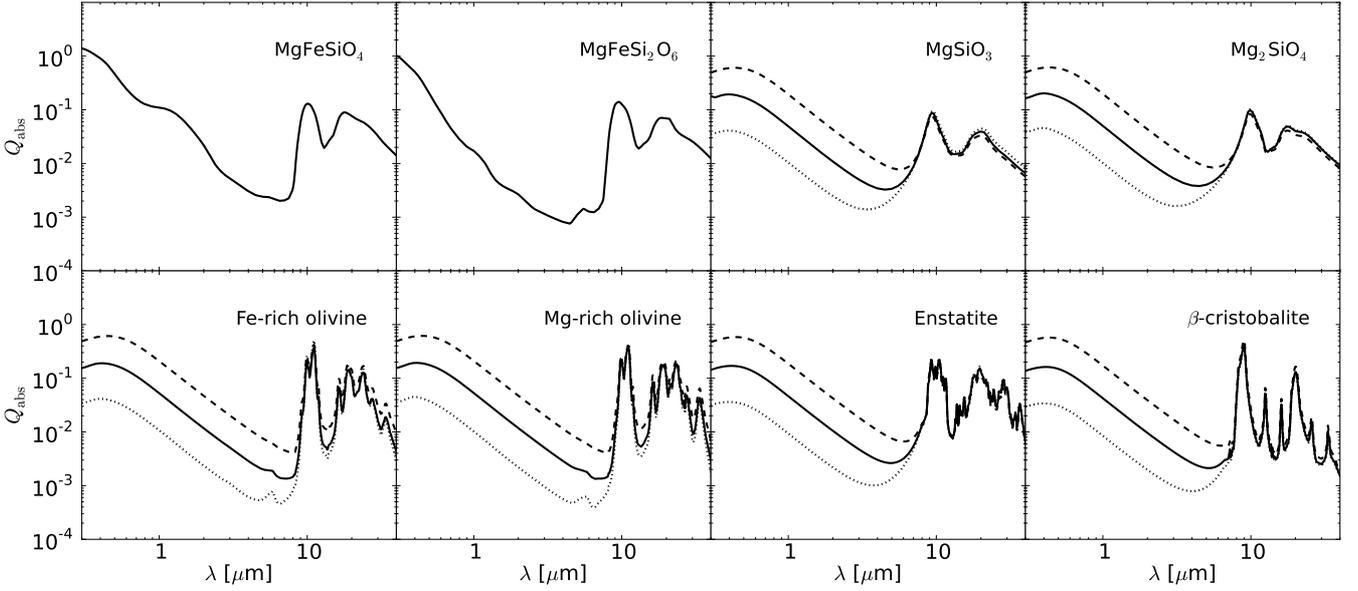}
\caption{\label{fig:Qabs_all}Absorption efficiencies ($s =$\,0.1\,$\mu$m) for the dust species used to model \textsc{Irs} spectra (except carbon) in the range 0.3--40\,$\mu$m. Top row shows absorption efficiencies for amorphous dust grains, bottom row shows $Q_{\mathrm{abs}}$ values for crystalline dust grains. With the exception of the first two panels, dotted, solid and dashed lines represent absorption efficiencies mixed with volume fractions of 1, 5 and 20\,\% carbon, respectively.}
\end{center}
\end{figure*}

The fitting process used in this study highly depends on the grain temperatures. As explained in Sect.\,\ref{sec:temperature}, these temperatures are determined using the $Q_{\mathrm{abs}}$ values (from optical to mm wavelengths) of the considered dust species and grain sizes. It is therefore necessary to have a precise knowledge of the absorption efficiencies. If the $Q_{\mathrm{abs}}$ values are too low (e.g., $\sim 10^{-4}$) in the optical and near-IR where the stellar emission peaks, the grains will become almost ``transparent" to stellar light. Consequently, at the same distance from the star, such grains will have a much smaller temperature than grains with higher $Q_{\mathrm{abs}}$ values in the near-IR. If all the dust species had similar absorption efficiencies in the near-IR this would simply translate into smaller distances $r$ to get appropriate temperatures. However, grains such as amorphous MgFeSiO$_4$ or MgFeSi$_2$O$_6$ have higher $Q_{\mathrm{abs}}$ values in the near-IR and are expected to be detected (as in Class\,II disks). If one consider the best fit will match crystalline features such as the ones from Mg- and Fe-rich olivines, in that case, the temperature of amorphous MgFeSiO$_4$ or MgFeSi$_2$O$_6$ grains, for which the $Q_{\mathrm{abs}}$ values are higher in the near-IR, will most likely be above the sublimation temperature.

To illustrate the problem, we model the spectrum of HD\,69830 using the dust species listed in Table\,\ref{tbl:dust}. For Mg- and Fe-rich crystalline olivine grains, we use the near-IR data from \citet{Zeidler2011} as described previously, which have small absorption efficiencies in the near-IR ($\sim 10^{-4}$, see Fig.\,\ref{fig:qfo}). In that case, the final temperatures of 0.1\,$\mu$m-sized Mg- and Fe-rich crystalline olivine grains are of about 380 and 475\,K, respectively, at $r = r_{\mathrm{in}} = 4.4 \times 10^{-2}$\,AU. The corresponding temperatures for 0.1\,$\mu$m-sized amorphous MgFeSiO$_4$ grains is $\sim$\,1970\,K, a temperature at which they should vaporize. If one limits the $r_{\mathrm{in}}$ value for which these grains will have a $T_{\mathrm{d}} \sim 1500$\,K, then the crystalline olivine grains will be too cold and will consequently not contribute to the emission, resulting in a poor fit to the spectrum. 

To circumvent this temperature issue, one can consider ``dirty" silicates. In other words, one can assume that grains contain a fraction of impurities, such as transition metal ions, or are in contact with other dust grains that have high absorption efficiencies in the near-IR (see, \citealp{Ossenkopf1992}). To account for this dirtiness, we use the Maxwell-Garnett effective medium theory to mix different fractions of carbon (\citealp[``cel600", ][]{Jager1998}) with Mg- and Fe-rich crystalline olivine grains. The carbon admixture was performed on the combined optical constant from \citet{Zeidler2011} and \citet{Sogawa2006}. We then replace the resulting $Q_{\mathrm{abs}}$ values with the normalized aerosol data for Mg- and Fe-rich olivine grains in the range $7 \leq \lambda \leq 40$\,$\mu$m (see last paragraph of Sect.\,\ref{sec:forst}). The choice of using carbon over other species (e.g., metallic iron) has very little effect on the results, since the primary goal is simply to heat up the grains. This is achieved by choosing featureless $Q_{\mathrm{abs}}$ values for the impurities. Similar low-temperature problems will affect enstatite, silica (see Sect.\,\ref{sec:sio2}), amorphous Mg$_2$SiO$_4$ or MgSiO$_3$ grains that have low $Q_{\mathrm{abs}}$ values in the near-IR. For these dust species, from now on, we will consider that grains can be mixed with different volume fractions of carbon: 1, 5 and 20\% to represent for low-, intermediate-  and high-temperature grains. The need for dirty grains can be justified by dynamical process at stake in debris disks. Dust grains have most likely encountered several collisions during their lifetime, which renders the assumption of pure dust grains unlikely. Consequently we exclude from the modeling the absorption efficiencies with no impurities for Mg-, Fe-rich olivines, enstatite, silica and Fe-free amorphous grains. However, as we have no available diagnostics for this effect, we allow different levels of dirtiness for the aforementioned dust grains, by mixing different fractions of carbon. 

Figure\,\ref{fig:Qabs_all} shows the final absorption efficiencies (for $s =$\,0.1\,$\mu$m) used in this study to model the \textsc{Irs} spectra. With the exception of MgFeSiO$_4$ and MgFeSi$_2$O$_6$, the dotted, solid and dashed lines represent carbon admixtures of 1, 5 and 20\,\%, respectively, to account for the dirtiness of dust grains in a collisional environment. Introducing several impurities fractions reveals some of the unfortunate degeneracies of SED modeling, especially on the disk geometry. Spectroscopic observations contain a finite amount of information; the ``star + disk" system is not spatially resolved by the \textsc{Irs} instrument. By modeling the emission features, one can determine typical temperatures of the dust grains. However, the distance $r$ from the star, and hence the geometry of the system, will depend on how efficiently these grains can absorb the stellar radiation. In other words, if one does not know the dirtiness of grains, then one should not claim to constrain the exact morphology of the disk. At that point, spatially resolved observations become mandatory to break down the number of degeneracies on the morphology of the system in order to refine the model.

This underlines the overall complexity of modeling mid-IR spectra of debris disks in the optically thin regime. The need for dirty silicates, which is valid for all the sources in our sample, is the first result of this study. Even if the dust grains are not in thermal contact (one single temperature $T_d(r)$ for all species and sizes), pure grains as opposed to grains mixed with impurities, cannot provide satisfying results, as it would result either in too weak contributions to the emission or, in temperatures above the sublimation radius for other dust species. 

\subsection{Need for Fe-rich olivine grains\label{sec:ferich}}

\begin{figure*}
\begin{center}
\includegraphics[angle=0,width=\columnwidth,origin=bl]{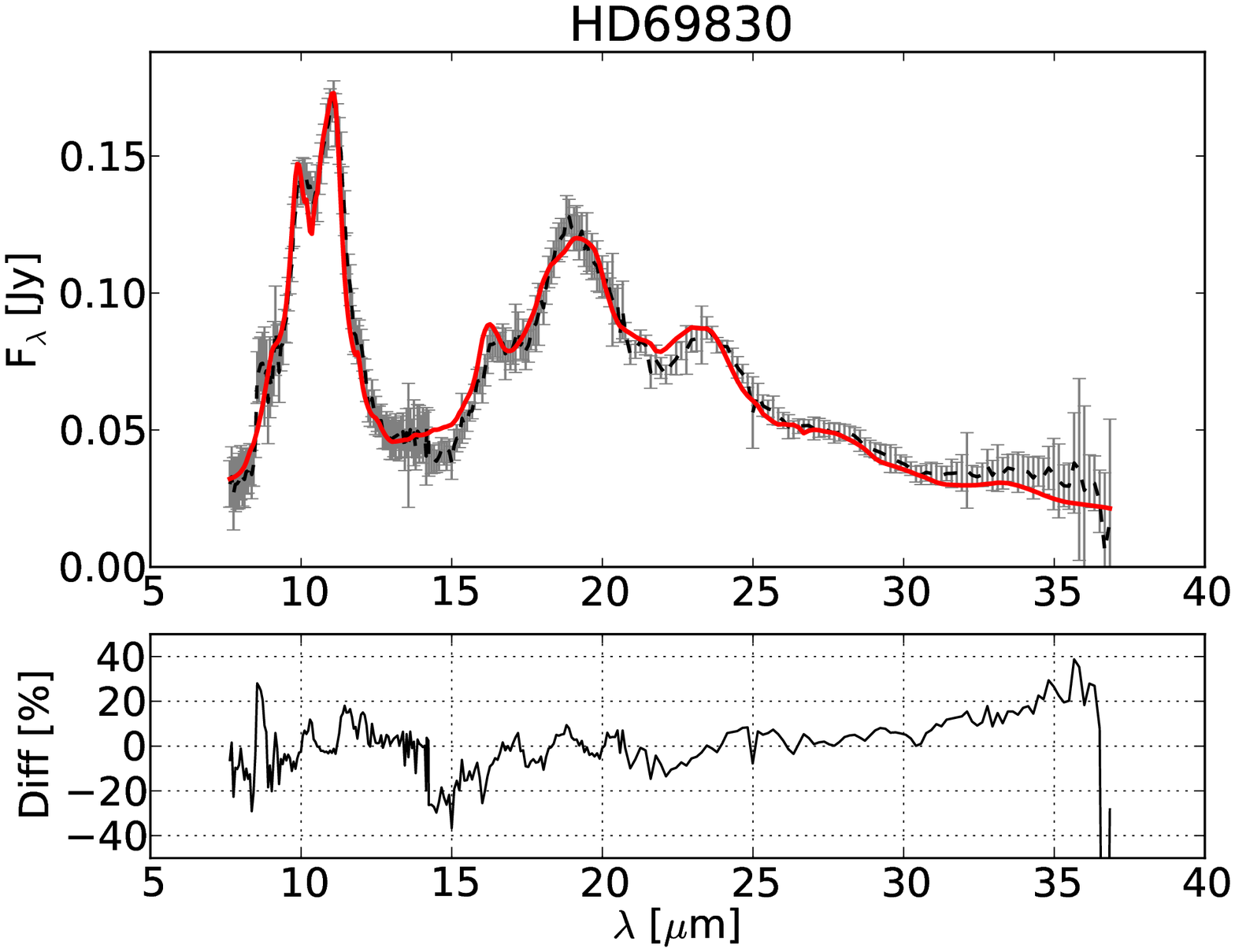}
\includegraphics[angle=0,width=\columnwidth,origin=bl]{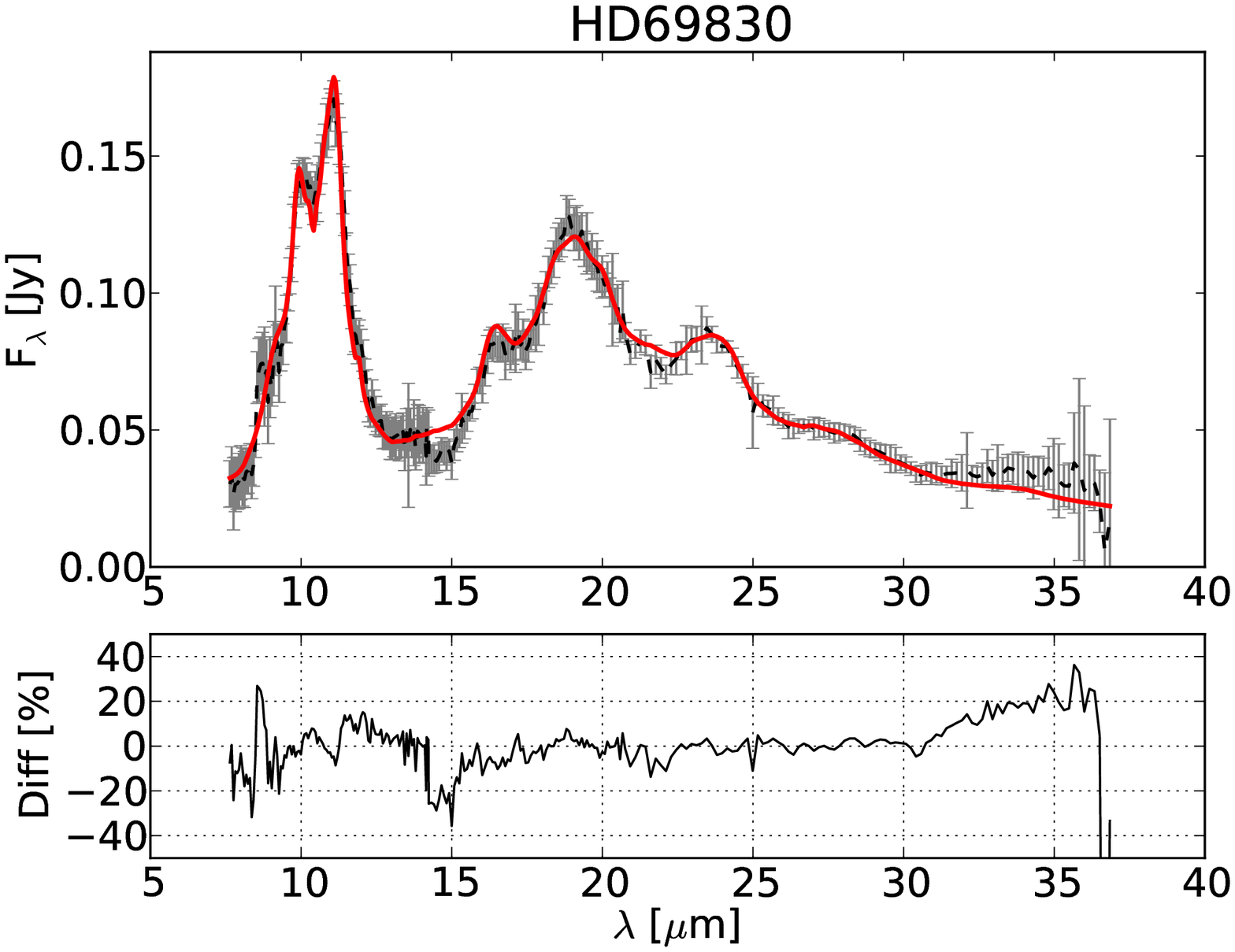}
\includegraphics[angle=0,width=\columnwidth,origin=bl]{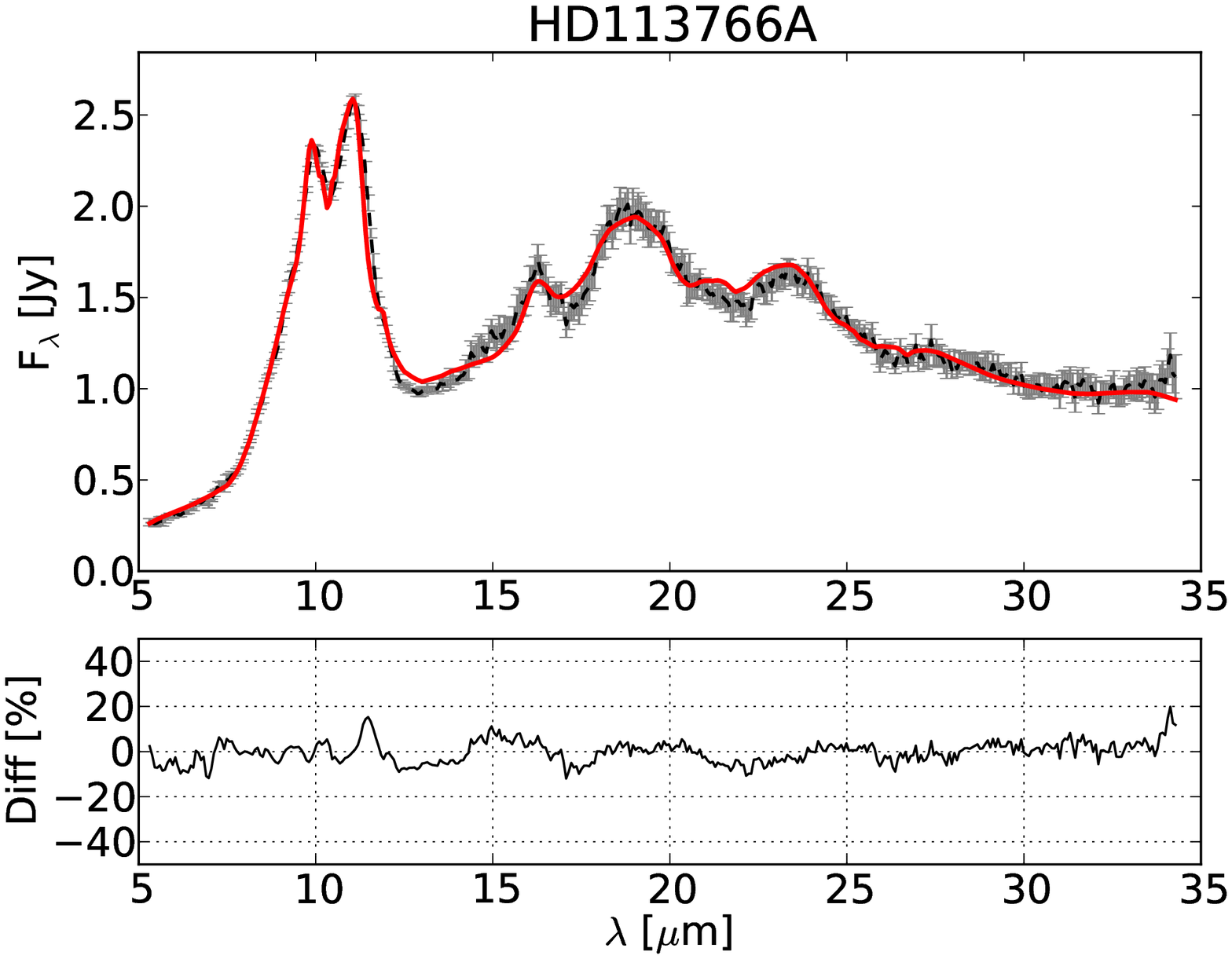}
\includegraphics[angle=0,width=\columnwidth,origin=bl]{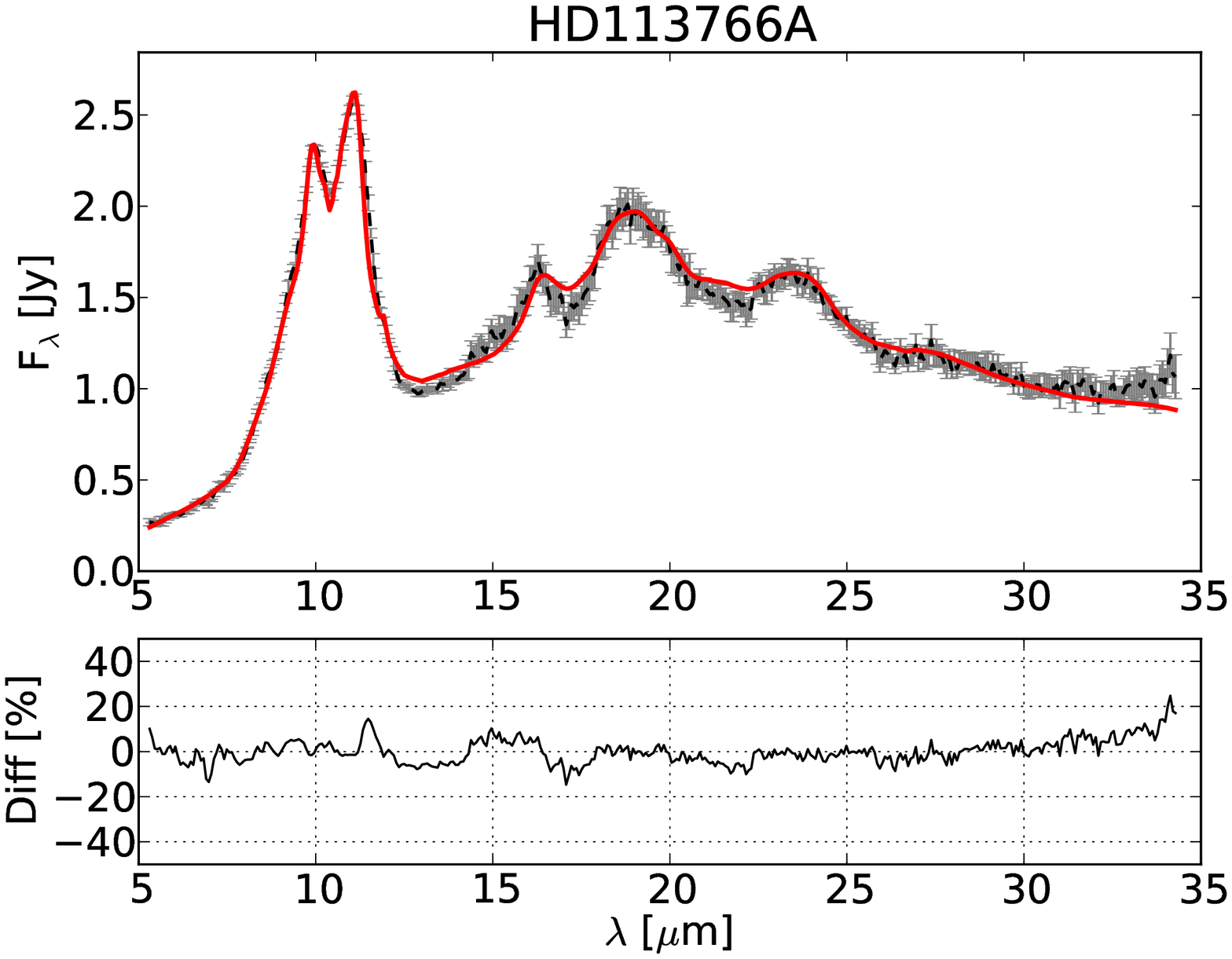}
\caption{\label{fig:ferich}{\it Left panels:} fits to stellar subtracted spectra of HD\,69830 and HD\,113766\,A ({\it top} and {\it bottom}, respectively), without Fe-rich olivine grains. {\it Right panels:} same, as left panels including Fe-rich olivine grains in the modeling. For each panel, residuals are shown at the bottom (see Sect.\,\ref{sec:indiv}).}
\end{center}
\end{figure*}

From Figure\,\ref{fig:qSoTa} and the corresponding discussion (Sect.\,\ref{sec:forst}), aerosol measurements of Mg- and Fe-rich olivine grains, seem to provide a better match to the observed \textsc{Irs} spectra (Fig.\,\ref{fig:11mu}). From several previous studies about dust mineralogy in various circumstellar environments, we know that crystalline olivine grains are usually Mg-rich rather than Fe-rich (\citealp{Bouwman2008}, \citealp{Juh'asz2010}, \citealp{Olofsson2010}). Under this assumption, we first run the \textsc{Debra} code including only Mg-rich olivine grains. This choice being motivated by the comparison of peak positions of several emission features (Fig.\,\ref{fig:peak}).  Results for the two most relevant examples, HD\,69830 and HD\,113766\,A, are shown in left panels of Fig.\,\ref{fig:ferich}. For these two sources, the peak positions and shapes of emission features at 19 and 24\,$\mu$m do not perfectly match the observed spectra. We therefore repeat this exercise, this time including the contribution of Fe-rich olivine grains, and the results are shown in right panels of Fig.\,\ref{fig:ferich}. Since the differences between laboratory data of Mg- and Fe-rich olivine grains are overall small (see Fig.\,\ref{fig:qSoTa}), the fits to the data do not change drastically. However, the peak positions and shapes of emission features at 19 and 24\,$\mu$m are much better reproduced when including Fe-rich olivine grains (for HD 69830 this has also been shown by \citealp{Tamanai2010}). 

\begin{figure}
\begin{center}
\includegraphics[angle=0,width=\columnwidth,origin=bl]{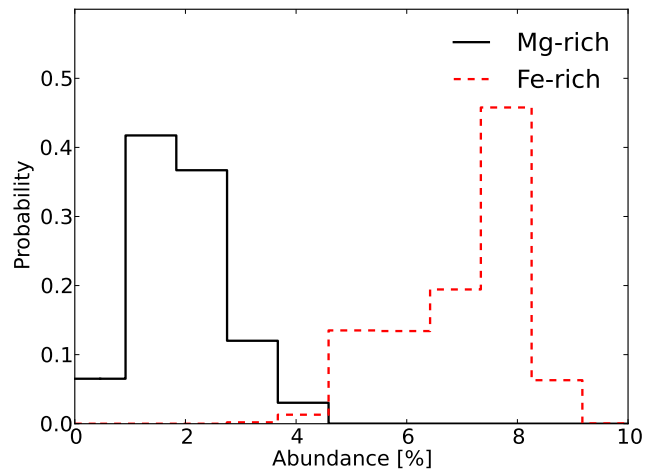}
\caption{\label{fig:fe_proba}Probability distributions from MCMC run for HD\,113766\,A, as a function of relative abundances of Mg- and Fe-rich olivine grains (solid black, and dashed red, respectively).}
\end{center}
\end{figure}

This result is further supported by the uncertainty estimation for the relative abundances. Figure\,\ref{fig:fe_proba} shows two probability distributions from the MCMC run for HD\,113766\,A. The solid black histogram shows the probability distribution as a function of the abundance of Mg-rich olivine grains, while the dashed red histogram shows the probability distribution as a function of the abundance of Fe-rich olivine grains. It immediately appears from Figure\,\ref{fig:fe_proba} that abundances of Fe-rich grains must be larger compared to abundances of Mg-rich grains in order to provide a good fit to the data. Following the results from this exercise, we decide to include aerosol measurements of Fe-rich olivine grains in the modeling of all the sources discussed in this study.

As shown in both Figures\,\ref{fig:qSoTa} and \ref{fig:peak}, aerosol measurements of synthesized forsterite and Mg-rich olivine grains are pretty similar. In order to distinguish between them, one would need a precision of a few tenths of $\mu$m ($\sim$\,0.1\,$\mu$m) on the peak positions of emission features. As explained previously, since we prefer to limit the number of degenerate parameters, we opt not to use laboratory measurements of synthesized forsterite grains (which seem less representative to the data, see Fig.\,\ref{fig:peak}). This choice has for consequence that we are not able to conclude on Fe fractions smaller than 7.25\%. To be as conservative as possible, we will only claim that crystalline olivine grains have been enriched in Fe if the relative abundances of Fe-rich olivine grains are larger than those of Mg-rich olivine grains.

\subsection{Continuum emission\label{sec:smax}}

When modeling the dust mineralogy of optically thick protoplanetary disks, a continuum component is included in the modeling (e.g., \citealp{Juh'asz2010}, \citealp{Olofsson2010}). This continuum is originally meant to account for the emission of optically thick regions of the disk (the emission features arising from the optically thin layers). As grains larger than $\sim$10\,$\mu$m do not show significant emission features, they are usually not included in the spectral decomposition procedure. Their actual presence in the proto-planetary disk is not questioned, but their contribution to the emission is assumed to be part of the continuum. This assumption no longer holds for debris disks that are optically thin at all wavelengths, therefore the influence of the maximum grain size has to be tested.

First, one should note that, at a given distance $r$ from the star, 100 or 1000\,$\mu$m-sized grains are colder than $\mu$m-sized grains. The emission from large grains will therefore be more important at longer wavelengths (since they are colder). As grains larger than typically 10\,$\mu$m have featureless spectra, if we aim at modeling the emission features detected in spectra, the simplest assumption is to opt for $s_{\mathrm{max}}=$\,10\,$\mu$m. In that case, preliminary tests have shown there are a couple of possible ways to compensate for the absence of large grains.  The first possibility is an increased abundance of carbon grains, which have featureless spectra as well and can serve as continuum. However, given that $Q_{\mathrm{abs}}$ values of amorphous carbon grains are rather high in the near-IR, such grains will have high temperatures and will mostly contribute at short wavelengths. Increasing the width (or $r_{\mathrm{out}}$) of the disk may circumvent this issue since the outermost carbon grains will be colder and can contribute as a ``colder" continuum. However it may provide a worst fit to the emission features. We assume the dust composition to be homogeneous over the entire disk (same dust composition at all radii), therefore if $r_{\mathrm{out}}$ increases by too much, some of the crystalline grains (e.g., Mg-, Fe-rich olivine grains) will be colder as well. This will boost their emission features at longer wavelengths (e.g., 33\,$\mu$m). For all the objects (except HD\,169666), such features are not detected, which practically rule out the presence of such cold crystalline grains.

\begin{figure}
\begin{center}
\hspace*{-0.5cm}\includegraphics[angle=0,width=\columnwidth,origin=bl]{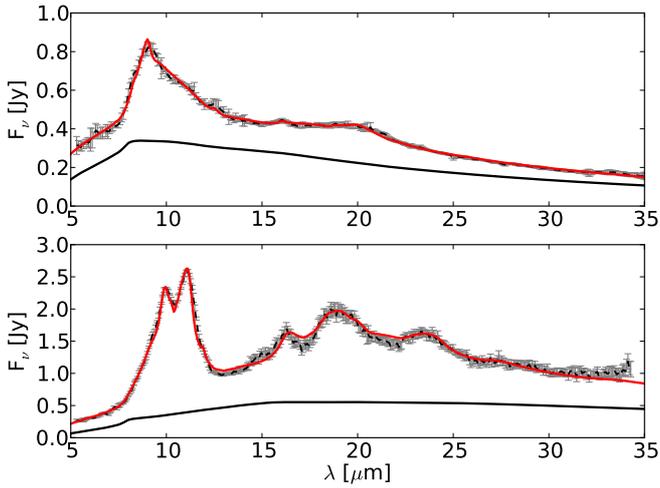}
\caption{\label{fig:smax}Fits to stellar subtracted spectra of HD\,15407\,A (top) and HD\,113766\,A (bottom). Fits are in red, data in dashed black. Contribution of grains larger than 10\,$\mu$m is shown in solid black.}
\end{center}
\end{figure}

The second possibility to compensate for the absence of large grains is to mimic their presence by a shallower grain size distribution exponent. One can understand $p$ as a competition between continuum emission and emission features. A steep grain size distribution will favor emission from small grains and their associated emission features, while a shallow $p$ will favor contribution from large grains and will produce more ``continuum" emission (large grains having featureless emission profiles). A shallow $p$ value will decrease the relative number of small $\mu$m-sized grains, and this will affect the total emission profile of amorphous dust grains. As an example, when fitting the spectrum of HD\,15407\,A with $s_{\mathrm{max}} =10$\,$\mu$m, we obtain a value of $p = -1.70$, which is far from $p=-3.5$ expected for a collisional disk (\citealp{Dohnanyi1969}). Such result would most likely require unusual constraints on the dynamical processes at stake in the disk. We therefore estimate the most straightforward (and less biased) solution is to simply increase $s_{\mathrm{max}}$ up to 1000\,$\mu$m for the silicate and carbon amorphous grains, so that such large grains can account for the continuum emission. Since we do not expect grain size distributions to be truncated for grain sizes $s \geq$ 10\,$\mu$m, this assumption appears to be a faithful description of grain size distribution in debris disks. In Figure\,\ref{fig:smax}, we show fits to stellar subtracted spectra of HD\,15407\,A and HD\,113766\,A (top and bottom panels, respectively). The solid black lines represent the flux contribution of silicate and carbon amorphous grains with sizes between $10 \leq s \leq 1000$\,$\mu$m. The flux fractions over the \textsc{Irs} range are of about $\sim$60\% and 38\% of the total flux, for both objects. The conclusion of this exercise is that grains larger than 10\,$\mu$m have to be included, when modeling disks in the optically thin regime. A value of $s_{\mathrm{max}} =$\,1000\,$\mu$m is sufficient and provide good results.

\section{Individual source descriptions\label{sec:indiv}}

\begin{figure*}
\begin{center}
\includegraphics[angle=0,width=0.95\columnwidth,origin=bl]{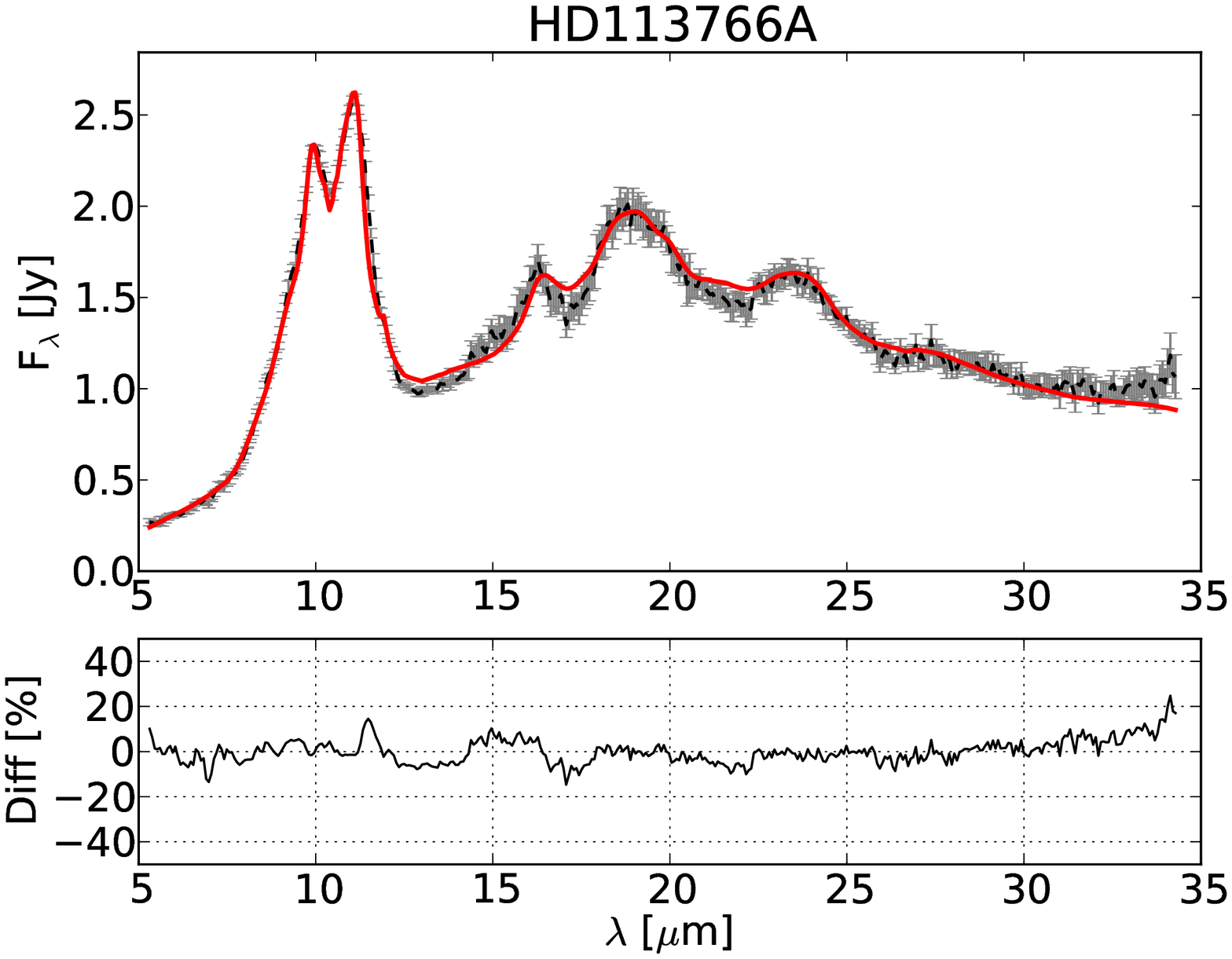}
\includegraphics[angle=0,width=0.95\columnwidth,origin=bl]{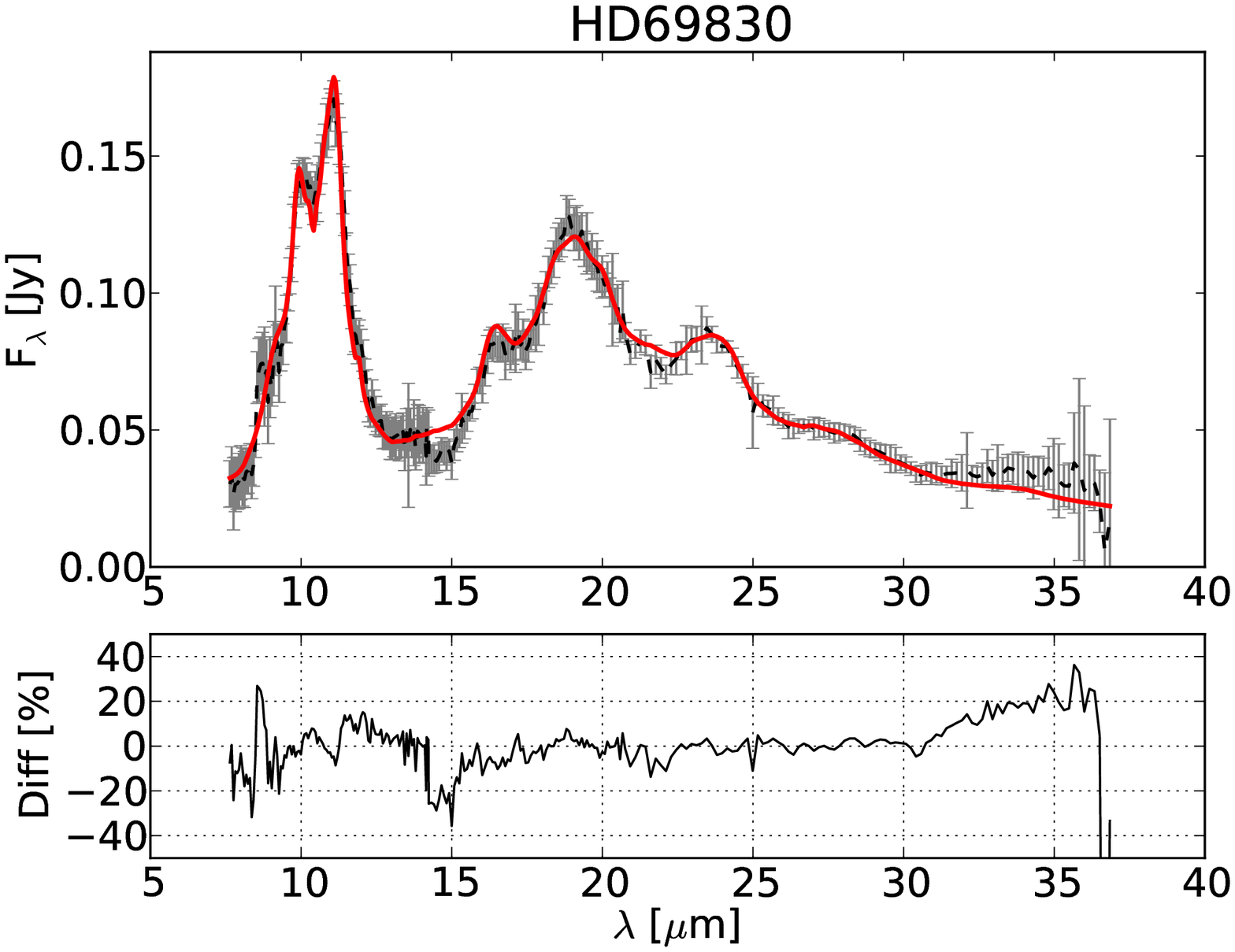}
\includegraphics[angle=0,width=0.95\columnwidth,origin=bl]{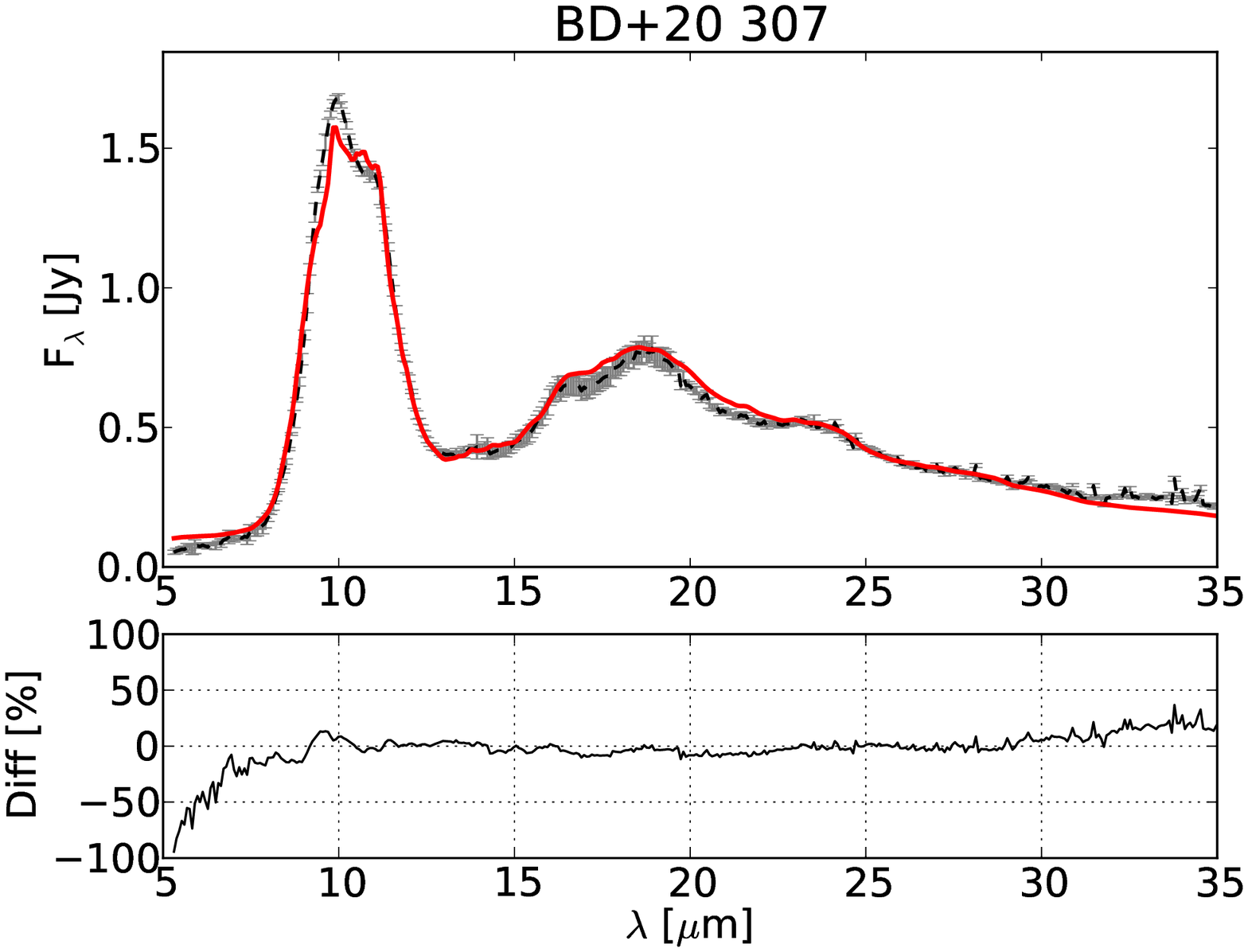}
\includegraphics[angle=0,width=0.95\columnwidth,origin=bl]{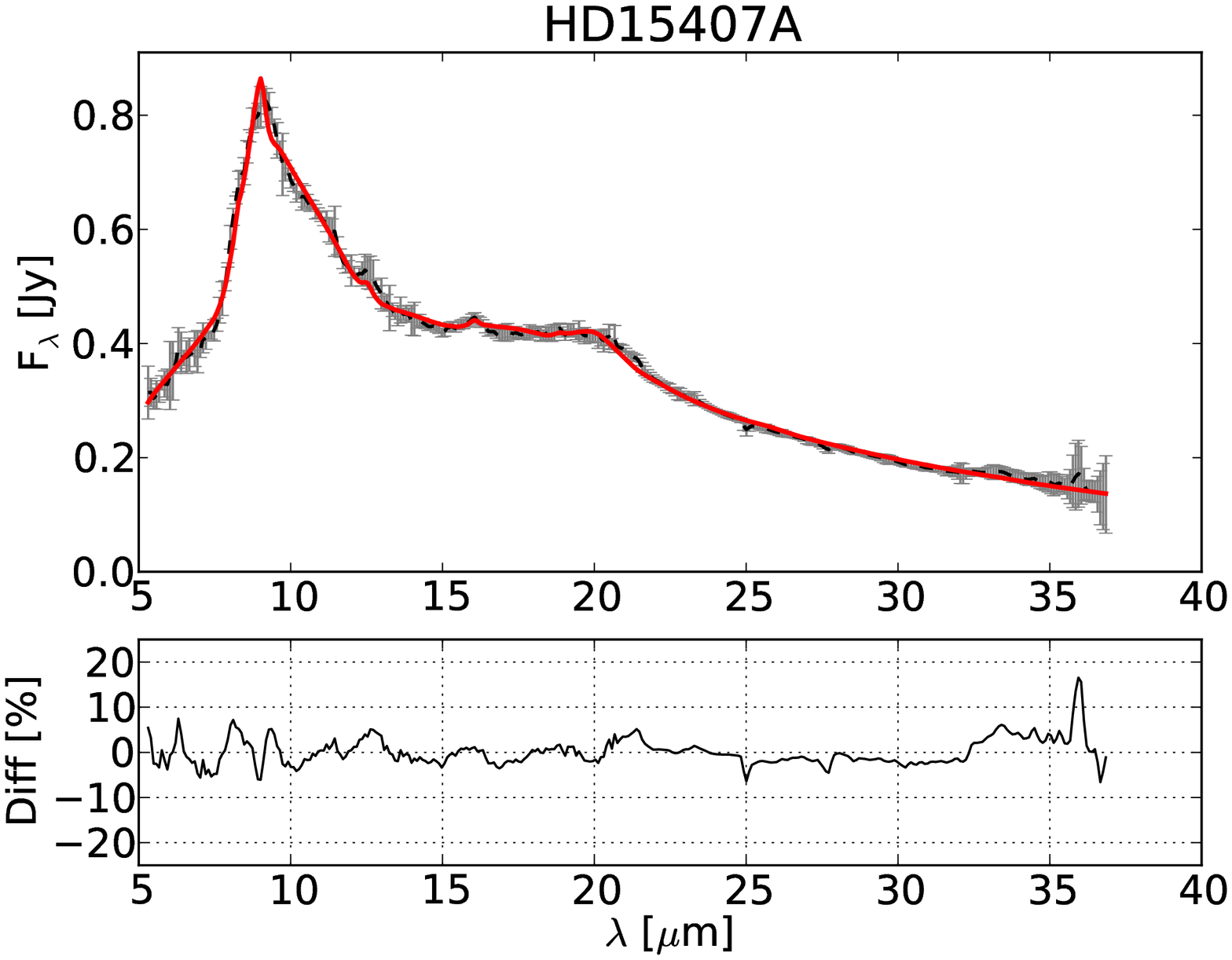}
\includegraphics[angle=0,width=0.95\columnwidth,origin=bl]{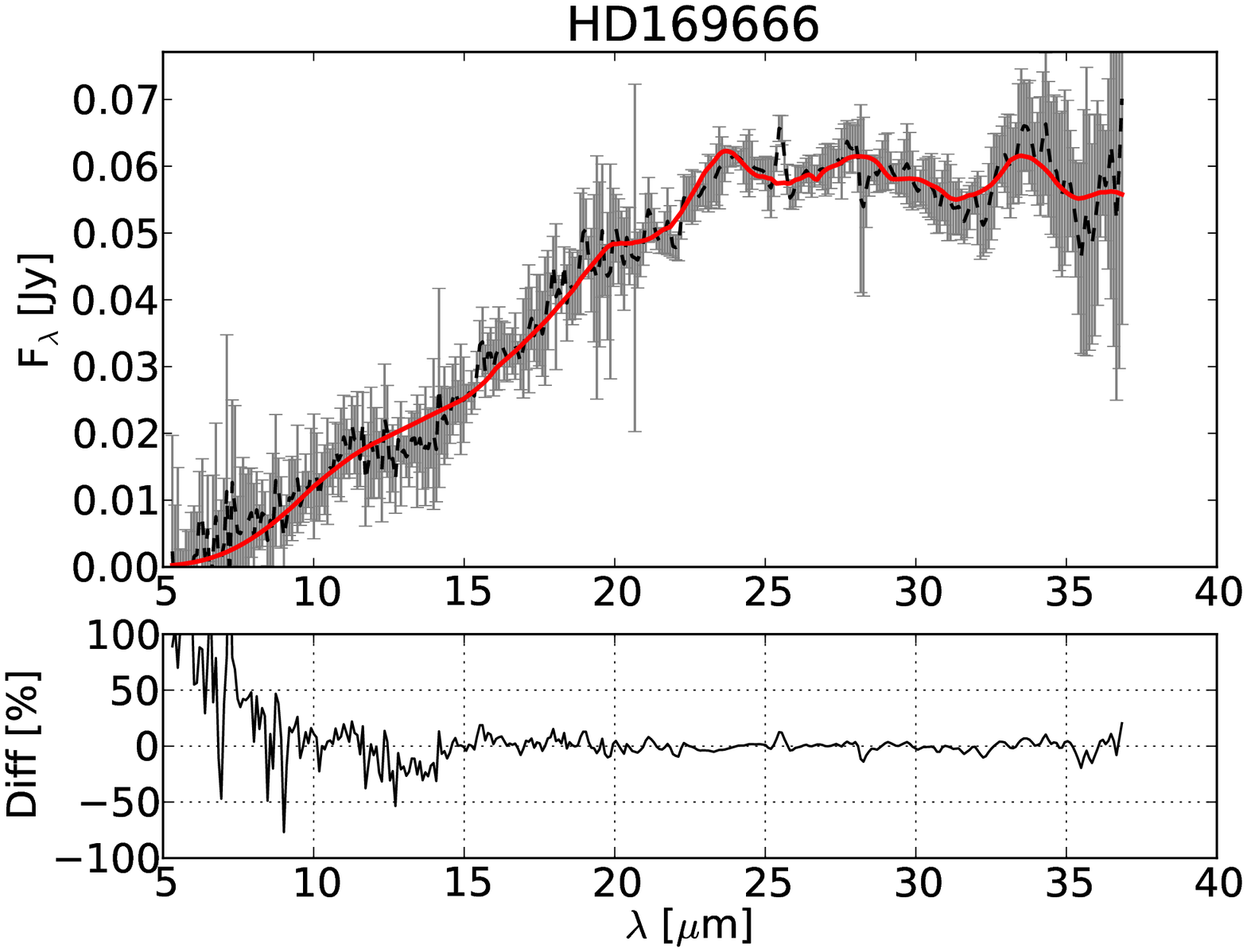}
\includegraphics[angle=0,width=0.95\columnwidth,origin=bl]{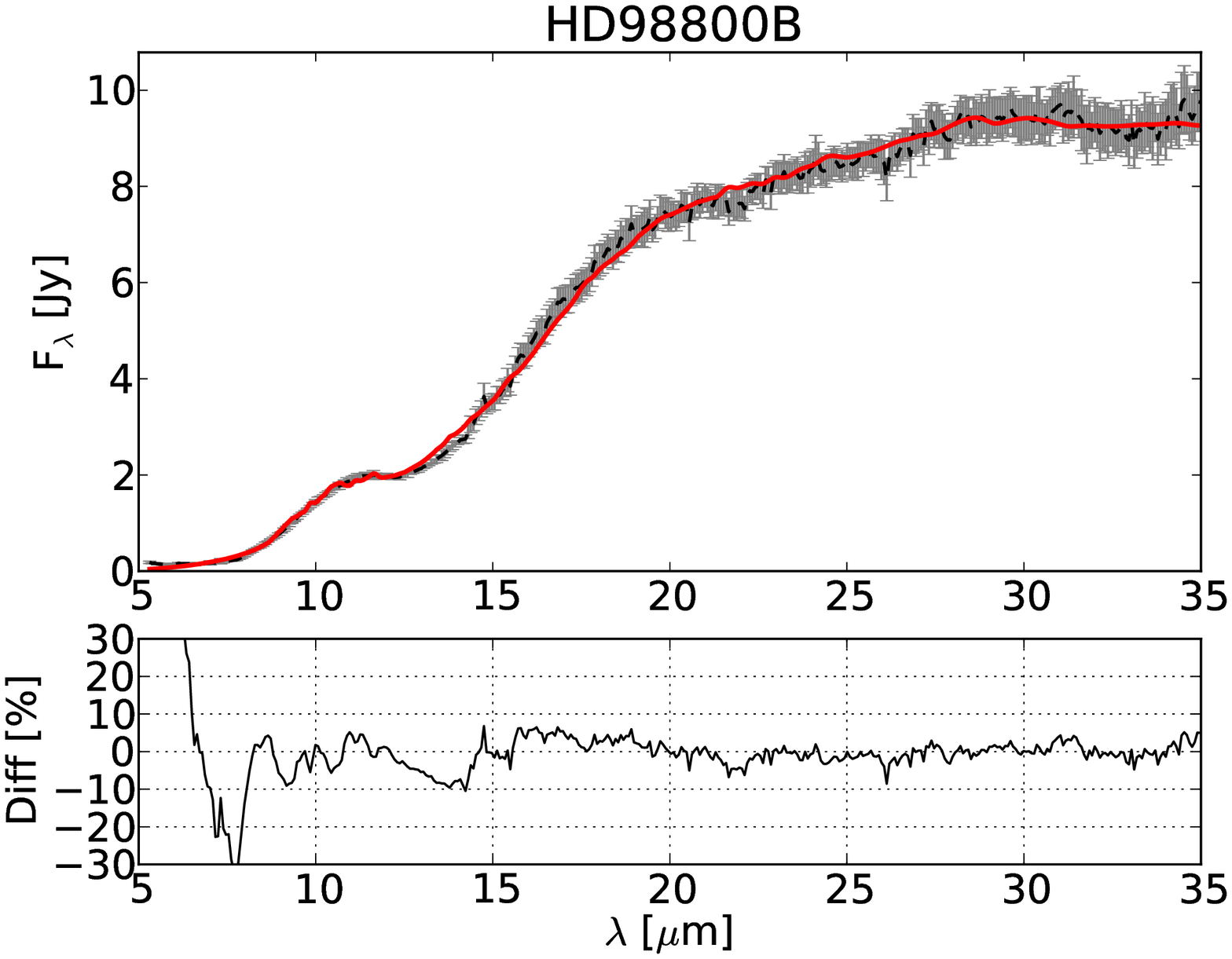}
\includegraphics[angle=0,width=0.95\columnwidth,origin=bl]{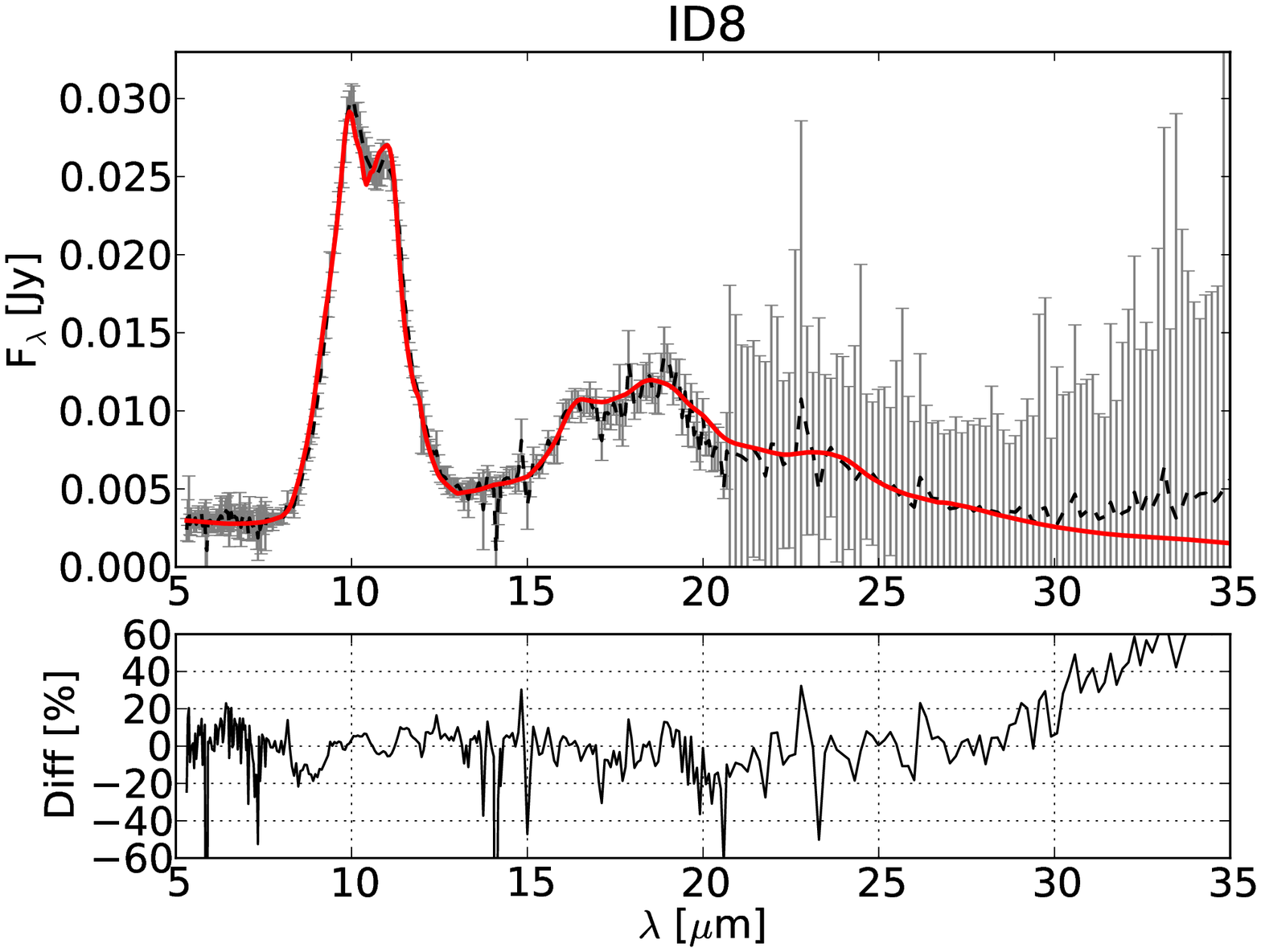}
\caption{\label{fig:fit_1}Best fits (in red) to the \textsc{Irs} spectra (dashed lines). Uncertainties are 3-$\sigma$. For each source bottom panels show the residuals.}
\end{center}
\end{figure*}

\begin{figure*}
\begin{center}
\includegraphics[angle=0,width=0.95\columnwidth,origin=bl]{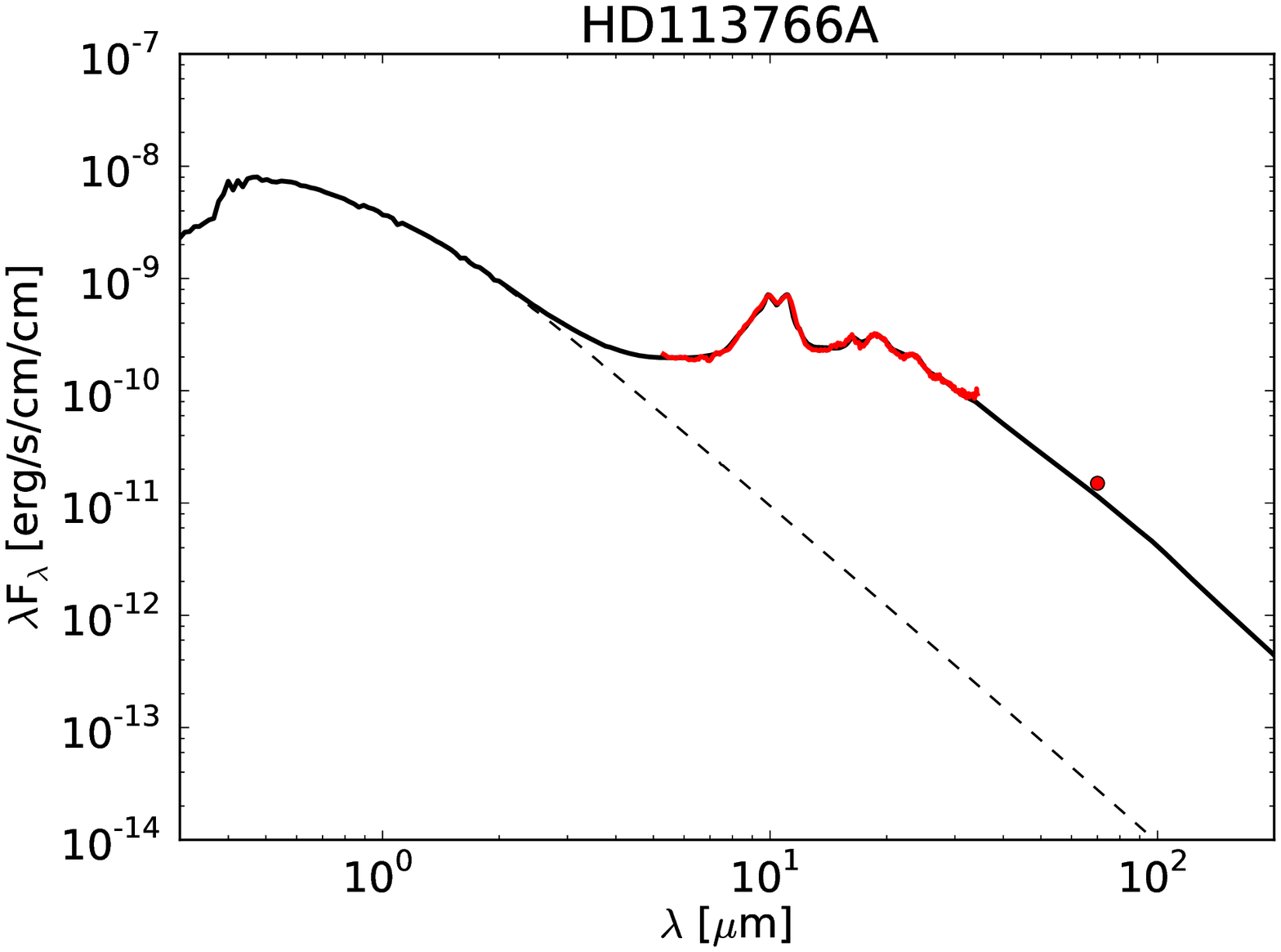}
\includegraphics[angle=0,width=0.95\columnwidth,origin=bl]{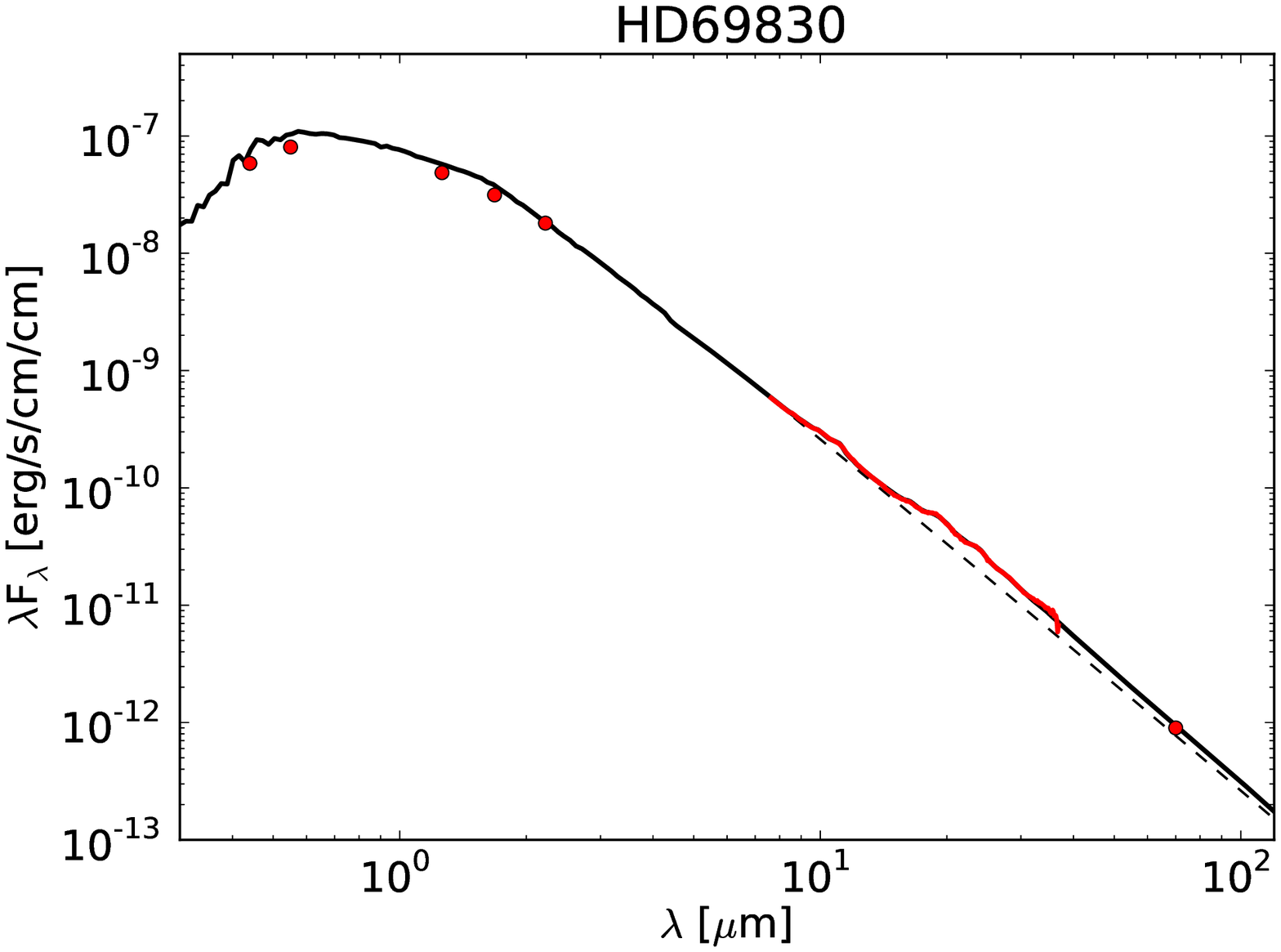}
\includegraphics[angle=0,width=0.95\columnwidth,origin=bl]{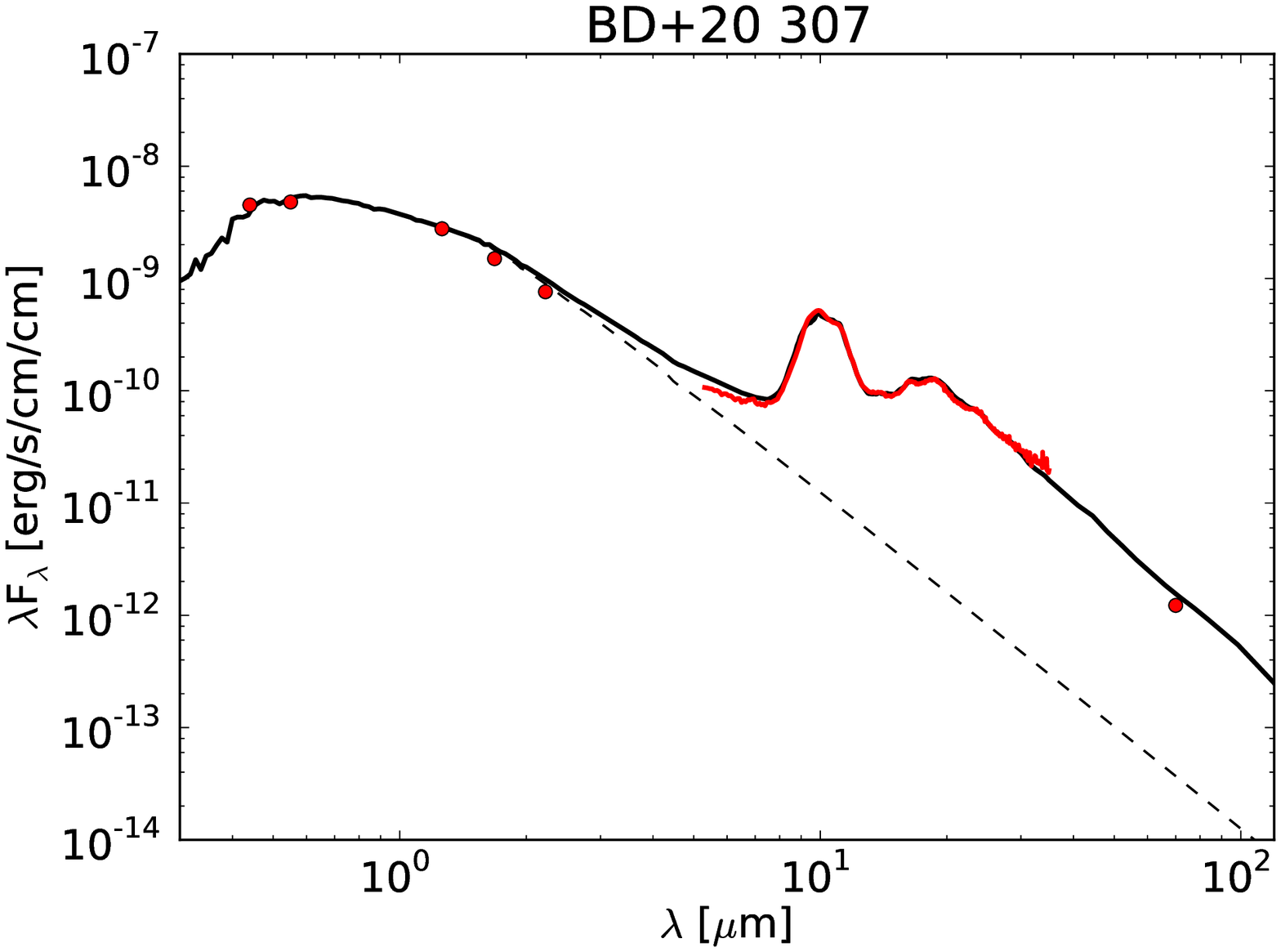}
\includegraphics[angle=0,width=0.95\columnwidth,origin=bl]{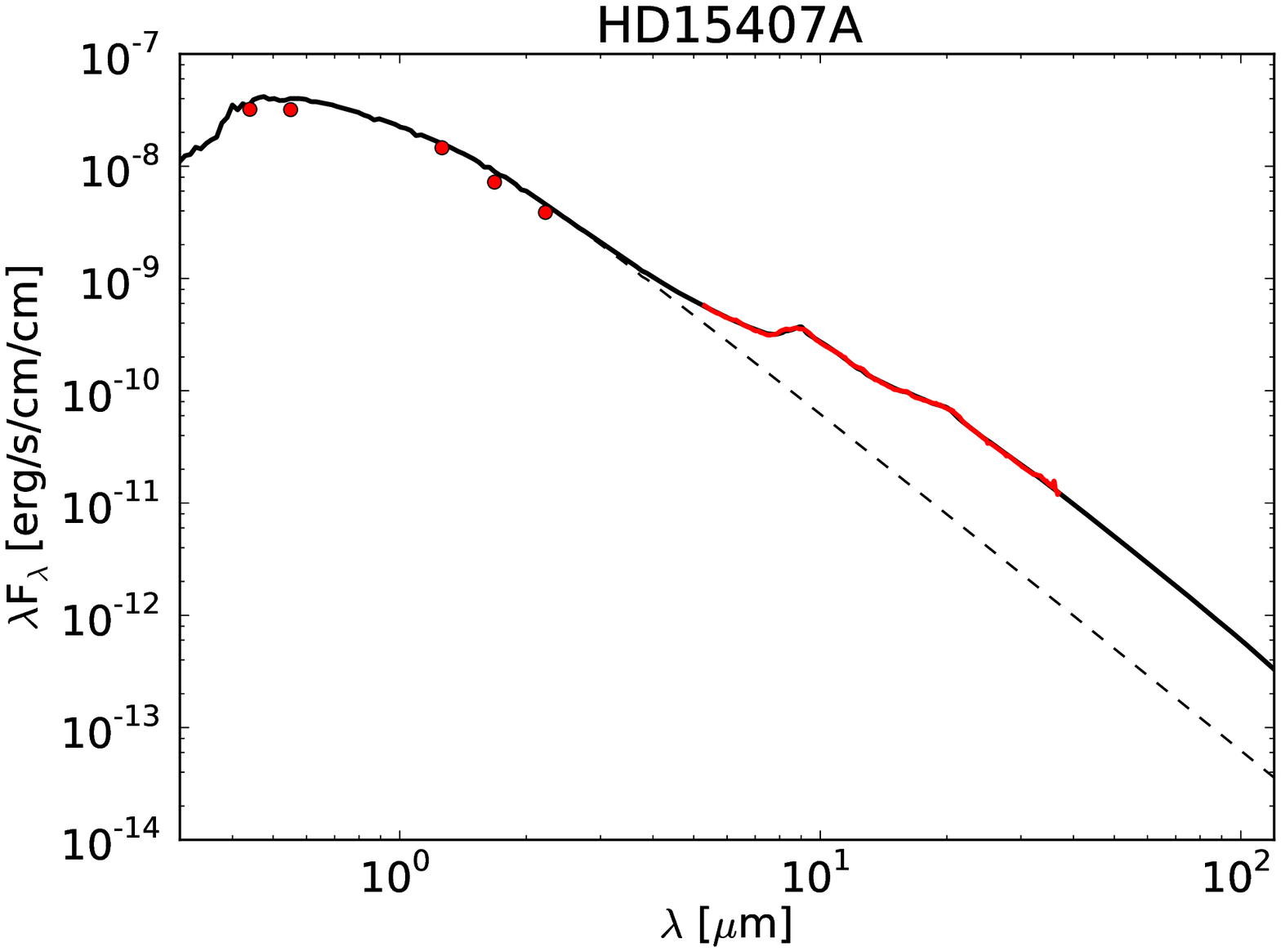}
\includegraphics[angle=0,width=0.95\columnwidth,origin=bl]{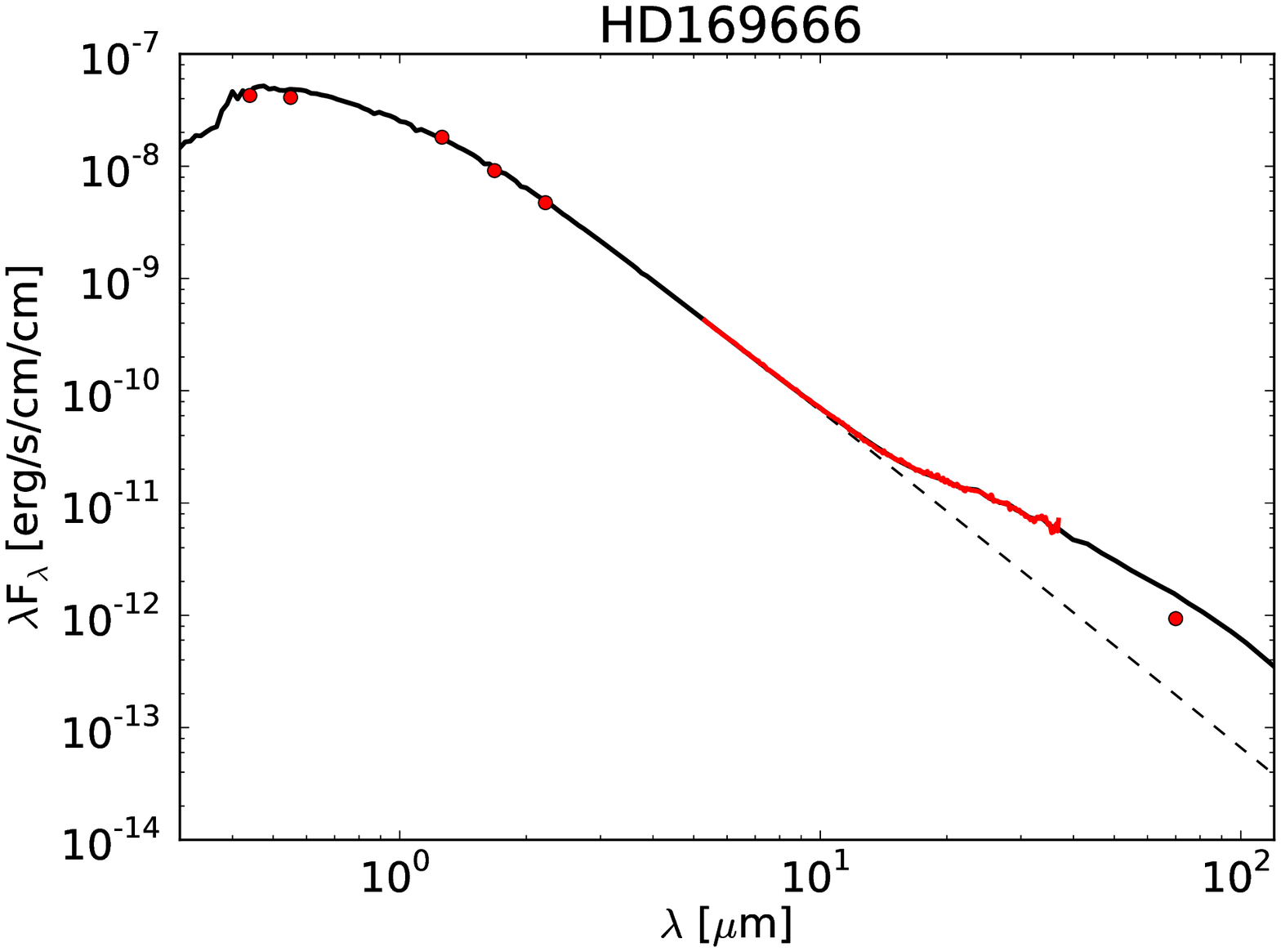}
\includegraphics[angle=0,width=0.95\columnwidth,origin=bl]{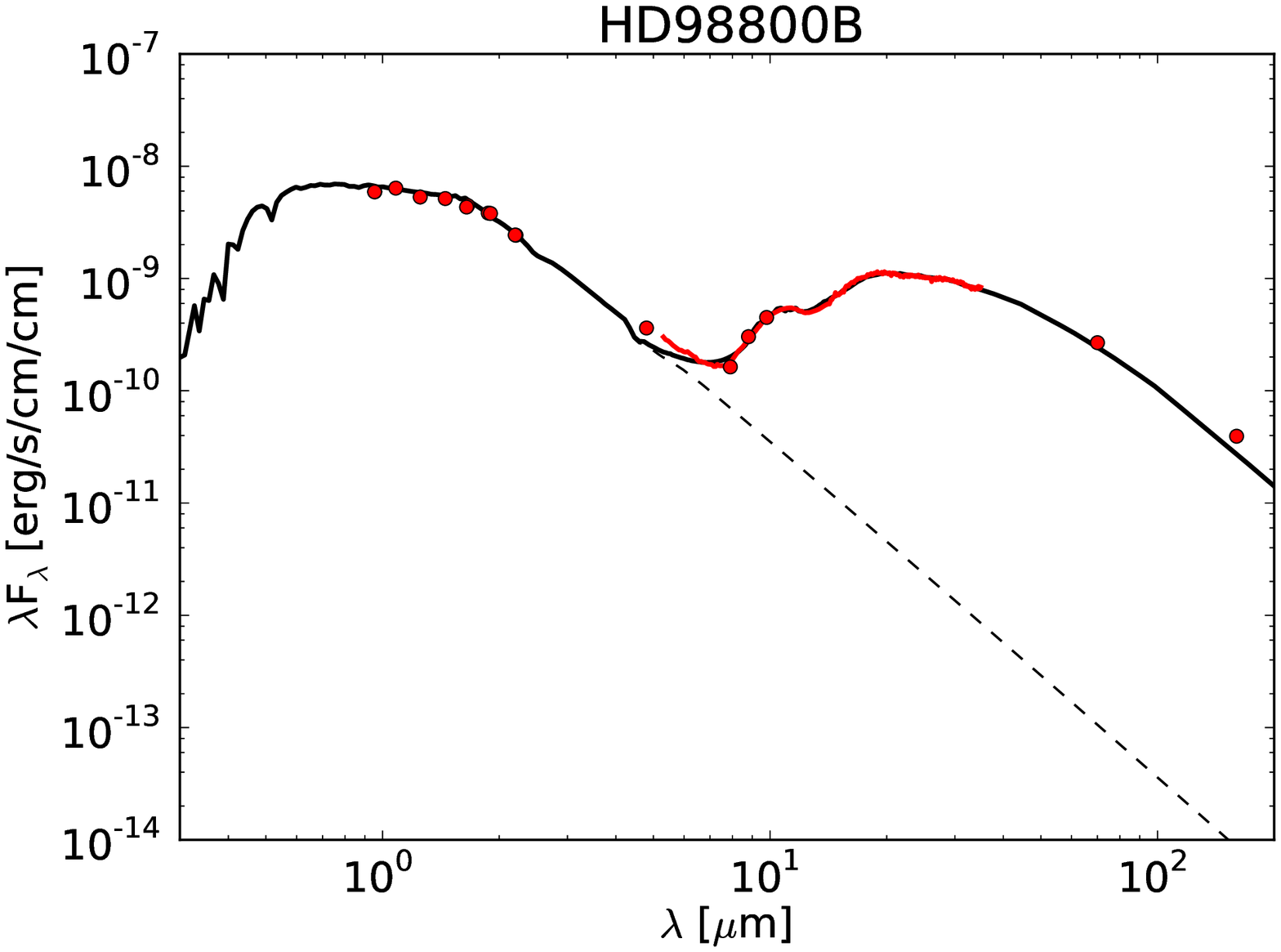}
\includegraphics[angle=0,width=0.95\columnwidth,origin=bl]{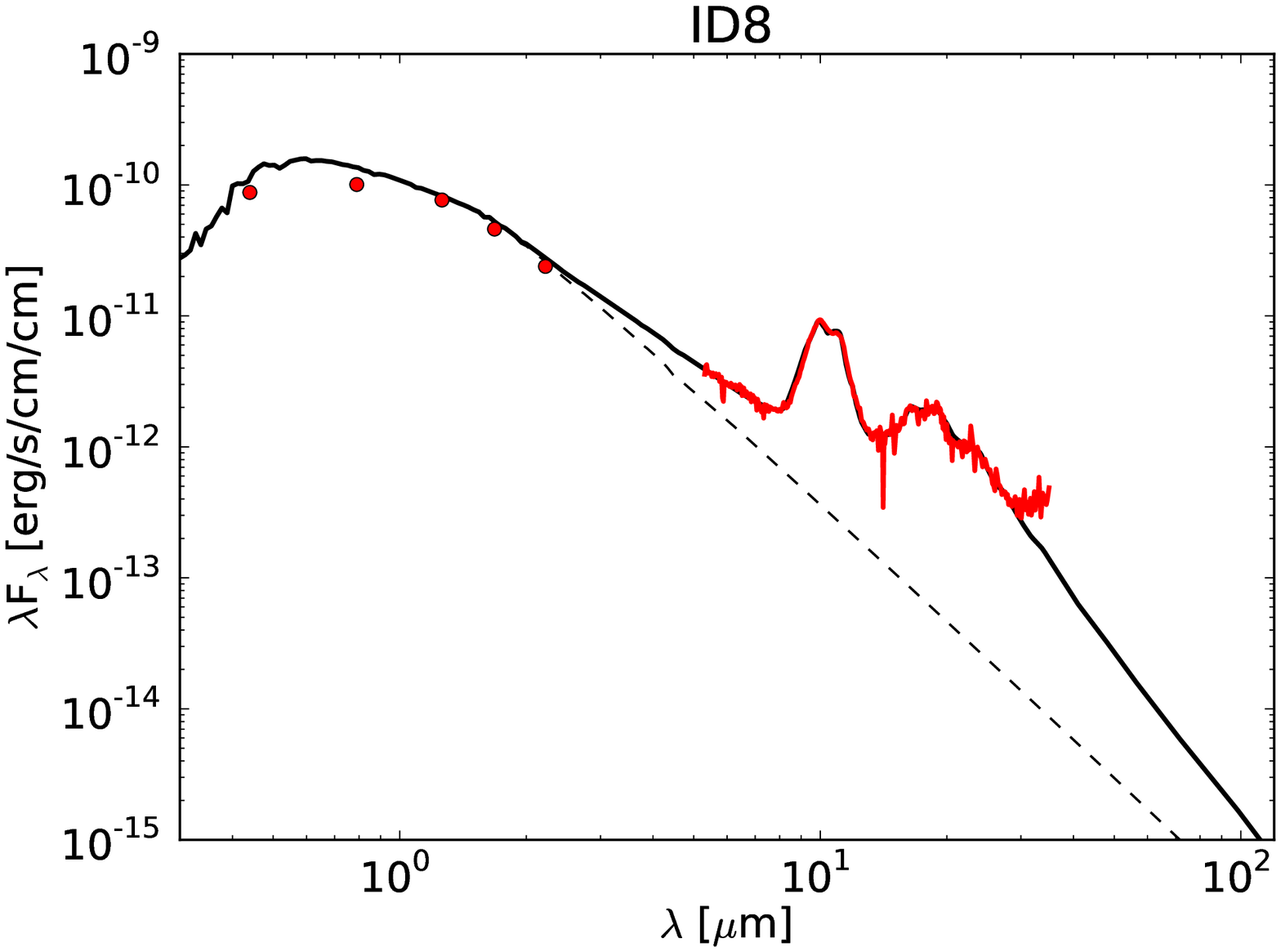}
\caption{\label{fig:SED_1}Complete SEDs for all the sources, with \textsc{Irs} spectra and photometric observations in red. Photospheres are in dashed black and modeled SEDs are in solid black.}
\end{center}
\end{figure*}

As several sources in our sample are well-studied debris disks, we present and discuss in the following the best fits we obtain for each source and compare our results to previous studies. One should note that we focus our study solely on Spitzer/\textsc{Irs} data and we do not include other measurements such as \textsc{Mips} photometric observations. But when available, we compare the flux predictions of our best model with these measurements. Tables\,\ref{tbl:param} and \ref{tbl:comp} show the disk parameters and relative abundances for the best models, respectively. Figure\,\ref{fig:fit_1} shows the fits to the stellar subtracted spectra as well as the residuals from the fits, which are computed as follows:
\begin{eqnarray}
\mathrm{Diff~[\%]} = 100 \times \frac{F_{\nu, \mathrm{obs}} - F_{\nu, \mathrm{model}}}{F_{\nu, \mathrm{obs}}},
\end{eqnarray}
with $F_{\nu, \mathrm{obs}}$ and $F_{\nu, \mathrm{model}}$ being the observed and modeled fluxes, respectively. Figures\,\ref{fig:SED_1} shows the entire SEDs as modeled using the \textsc{Debra} code. We check the optical depth is always smaller than unity meaning all sources can be modeled within the optical thin regime. 

\begin{table*}
  \begin{center}
    \caption{\label{tbl:param}Disk parameters, and corresponding uncertainties, for the best fits, as well as the maximum temperature reached}
    \begin{tabular}{lcccccc}
      \hline \hline
      Star & $r_{\mathrm{in}}$ & $r_{\mathrm{out}}$ & $\alpha$ & $p$ & $M_{\mathrm{dust}}$ & $T_{\mathrm{max}}$ \\
       & [AU] & [AU] &  & & [$M_{\oplus}$] & [K] \\
      \hline
HD\,113766\,A  &  0.6\,$\pm$\,0.1  &  11.8\,$\pm$\,2.8  &  $-1.77$\,$\pm$\,0.1  &  $-3.17$\,$\pm$\,0.1  &  7.05$\times 10^{-3 }$ &  1171  \\
HD\,69830  &  0.7\,$\pm$\,0.1  &  1.5\,$\pm$\,0.2  &  $-2.47$\,$\pm$\,0.8  &  $-3.54$\,$\pm$\,0.1  &  3.70$\times 10^{-7 }$ &  684  \\
BD+20\,307  &  0.2\,$\pm$\,0.1  &  18.2\,$\pm$\,2.4  & $-2.41$\,$\pm$\,0.1  &  $-3.38$\,$\pm$\,0.1  &  3.75$\times 10^{-4 }$ &  1405  \\
HD\,15407\,A  &  0.4\,$\pm$\,0.15  &  19.2\,$\pm$\,5.3  &  $-3.00$\,$\pm$\,0.8  &  $-3.12$\,$\pm$\,0.1  &  7.73$\times 10^{-5 }$ &  1531  \\
HD\,169666  &  52.7\,$\pm$\,18.7  &  68.3\,$_{-15.6}^{+16.3}$  &  $-2.44$\,$\pm$\,0.9  &  $-3.59$\,$\pm$\,0.3  &  1.19$\times 10^{-3 }$ &  215  \\
HD\,98800\,B  &  1.3\,$\pm$\,0.3  &  7.2\,$\pm$\,1.9  &  $-1.12$\,$\pm$\,0.5  &  $-2.99$\,$\pm$\,01  &  5.76$\times 10^{-2 }$ &  536  \\
ID\,8  &  0.39$\pm$\,+0.07 &  0.52\,$\pm$\,0.12  &  $-2.2$\,$\pm$\,0.9  &  $-4.01$\,$\pm$\,0.2  &  2.37$\times 10^{-6 }$ &  864  \\
      \hline
   \end{tabular}
  \end{center}
\end{table*}

\begin{table*}
  \begin{center}
    \caption{\label{tbl:comp}Relative abundances (\%) and corresponding uncertainties for the best fits}
    \begin{tabular}{lccccccccc}
      \hline \hline
      Star & MgFeSiO$_4$ & MgFeSi$_2$O$_6$ & MgSiO$_3$ & Mg$_2$SiO$_4$ & Fe-rich & Mg-rich & Enstatite & $\beta$-cristobalite & C \\
      & amorphous & amorphous & amorphous & amorphous & olivine & olivine & & & \\
      \hline
HD\,113766\,A & 0\,$+$\,0.6 & 45.6\,$\pm$\,3.1 & 0\,$+$\,0.7 & 29.4\,$\pm$\,1.8 & 7.6\,$\pm$\,1.1 & 1.7\,$\pm$\,0.9  &  0.8\,$\pm$\,0.8 & 0\,$+$\,0.7 & 14.9\,$\pm$\,2.6 \\
HD\,69830 & 24.3\,$\pm$\,5.9 &  29.9\,$\pm$\,0.7 & 0\,$+$\,0.7 & $3.7^{+4.2}_{-3.7}$ & 6.5\,$\pm$\,1.8 & 1.4\,$\pm$\,0.6 & 2.1\,$\pm$\,0.7 & 2.6\,$\pm$\,2.6 & 29.5\,$\pm$\,6.2 \\
BD+20\,307 & 68.6\,$\pm$\,2.5 & 15.4\,$\pm$\,0.9 & 0\,$+$\,0.6 & 7.3\,$\pm$\,1.7 & 4.0\,$\pm$\,0.7 & $0.6\,_{-0.6}^{+0.7}$ & 4.1\,$\pm$\,0.8 & 0\,$+$\,0.7 & 0\,$+$\,0.7 \\
HD\,15407\,A & 0\,$+$\,0.6 & 25.0\,$\pm$\,5.4 & 28.0\,$\pm$\,7.2 & 0\,$+$\,0.6 & 0\,$+$\,0.5 & 0\,$+$\,0.8 & 0\,$\pm$\,0.6 & 17.3\,$\pm$\,2.1 & 29.7\,$\pm$\,12.8 \\
HD\,169666 & 0\,$+$\,0.6 & 0\,$+$\,0.7 & 0\,$+$\,0.7 & 65.1\,$\pm$\,0.7 & 0\,$+$\,0.8 & 6.9\,$\pm$\,3.4 & 7.7\,$\pm$\,1.2 & 3.6\,$\pm$\,0.7 & 16.7\,$\pm$\,6.5 \\
HD\,98800\,B & 25.9\,$\pm$\,0.6 & 28.0\,$\pm$\,0.7 & 0\,$+$\,0.5 & 22.6\,$\pm$\,6.7 & 0\,$+$\,0.8 & 0.2\,$^{+0.8}_{-0.2}$ & 5.2\,$\pm$\,1.5 & 0\,$+$\,0.7 & 18.2\,$\pm$\,6.5 \\
ID\,8 & 26.4\,$\pm$\,12.0 & 0\,$+$\,0.6 & 0\,$+$\,0.7 & 60.8\,$\pm$\,13.2 & 2.0\,$\pm$\,1.1 & 0.3\,$^{+0.9}_{-0.3}$ & 0\,$+$\,0.5 & 0\,$+$\,0.6 & 10.5\,$\pm$\,3.1 \\
	\hline
   \end{tabular}
  \end{center}
\end{table*}

\subsection{HD\,113766\,A\label{sec:HD11}}

{\it Previous studies: }The \textsc{Irs} spectrum of HD\,113766\,A (the primary of the binary system) shows several strong emission features attributed to crystalline grains (9.95, 11.0, 16.25 and 23.5\,$\mu$m). The circumprimary dust orbiting the F3-type star has been studied in detail by \citet{Lisse2008}. To model the Spitzer data, the authors used several species among a vast library of 80 different chemical components. Among them, they found two types of Mg$_2$SiO$_4$ forsterite, labeled as ``ForsteriteKoike" and ``Forsterite038". Unfortunately, the authors do not discuss how different these measurements are (in \citealp{Lisse2008} or in the supporting online material of \citealp{Lisse2006}), or the implications of why both species are required. In addition to olivine and pyroxene dust species, the authors state they need additional dust species to model the spectrum, such as ferrosilite (Fe$_2$Si$_2$O$_6$), diopside (CaMgSi$_2$O$_6$), phyllosilicates (smectite notronite), water ice and metal sulfides (ningerite Mg$_{10}$Fe$_{90}$S). Concerning the latter dust species, to our knowledge, there is a major gap in the wavelength range of laboratory measurements. Optical constant are available between 0.3 and 1.1\,$\mu$m (\citealp{Egan1977}) and between 10 and 500\,$\mu$m (\citealp{Begemann1994}). The $Q_{\mathrm{abs}}$ values remain unknown both in the near-IR (rendering the temperature determination uncertain), and in the 5--10\,$\mu$m range, i.e., in the \textsc{Irs} spectral range. Additionally, in the range 10-40\,$\mu$m only one emission feature is visible and is very broad ($\sim$26--35\,$\mu$m). We estimate that the lack of a clear, narrow emission feature makes the identification of this particular dust species difficult, the risk being to use metal sulfides as a proxy for continuum emission (see Sect.\,\ref {sec:smax}). Finally, concerning the geometry of the dust belt, \citet{Lisse2008} found the dust to be located at 1.8\,AU from the star, with a width of 0.4\,AU with typical temperatures between 420 and 490\,K. The grain size distribution for their best fit has an exponent of $-3.5 \pm$0.2 for grain sizes between 0.1 and 20\,$\mu$m, and finally the authors found a mass of $\sim 5 \times 10^{-5}$\,$M_{\oplus}$ for grain sizes up to 20\,$\mu$m.

{\it The amorphous phase: }According to our model, the abundances of amorphous grains with olivine and pyroxene stoichiometries are of the same order of magnitude (46 versus 29\%, respectively), with an edge for grains of the pyroxene group, a result different from the analysis of \citet{Lisse2008} who found a strong depletion of amorphous grains with pyroxene stoichiometry. We find that amorphous grains of the pyroxene group are all Fe-bearing while grains from the olivine group are Fe-free (no MgFeSiO$_4$). The abundance of amorphous carbon grains is of about 15\%. Grains between 10 and 1000\,$\mu$m in size contribute to about 38\% of the total emission over the entire \textsc{Irs} range.

{\it The crystalline phase: }According to our model, the crystalline content simply consists of Fe-rich olivine grains ($\sim$8\%) and a small amount of Mg-rich olivine grains ($\sim$2\%), both species being mixed with 20\% of carbon. Enstatite is tentatively detected at 0.8\% ($\pm$0.8). No silica grains are used (and needed) in the best fit model. Inspecting the residuals of the best fit, we do not see any conclusive evidences for additional species such as the ones used by \citet{Lisse2008}.

{\it Disk geometry: } The dust belt is found to be rather extended, with $r_{\mathrm{in}} =$\,0.6\,AU and $r_{\mathrm{out}} =$\,11.8\,AU. The grain size distribution follows a power-law with an exponent of $p = -3.17$ and we obtain a total mass of dust of $\sim 7 \times 10^{-3}$\,$M_{\oplus}$ for grain sizes up to 1000\,$\mu$m. 

{\it Additional observations and comments: } In a recent study, \citet{Smith2012} presented VLTI/\textsc{Midi} 10\,$\mu$m observations of HD\,113766\,A. Modeling the observed visibilities, the authors find the debris disk to be partially resolved and their observations are consistent with a Gaussian profile with a FWHM of about 1.2--1.6\,AU (hence, $r$\,$\sim$\,0.7\,AU). Unfortunately, since the interferometric observations are partially resolved, the exact location of the inner radius $r_{\mathrm{in}}$ cannot be further constrained. \textsc{Visir} images from \citet{Smith2012} additionally place an upper limit on the extension of the disk at 17\,AU from the central star, a result consistent with our best model ($r_{\mathrm{out}} = 12$\,AU).

Our best fit returns a 70\,$\mu$m flux of 269\,mJy, to be compared to the \textsc{Mips} flux of 390$\pm$28\,mJy (\citealp{Chen2011}, revised from 350$\pm$35\,mJy originally reported in \citealp{Chen2005}). The $\sim$\,120\,mJy difference can be explained in two different ways. First, one may consider time variability between \textsc{Irs} and \textsc{Mips} observations. In the case of HD\,113766\,A, this explanation is quite unlikely as only a few days separate the two observations (23$^{\mathrm{rd}}$ of February 2004 and 1$^{\mathrm{st}}$ of March 2004 for \textsc{Mips} and \textsc{Irs}, respectively). Second, an outer dust belt might surround the inner warm dust belt. \citet{Lisse2008} considered an equivalent scenario as their model suggests the presence of two additional dust belts, one as far out as 9\,AU where water ice may be located and another one at 30--80\,AU. We note that according to our best model, the dust belt is rather extended (up to 12\,AU), suggesting the need for cold dust grains, which may support the possibility of an outer belt.

\subsection{HD\,69830}

{\it Previous studies: } In addition of showing emission features in its \textsc{Irs} spectrum, HD\,69830 is known to host 3 Neptune-sized planets (\citealp{Lovis2006}), with semi-major axis of $\sim$0.08, 0.186 and 0.63\,AU, with respective minimum masses of 10.2, 11.8 and 18.1\,$M_{\oplus}$. The mineralogy of the dust content has been studied in detail by \citet{Lisse2007}, who modeled the \textsc{Irs} spectrum in a similar way as the one of HD\,113766\,A. According to their model, the authors detected several species. Similarly to \citet{Lisse2008}, they used two types of Mg$_2$SiO$_4$ forsterite (``ForsteriteKoike" and ``Forsterite038"). Fayalite, the Fe-rich end member of the olivine group was also included in their model. No amorphous grains with a pyroxene stoichiometry were detected, but bronzite (Mg$_{1-x}$Fe$_x$Si$_2$O$_6$) ferrosilite (Fe$_2$Si$_2$O$_6$) and diopside (CaMgSi$_2$O$_6$) were necessary for their model to reproduce the data. Finally, the authors state they detected several carbonate species, including magnesite (MgCO$_3$), dolomite (CaMgC$_2$O$_6$), water ice and amorphous carbon. \citet{Lisse2007} concluded the dust belt is located at 1.04$\pm$0.12\,AU from the central star. According to the stability analysis from \citet{Lovis2006}, this value is consistent with the 2:1 and 5:2 orbital resonances with the outer Neptune-sized planet. The typical temperature for the dust is $\sim$\,400\,K and the grain size distribution exponent for grain sizes between 0.1 and 10\,$\mu$m is $-3.9 \pm$0.2, with a dust mass for these grain sizes of $\sim 5 \times 10^{-8}$\,$M_{\oplus}$. 

In a recent study, \cite{Beichman2011} investigated the time variability of the dust around HD\,69830. With new spectroscopic observations and using the model from \citet{Lisse2007}, they led a similar mineralogical analysis. Their model confirmed the previous detections of amorphous grains with olivine stoichiometry, both species of ``ForsteriteKoike" and ``Forsterite038" and fayalite. For the pyroxene group, only bronzite was detected in the new analysis (while ferrosilite and diopside went from ``detected" to ``marginally detected" in the new analysis). The carbonates species of magnesite and dolomite were also considered as marginally detected, as opposed to the best fit results of \citet{Lisse2007}. Sulfides (Fe$_{90}$Mg$_{10}$S) for which an upper limit were found in the 2007 analysis was detected in the latest best fit model. According to \cite{Beichman2011}, the most significant difference between the two models, is that water ice was no longer needed to reproduce the spectrum, therefore concluding the debris disk to be really dry. Based on the main mineralogical markers from both analysis, the authors state that the dust has not evolved significantly in time, but instead that the aforementioned differences arise from higher signal-to-noise ratio (SNR hereafter) of the new observations and a better data calibration for the \textsc{Irs} spectra. This comforts us in our conclusion that spectral decomposition is a non-trivial problem, strongly dependent on several aspects, from SNR in the data, to imperfect knowledge of absorption efficiencies. This is why we prefer to use less dust species but focus on those that have been intensively studied in laboratory experiments, and have a good coverage of their optical data. Finally, \cite{Beichman2011} mentioned that no gas lines could be convincingly detected in their data.

{\it The amorphous phase: }According to our model, the amorphous dust content is best described by Fe-bearing amorphous grains with olivine stoichiometry (24\%) and Fe-bearing amorphous grains with pyroxene stoichiometry (30\%). No MgSiO$_3$ amorphous grains are required in the best fit model, while a small fraction ($\sim$\,4\%) of Mg$_2$SiO$_4$ grains are found. The abundance of amorphous carbon is of about 30\%. Grains with sizes between 10 and 1000\,$\mu$m amounts for 15\% of the total flux in the \textsc{Irs} range.

{\it The crystalline phase: }The total crystallinity fraction of $\sim$13\% is the sum of contributions from Fe-rich olivine grains ($\sim$7\%, mixed with 20\% carbon), Mg-rich olivine grains ($\sim$1\%, mixed with 1\% carbon), enstatite (2\%, mixed with 5\% carbon), and $\beta$-cristobalite (3\%, most of it being mixed with 1\% carbon). Inspecting the residuals between the modeled and observed fluxes, we do not see evidences for other emission features. 

{\it Disk geometry: }The dust belt is located between 0.7 and 1.5\,AU, values that are not in contradiction with the position of the outermost Neptune-sized planet. Since we assume the disk to be circular (eccentricity $e =0$), the solution we find is in agreement with the stability analysis from \citet{Lovis2006}. The grain size distribution exponent is $p = -3.54$, with a total dust mass $M_{\mathrm{dust}} = 3.7 \times 10^{-7}$\,$M_{\oplus}$ for sizes up to 1000\,$\mu$m. 

{\it Additional observations and comments: }\citet{Beichman2005} reported a Spitzer/\textsc{Mips} flux of 19$\pm$3\,mJy at 70\,$\mu$m. Our model returns a slightly higher flux of 22\,mJy, within the uncertainties of the 70\,$\mu$m observations. This suggests there is no colder material that contribute up to 70\,$\mu$m and that the system can be successfully described with one single dust belt close to the star. Our best fit model is consistent with the VLTI/\textsc{Midi} observation and VLT/\textsc{Visir} imaging of \citet{Smith2009}, who constrain the warm dust in HD\,69830 system to lie within 0.05--2.4\,AU.

\subsection{BD+20\,307\label{sec:BD}}

{\it Previous studies: }The dust in the debris disk of BD+20\,307 surrounds a spectroscopic binary of similar F-type stars, as explained by \citet{Weinberger2008}. \citet{Zuckerman2008a} estimated an age of a few Gyr for this binary system. The \textsc{Irs} spectrum shows prominent feature at 10\,$\mu$m, with a shoulder at 11\,$\mu$m. At longer wavelengths, smaller emission features are visible at 16.5, 18.7 and 23.5--24\,$\mu$m. No detectable feature is seen around 33\,$\mu$m. \citet{Weinberger2011} modeled the spectrum of this object, using amorphous grains with olivine (MgFeSiO$_4$) and pyroxene (MgFeSi$_2$O$_6$) stoichiometries plus forsterite grains. According to their modeling, the dust is located at 0.85\,AU from the binary, with typical temperatures between 450 and 500\,K for the dust grains. The inferred crystallinity fraction is $\sim$\,20\% in forsterite grains. Finally, their model slightly over-predicts the \textsc{Mips} flux at 70\,$\mu$m (measured: 27.8\,mJy, against 40.2\,mJy for their model).

{\it The amorphous phase: } The amorphous dust content is dominated by Fe-bearing grains from the olivine group (69\%), but also consists of Fe-free amorphous grains with olivine stoichiometry (7\%) and Fe-bearing grains with pyroxene stoichiometry (15\%). No amorphous carbon grains are required in the best fit model. Grains with sizes between 10 and 1000\,$\mu$m contribute to 20\% of the total flux in the 5--35\,$\mu$m range.

{\it The crystalline phase: } The remaining relative abundances are divided between Fe-rich and Mg-rich olivine grains (4 and 0.6\%, respectively) both with a 20\% mixed carbon fraction and warm enstatite (4\%) in contact with 20\% of carbon. 

{\it Disk geometry: } We find the dust belt to be located between $r_{\mathrm{in}}=$0.2\,AU and $r_{\mathrm{out}}=$18\,AU. According to \citet{Weinberger2008}, the spectroscopic binary has a separation of 0.06\,AU at present days, meaning our estimation of $r_{\mathrm{in}}$ is compatible with the geometry of the system. The grain size distribution exponent $p$ is $-3.38$ and the total dust mass is of about $3.7 \times 10^{-4} M_{\oplus}$ for sizes up to 1000\,$\mu$m.

{\it Additional observations and comments: }Our model slightly over-predicts the \textsc{Mips} observation, as we obtain a 70\,$\mu$m flux of 36\,mJy, to be compared with 27.8\,mJy. Interestingly, our fit to the 10\,$\mu$m feature does not match perfectly the data, as in \citet{Weinberger2011}. The feature at 10\,$\mu$m looks ``primitive" (or pristine, quite similar to ISM dust), with a shoulder around 11\,$\mu$m but no clear emission features on top of it, while our best fit model shows small features caused by the presence of crystalline olivine grains and enstatite. The inferred dust composition contains Fe-rich olivine grains because of the features detected at longer wavelengths (16.5 and 24\,$\mu$m). In other words, the spectrum seems to show two different dust compositions, a first one with very low crystallinity and a second one with an increased crystallinity fraction. To reproduce the strong, almost pristine 10\,$\mu$m feature, amorphous grains need to be warm enough to reproduce the data, therefore they need to be located close to the star (0.2\,AU in our case, 0.85\,AU according to \citealp{Weinberger2011}). As we assume the disk to be homogenous, as soon as crystalline olivine grains are present they will contribute around 10\,$\mu$m. One may argue that the temperature of the olivine grains could in principle be lower and thus contribute only at longer wavelengths. In our best fit, the mixed carbon fraction for the Fe-rich olivine grains is 20\%, while we allow it to be 1 or 5\% (which would provide smaller temperatures). Such solution is apparently not providing a better fit to the data. Allowing smaller fractions of carbon would basically mean that olivine grains are in their pure form, an unlikely prediction for dust grains that have most likely encountered collisions. The small uncertainties on the relative abundances of Fe-rich grains confirm that the fit cannot be improved by much, with the assumption of one single belt. This overall means we should consider alternative solutions.

As discussed previously, spectroscopic observations are unresolved and do not contain spatial informations. The simplest solution is always to assume that the emission probed by \textsc{Irs} is arising from one single dust belt. With this assumption, it appears the fit to the data could be improved. Additionally, the shape of the emission features, and thus the mineralogy suggests a two-components dust composition. Even though the 70\,$\mu$m \textsc{Mips} observations do not suggest the presence of colder material, one can imagine the emission in excess is arising from two separated dust belts: a warm inner dust belt, containing mainly amorphous grains and a second, slightly colder dust belt with a higher crystallinity fraction. This possible scenario is also supported by the extension of the dust belt, up to $\sim$18\,AU, which suggests the presence of rather cold dust, further away from the star. 

\subsection{HD\,15407\,A\label{sec:HD15}}

HD\,15407\,A is the primary of a binary system (21.25$''$ separation), with an age estimate of $\sim$\,80\,Myr (\citealp{Melis2010}). The \textsc{Irs} spectrum mostly displays one strong emission feature peaking at $\sim$\,9.1\,$\mu$m that can be attributed to $\beta$-cristobalite SiO$_2$ grains.

{\it The amorphous phase: }No amorphous grains with olivine stoichiometry are required in order to reproduce the spectrum of HD\,15407\,A. Fe-bearing amorphous grains from the pyroxene group amount for 25\%, and Fe-free grains represent 28\% of the total abundance. The abundance of amorphous carbon is 29.7\%. Grains with sizes larger than 10\,$\mu$m contribute significantly to the total emission (60\%, see Fig.\,\ref{fig:smax}).

{\it The crystalline phase: } Beside the aforementioned species, the only remaining dust composition required to reproduce the data is $\beta$-cristobalite, for which we find an abundance of 17\% with a mixed carbon fraction of 1\%. Even though the emission feature at 12.5\,$\mu$m is slightly under-estimated (but still visible in the model), the 16\,$\mu$m feature is well-reproduced, as well as the broader feature at 20\,$\mu$m. As shown in Fig.\,\ref{fig:qsio} (as well as Fig.\,2 to 4 of \citealp{Sargent2009a}), $\beta$-cristobalite (annealed silica) is one of the few polymorphs of SiO$_2$ that display an emission feature at 16\,$\mu$m (coesite also displays a similar emission feature at these wavelengths). Therefore, it is difficult to firmly conclude on the exact nature of the SiO$_2$ polymorph we detect in the disk around HD\,15407\,A, but we can narrow down possible polymorphs either to $\beta$-cristobalite or coesite. No additional crystalline grains are found in the best fit, and the residuals do not show evidences for other emission features.

{\it Disk geometry: }For the best model, the dust is located in an extended belt, between 0.4 and 19.2\,AU. The grain size distribution follows a power-law distribution with a slope in $p=-3.1$, and the total dust mass is $\sim 7.7 \times 10^{-5}$\,$M_{\oplus}$, for grain with sizes smaller than 1000\,$\mu$m.

{\it Additional observations and comments: }The large extension of the dust belt (up to $\sim$19\,AU) underlines the need for cold dust. 

\subsection{HD\,169666\label{sec:HD16}}

\begin{figure*}
\begin{center}
\hspace*{-0.cm}\includegraphics[angle=0,width=\columnwidth,origin=bl]{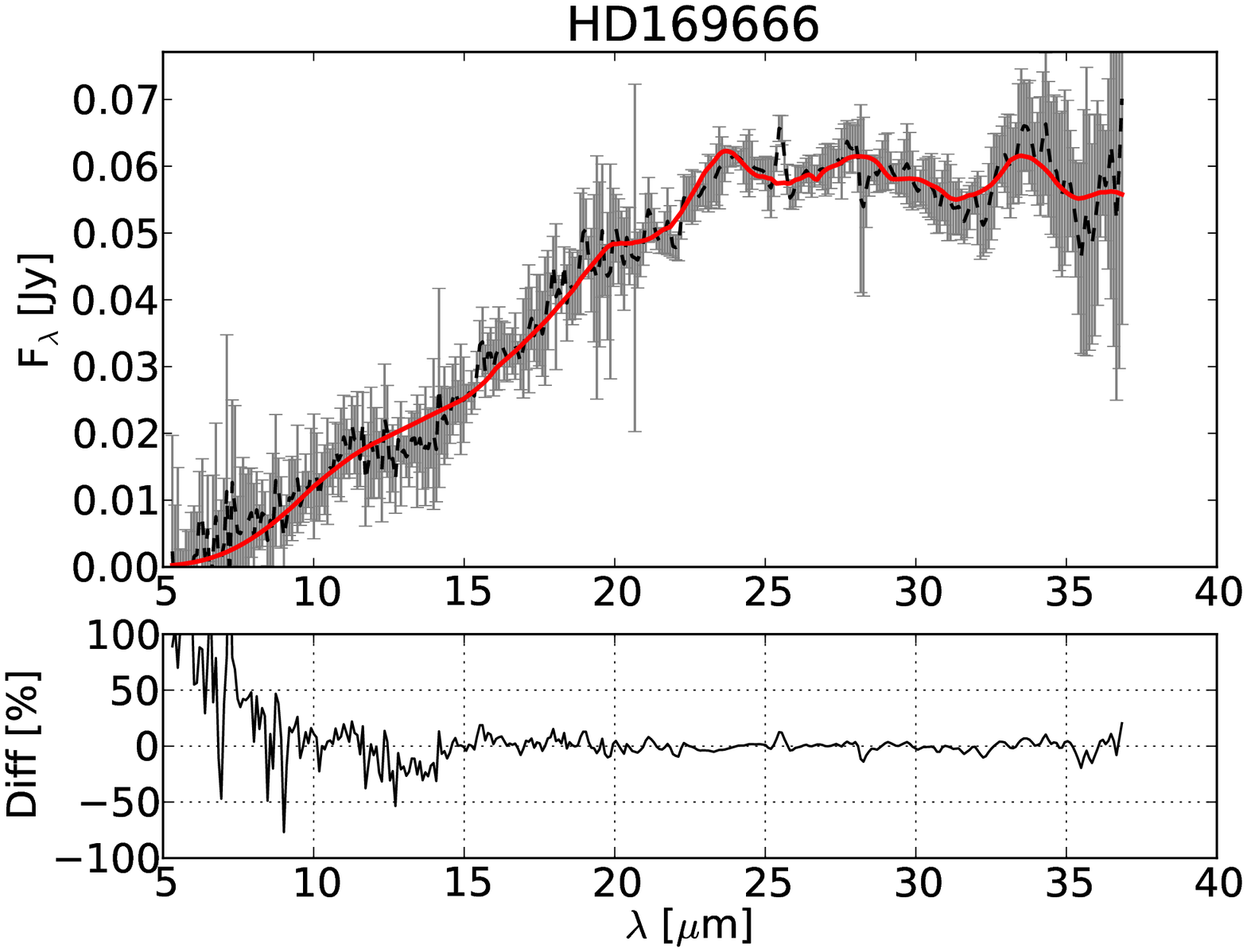}
\hspace*{-0.cm}\includegraphics[angle=0,width=\columnwidth,origin=bl]{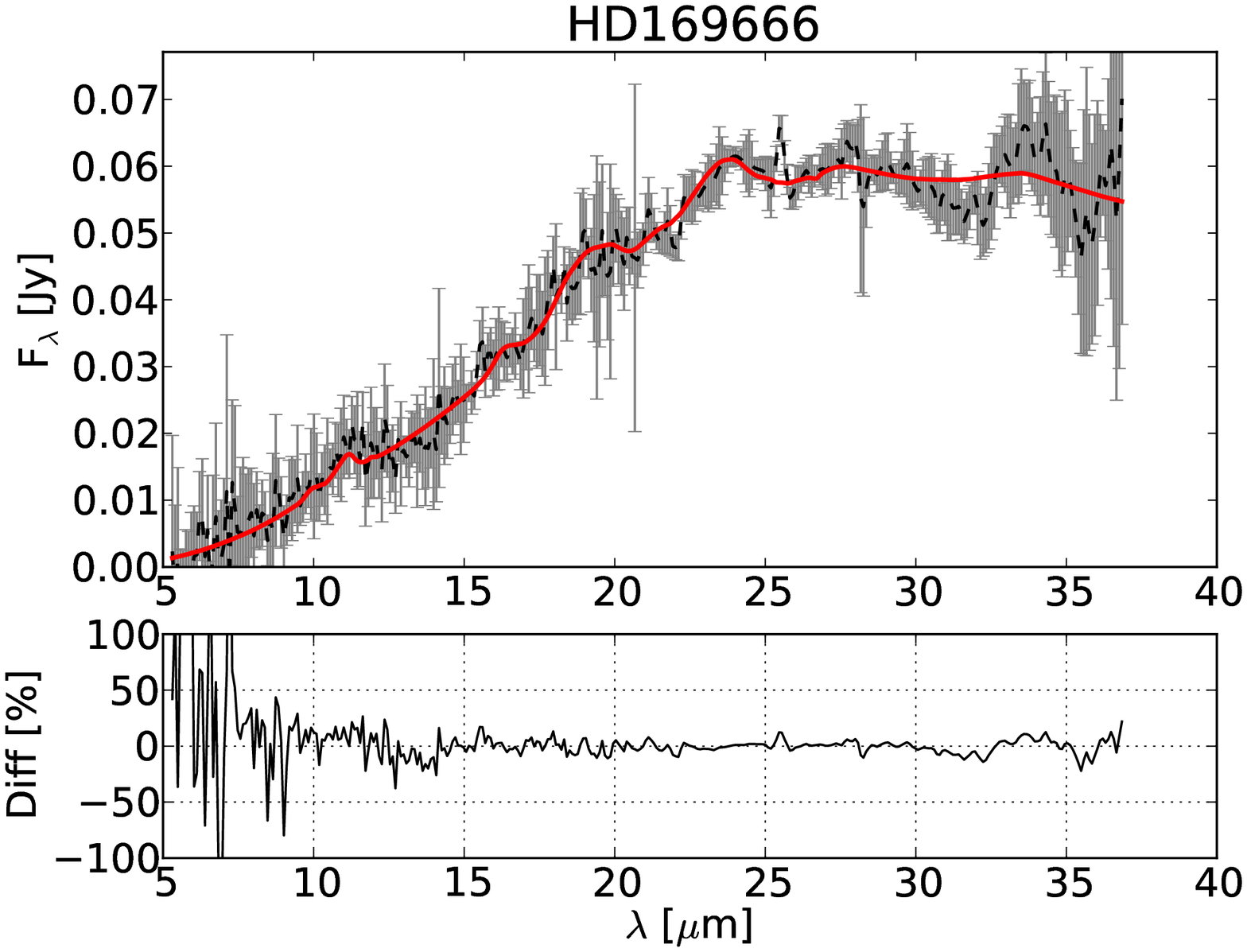}
\caption{\label{fig:HD16_2}Fits to stellar subtracted spectra of HD\,169666 (fits in red, data in dashed black), for the ``cold" (left panel, $T_{\mathrm{max}} \sim 215$\,K, $r \sim 53$\,AU) and ``warm" solutions (right panel, $T_{\mathrm{max}} \sim 380$\,K, $r \sim 13$\,AU).}
\end{center}
\end{figure*}

The emission in excess seen in the SED of HD\,169666 starts at longer wavelengths ($\sim$\,10\,$\mu$m) compared to the other objects of our sample, but still present emission features characteristic of small grains, especially at 24, 28 and 33\,$\mu$m, that can be attributed to crystalline olivine grains. \cite{Mo'or2009} reported a \textsc{Mips} flux of 21.8$\pm$4.0\,mJy at 70\,$\mu$m. 

{\it The amorphous phase: }The composition of the amorphous content is solely represented by Fe-free grains with olivine stochiometry (65\%) mixed with 1 and 20\% carbon (abundances of 50 and 15\%, respectively). The abundance of amorphous carbon grains is of about 17\%. Grains with sizes larger than 10\,$\mu$m contribute to 5\% of the total emission in the 5--35\,$\mu$m range.

{\it The crystalline phase: }Emission features seen at 24, 28 and 33\,$\mu$m can be reproduced using Mg-rich olivine grains, with an abundance of 7\%. The rather broad features around 28\,$\mu$m are best matched with a additional contribution of enstatite (8\%) and finally $\sim$4\% of $\beta$-cristobalite are used in order to reproduce the shape around 20\,$\mu$m. All these three crystalline species are mixed with 1\% carbon. 

{\it Disk geometry: }According to our best model, the dust is located at rather large distances from the star, with $r_{\mathrm{in}} = $53\,AU and $r_{\mathrm{out}} = $68\,AU. The grain size distribution exponent is $p = -3.6$ and we find a total dust mass of $1.2 \times 10^{-3} M_{\oplus}$ for grains with sizes smaller than 1000\,$\mu$m. However, the results of the uncertainties estimation show that modeling the \textsc{Irs} spectrum of HD\,169666 is an extremely degenerate problem. Given the low SNR over the entire spectral range, probability distributions for both $r_{\mathrm{in}}$ and $r_{\mathrm{out}}$ are broad ($\sigma \sim$20 and 15\,AU, respectively). This means that the geometry of the debris disk cannot be firmly constrained with existing data.

{\it Additional observations and comments: }Our best model predicts a flux at 70\,$\mu$m of 36\,mJy, which does not compare so well with the \textsc{Mips} measurement of 21.8$\pm$4\,mJy reported by \cite{Mo'or2009}. The current best fit reproduces well the emission features detected at longer wavelengths, however it does not match properly other, smaller emission features, especially those around 11 and 16\,$\mu$m. Among all the models tested by the genetic algorithm, we find another solution that reproduces these two specific features slightly better (but has a higher $\chi^2_{\mathrm{r}}$). For this model, the disk is located much closer to the star ($r_{\mathrm{in}} = $13.4, $r_{\mathrm{out}}$=16.6\,AU versus 53 and $\sim$68\,AU), therefore the olivine grains are warmer and can better reproduce the emission features around 11--16\,$\mu$m. To distinguish between both solutions, we refer to this one as the ``warm" solution (with $T_{\mathrm{max}} \sim 380$\,K) and to the original one as the ``cold" solution (with $T_{\mathrm{max}} \sim 215$\,K), and both fits to the data are shown in Figure\,\ref{fig:HD16_2}. As can be seen on the right panel of Fig.\,\ref{fig:HD16_2}, the warm solution does not provide a good fit to the 33\,$\mu$m feature, which is the most prominent feature in the spectrum, and this model still over-predicts the \textsc{Mips} flux at 70\,$\mu$m (32 versus 21.8\,mJy). Compared to the cold model, the $\chi_{\mathrm{r}}^2$ increases by $\sim$7\%. In the rest of the study, we will still consider the cold solution as the best fit model, since it provides the smallest $\chi_{\mathrm{r}}^2$ and reproduces well the strongest emission features. But, this simple exercise underlines the crucial importance of SNR when modeling spectroscopic observations. HD\,169666 is the source with the lowest SNR once the stellar spectrum is subtracted, and this renders both the analysis and interpretations difficult. 

\subsection{HD\,98800\,B\label{sec:HD98}}

{\it Previous studies: } As the title of \citet{Verrier2008} suggests, HD\,98800\,B is an unusual debris disk. The system is composed of two bound spectroscopic binaries A and B. Both systems were estimated to be 10\,Myr old (\citealp{Soderblom1998}). With their Keck diffraction-limited observations, \citet{Prato2001} found that the dust belt surrounds the second spectroscopic binary HD\,98800\,B, and that the dust belt is located between $\sim$\,2 and 5\,AU. The structure of the dusty component has also been studied by several authors, for instance, \cite{Koerner2000} estimated the disk to be located at $r_{\mathrm{in}} = 5 \pm 2.5$\,AU with a width $r_{\mathrm{out}} - r_{\mathrm{in}}$ of 13$\pm$8\,AU. \citet{Furlan2007} also studied the dust content around HD\,98800\,B with a two components scenario: an optically thin disk located between 1.5 and 2\,AU from the spectroscopic binary and an optically thick disk at a distance of 5.9\,AU. \citet{Low2005} reported \textsc{Mips} fluxes of 6.26$\pm$1.25\,Jy and 2.1$\pm$0.42\,Jy at 70 and 160\,$\mu$m, respectively.

{\it The amorphous phase: }A broad feature is seen at around 10\,$\mu$m, but no other emission features can be firmly detected in the \textsc{Irs} spectrum. Fe-bearing amorphous grains with pyroxene stoichiometry accounts for 28\% of the total abundance. For the olivine group, Fe-bearing grains represent 26\% of the total abundance, and $\sim$23\% of Fe-free amorphous grains with olivine stoichiometry are used to reproduce the data. The abundance of amorphous carbon grains is 18\%. Grains with sizes between 10 and 1000\,$\mu$m contribute to $\sim$60\% of the total emission in the \textsc{Irs} range.

\begin{figure}
\begin{center}
\hspace*{-0.cm}\includegraphics[angle=0,width=\columnwidth,origin=bl]{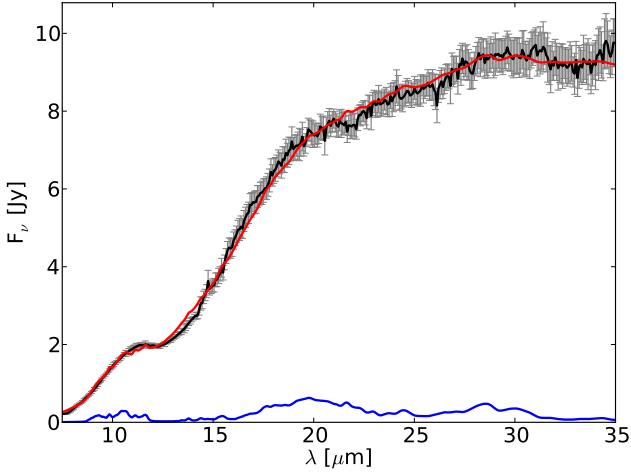}
\caption{\label{fig:HD98_ens}Fits to stellar subtracted spectra of HD\,98800\,B (fits in red, data in dashed black). Blue line shows the contribution of enstatite grains.}
\end{center}
\end{figure}

{\it The crystalline phase: }Even though no emission features attributed to crystalline grains are detected in the spectrum, the crystallinity fraction is not null. We tentatively detect Mg-rich olivine grains with a 0.2$^{+0.8}_{-0.2}$\% abundance. The presence of enstatite grains appears to be better constrained (5.2$\pm$\,1.5\%). The contributions of both crystalline species peak around 19-20\,$\mu$m in order to try to reproduce the rather broad band between 18 and $\sim$21.5\,$\mu$m. Enstatite grains additionally contribute around 28--30\,$\mu$m in order to match the other broad feature at the end of the spectrum. Figure\,\ref{fig:HD98_ens} shows the fit to the stellar subtracted spectrum as well as the contribution of enstatite grains to the total emission (in blue). We try to fit the spectrum without crystalline dust species. The overall shape of the spectrum is reproduced, however the best fit cannot account for the broad features around 20 and 30\,$\mu$m. Since the presence of crystalline grains is not supported by the detection of strong emission features above the continuum, we prefer to consider the presence of Mg-rich olivine and enstatite grains as tentative.

{\it Disk geometry: }We find the dust to be located between $r_{\mathrm{in}} = $1.3\,AU and $r_{\mathrm{out}} \sim $7\,AU, values consistent with previous analysis led on this object. The grain size distribution has an exponent $p = -3$, and we find a total dust mass of $6 \times 10^{-2}$\,$M_{\oplus}$, for grain with sizes smaller than 1000\,$\mu$m.

{\it Additional observations and comments: } Our model predicts fluxes of 5.69 and 1.45\,Jy at 70 and 160\,$\mu$m, respectively, to be compared with the values from \citet{Low2005} of 6.26$\pm$1.25\,Jy and 2.1$\pm$0.42\,Jy at the same wavelengths. The small differences may be the consequence of even larger grains contributing to the emission, but obtaining a better match of the photometric measurement at 160\,$\mu$m is out of the scope of this study.

\subsection{[GBR2007] ID\,8\label{sec:ID8}}

Emission in excess at mid-IR wavelengths for both [GBR2007]\,ID\,8 and [GBR2007]\,ID\,9 (ID\,8 and ID\,9 hereafter) was first identified by \citet{Gorlova2007} while studying the NGC\,2547 stellar cluster ($d = 361$\,pc, 38$^{+3.5}_{-6.5}$\,Myr, \citealp{Naylor2006}). Both sources display departure from photospheric emission at $\sim$4 and $\sim$2\,$\mu$m, respectively. Based on SEDs, \citet{Gorlova2007} proposed three possibilities to explain the measured excess. First, disks are transiting between primordial toward debris disks, second a collision between two planetesimals occurred, producing large excesses and finally both objects could be background post-MS stars experiencing mass loss. Based on complementary UVES observations, ID\,8 was confirmed to be a member of the  NGC2547 cluster, while ID\,9 appears to be a background object rather than a member of the stellar cluster (Gorlova et al. in prep). In this study, we therefore focus only on ID\,8. The \textsc{Irs} spectrum of this object shows a strong emission feature at 10\,$\mu$m as well as at 16 and 19\,$\mu$m. Unfortunately, as the source is faint, the Long Low module suffers from larger uncertainties, but a tentative feature at around $\sim$\,24\,$\mu$m is detected. 

{\it The amorphous phase: } Most of the dust in the disk around ID\,8 is in its amorphous form ($\sim$98\%). Fe-free amorphous grains with olivine stoichiometry clearly dominate the composition, with a total abundance of 61\%. Fe-bearing amorphous grains from the pyroxene group are also required to reproduce the data, with an abundance of 26\%. Amorphous carbon grains have an abundance of about 11\%. Grains with sizes between 10 and 1000\,$\mu$m contribute to 4\% of the total emission.

{\it The crystalline phase: } The $\sim$2\% crystallinity fraction obtained for the dust composition is represented by Fe-rich and Mg-rich olivine grains (2 and 0.3\,\%, respectively), both mixed with 20\% of carbon. The use of Fe-rich olivine grains instead of Mg-rich olivine mostly comes from the peak position of the feature around 24\,$\mu$m. Given the low SNR in this spectral range, the detection of Fe-rich instead of Mg-rich olivine grains is considered as tentative. The uncertainties show that Mg-rich olivine grains can also provide a satisfying fit to the data.

{\it Disk geometry: }According to our best model, the dust is located close to the star with $r_{\mathrm{in}} = 0.4$\,AU and $r_{\mathrm{out}} = 0.5$\,AU. The grain size distribution is really steep, with a slope in $-4.01$ and the total dust mass is of about $2 \times 10^{-6} M_{\oplus}$ for grain with sizes smaller than 1000\,$\mu$m. The steep grain size distribution exponent can be explained by the rather unusual strength of the 10\,$\mu$m emission feature, which requires that small grains dominate the emission.

{\it Additional comments: } Interestingly, the spectrum of ID\,8 compares well with the spectrum of BD+20\,307. Both spectra show a strong 10\,$\mu$m emission feature, and a smaller flux level at longer wavelengths. Emission features from crystalline olivine grains are detected at 16, 19 and 24\,$\mu$m. The main difference between the two spectra is the emission feature at 11\,$\mu$m on top of the emission feature from the amorphous grains for ID\,8, and this is the reason why we obtain a good fit to the spectrum of this source and not for BD+20\,307.

\section{Dust survival and origin\label{sec:survival}}

In this Section, we discuss the possible origin of the dust belts using diagnostics for their luminosities and we investigate the main process that can be responsible for the evacuation of the smallest grains, in order to better constrain their survival and thus their origin. We also discuss the eventual presence of outer dust belts located further away from the star in order to have a better description of the systems.

\subsection{Steady-state evolution or transient dust ?}

\begin{table}
\caption{Observed ($f_{\mathrm{obs}}$) and maximal ($f_{\mathrm{max}}$) fractional luminosities for our target list.\label{tab:fobs}}
\begin{center}
\begin{tabular}{@{\excs}lcccc}
\hline \hline
Star & $f_{\mathrm{obs}}$ & $f_{\mathrm{max}}$ & $f_{\mathrm{obs}}/f_{\mathrm{max}}$ & Transient \\
\hline
HD\,113766\,A  &  7.2\,$\times \, 10^{-2 }$ &  4.3\,$\times \, 10^{-5 }$ & $> 10^3$ & $\checkmark$ \\
HD\,69830  &  2.8\,$\times \, 10^{-4 }$ &  5.9\,$\times \, 10^{-8 }$ & $> 10^3$ & $\checkmark$ \\
BD+20\,307  &  5.2\,$\times \, 10^{-2 }$ &  2.6\,$\times \, 10^{-7 }$ & $> 10^3$ &$\checkmark$ \\
HD\,15407\,A  &  7.0\,$\times \, 10^{-3}$ &  7.0\,$\times \, 10^{-6 }$ & $\sim 10^3$ &$\checkmark$ \\
HD\,169666  &  1.8\,$\times \, 10^{-4 }$ &  1.6\,$\times \, 10^{-4 }$ & $\sim$1 & $\times$ \\
HD\,98800\,B  &  1.5\,$\times \, 10^{-1 }$ &  2.8\,$\times \, 10^{-4 }$ & $\sim$700 & ? \\
ID\,8  &  3.2\,$\times \, 10^{-2 }$ &  4.0\,$\times \, 10^{-7 }$ & $> 10^3$ & $\checkmark$ \\
\hline
\end{tabular}
\end{center}
\end{table}

As explained in \citet{Wyatt2007}, the fractional luminosity of the debris disk $f_{\mathrm{obs}} = L_{\mathrm{IR}} / L_{\star}$ is designed to be a proxy for the disk mass, in the optically thin case. If the observed debris disk is the result of a steady-state evolution, and assuming the disk has the same age as the host star, one can determine a maximum disk mass $M_{\mathrm{max}}$ (and its corresponding maximal fractional luminosity $f_{\mathrm{max}}$) that can survive up to the age $t_{\mathrm{age}}$ of the system. To compute $f_{\mathrm{max}}$, we start from Eqn. 20 of \citet{Wyatt2007}:
\begin{eqnarray}
f_{\mathrm{max}} = 0.58 \times 10^{-9} r^{7/3} (dr/r) D_{c}^{0.5}Q_{D}^{\star 5/6}e^{-5/3}M_{\star}^{-5/6} L_{\star}^{-0.5} t_{\mathrm{age}}^{-1},
\end{eqnarray}
in which we inject the stellar parameters ($t_{\mathrm{age}}$, $M_{\star}$ and $L_{\star}$), as well as our modeling results ($r=r_{\mathrm{in}}$, $dr = r_{\mathrm{out}} - r_{\mathrm{in}}$, both in AU) and use their prescribed values for the maximal size of planetesimals in cascade ($D_c = 2000$\,m), planetesimal strength ($Q_D^{\star} = 200$\,J.kg$^{-1}$), and planetesimal eccentricity ($e = 0.05$). The authors defined a threshold value of $100$ for the ratio $f_{\mathrm{obs}}/f_{\mathrm{max}}$: below this value, a steady-state evolution can explain the infrared luminosity of the debris disk. Table\,\ref{tab:fobs} summarizes the values we obtain for $f_{\mathrm{obs}}$, $f_{\mathrm{max}}$ and their ratios. First, the case of HD\,98800\,B is difficult to assess, the ratio being of about $\sim$700. Given the relatively young age of the object, \cite{Wyatt2007} concluded the collisional cascade has probably started recently, meaning the observed debris disk could possibly be the result of a steady-state evolution. Concerning HD\,169666, we find a ratio of about $\sim$1 for the best model (``cold" solution as discussed in Sect.\,\ref{sec:HD16}) and a ratio of about 40 for the ``warm" solution. As \cite{Mo'or2009} found a smaller value for $r_{\mathrm{in}}$ ($\sim$ 4\,AU), their ratio $f_{\mathrm{obs}}/f_{\mathrm{max}}$ is above the threshold value of 100 and the authors concluded the amount of dust is too important compared to what is expected for a steady-state model. Since the SNR is small for this source, we cannot easily conclude on the evolutionary status of this source since $f_{\mathrm{max}}$ depends on the modeling results. Finally, the remaining five objects are too old for their high infrared luminosities to be explained by a steady-state evolution and are therefore considered as being in a transient phase. A dynamical event, responsible for the production of small dust grains, must have occurred so that we can observe such high fractional luminosities in the mid-IR. 

It is of interest to mention that most of the sources are either in binary systems or have known planets orbiting the central star. The disks around HD\,113766\,A and HD\,15407\,A surround the primary of a binary system, and the dust belt around HD\,69830 is found to encompass the orbits of three Neptune-sized planets. In the case of BD+20\,307, the dust surrounds the spectroscopic binary, while the system of HD\,98800\,B is slightly more complicated. The disk is located around the secondary of the HD\,98800\,A--B binary system and surrounds the spectroscopic binary HD\,98800\,Ba--Bb. The multiplicity of ID\,8 remains unknown since the star was only recently discovered. Stellar companions or massive planets could be responsible for the stirring of smaller planetesimals (see e.g., \citealp{Matranga2010}). Therefore, we may be witnessing the consequences of dynamical instabilities triggered by massive objects in the vicinity of the central stars.

\subsection{Radiation pressure and time variability}

\begin{figure}
\begin{center}
\includegraphics[angle=0,width=\columnwidth,origin=bl]{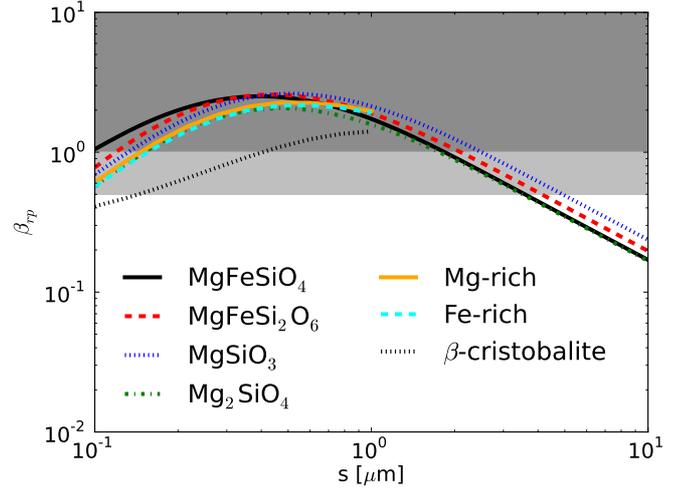}
\caption{\label{fig:beta}Radiation pressure efficiency ($\beta_{\mathrm{rp}}$) for grains around HD\,15407\,A, as a function of grain size $s$ for amorphous MgFeSiO$_4$ (solid black), MgFeSi$_2$O$_6$ (red dashed), MgSiO$_3$ (blue dotted), Mg$_2$SiO$_4$ grains (green dotted-dashed), crystalline Mg-rich and Fe-rich olivine grains (solid magenta, and dashed cyan) and $\beta$-cristobalite grains (dashed black). Shaded areas show regions where grains are gravitationally unbound to the star.}
\end{center}
\end{figure}

Most of the debris disks of our sample are in transient phase, one may therefore wonder about the survival of the fresh, recently produced dust we observe. In Section\,\ref{sec:indiv}, we have not discussed the effect of radiation pressure of the stellar radiation field on the small grains. This effect acts as an apparent reduction of the stellar mass and is the fastest way to evacuate $\mu$m-sized grains from dust belts. Using the absorption and scattering efficiencies, and assuming isotropic scattering we compute the unitless efficiency ratio $\beta_{\mathrm{rp}}$, as detailed in \cite{Burns1979} for different grain sizes and different dust compositions. For a given grain size $s$ and dust composition, $\beta_{\mathrm{rp}}$ is the ratio $F_{\mathrm{rp}} / F_{\mathrm{g}}$ between radiation pressure and gravitational forces, respectively, and quantifies the competition between them. As both forces decrease as $r^{-2}$, this ratio no longer depends on $r$. A $\beta_{\mathrm{rp}}$ value greater than 1 means that radiation pressure overcome gravitational forces and such grains will be evacuated from the system. In the range $0.5 < \beta_{\mathrm{rp}} < 1.0$, grains are in the unbound regime with hyperbolic orbits (see \citealp{Krivov2010}), and will be evacuated from the system as well. Figure\,\ref{fig:beta} shows an example of the $\beta_{\mathrm{rp}}$-ratio for HD\,15407\,A, with $s$ ranging between 0.1 and 10\,$\mu$m for several dust species. An immediate result is that grains smaller than typically 3.5--4\,$\mu$m are rapidly ejected from the system ($\beta_{\mathrm{rp}} > 0.5$). If a grain is subjected to a $\beta_{\mathrm{rp}}$ ratio greater than 0.5, then the eccentricity of its orbit will increase and the grain will quickly be evacuated from the system (comparable to the dynamical timescale). In the case of HD\,15407\,A, the evacuation of grains smaller than $\sim$\,3.5\,$\mu$m is expected to be very efficient, of the order of a few months.

Among the sources studied in our sample, grains around HD\,98800\,B and ID\,8 are not subjected to evacuation by radiation pressure. As discussed in \citet{Beichman2011} for HD\,69830, only grains with sizes in a narrow range ($0.4 < s < 0.7$\,$\mu$m) are affected by radiation pressure. Concerning the spectroscopic binary BD+20\,307 ($M_{\star} = 2.2 M_{\odot}$), grains with sizes in the range 0.3--0.7\,$\mu$m (0.4--0.9\,$\mu$m, depending on the composition) are expected to be evacuated from the system. This means that grains smaller than $\sim$0.3\,$\mu$m may survive to radiation pressure around the binary system, and these small grains can contribute to the emission of the strong and sharp 10\,$\mu$m feature. Overall for both HD\,69830 and BD+20\,307, since the ranges of grain sizes with $\beta_{\mathrm{rp}} > 0.5$ are narrow, we consider this effect to be of second order for the purpose of this study. Additional constraints are necessary to lower the degeneracies in the problem in order to refine the model and assess the influence of radiation pressure. For the four sources discussed here, the survival of the dust belt in time (and hence their detection in the mid-IR) is less constraining on the mechanisms responsible for the production of small $\mu$m-sized grains, compared to the other three sources. Since the timescale for evacuation by P-R drag is much longer ($t_{\mathrm{P-R}} \sim 400\,\beta_{\mathrm{rp}}^{-1}\,r^2\,M_{\star}^{-1}$\,yrs, with $r$ in AU and $M_{\star}$ in $M_{\odot}$, \citealp{Wyatt2008}) than the timescale for evacuation by radiation pressure ($\leq$\,1\,yr), these small grains can survive much longer being gravitationally bound to the central star.

In the following, we discuss the effect of radiation pressure for dust grains in the debris disks around the remaining three sources (HD\,15407\,A, HD\,113766\,A and HD\,169666). For each sources, we discuss which grains are in the unbound regime, depending on their sizes $s$. It is however out of the scope of this paper to include the effect of radiation pressure in the modeling approach. It would indeed require to have robust prescriptions about complex, and time-dependent, dynamical processes such as dust evacuation and dust fragmentation. Instead we discuss timescales about variability and subsequent constraints on the production of small $\mu$m-sized dust grains.

\subsubsection{Time variability of HD\,15407\,A}

\begin{table*}
\caption{Photometric observations (columns ``Flux") of HD\,15407\,A, HD\,113766\,A and HD\,169666 and flux predictions from their \textsc{Irs} spectra convolved with the corresponding photometric filters (columns ``P$_{\mathrm{\textsc{irs}}}$")\label{tab:HD15}. Second line of the Tabel give the epochs of the \textsc{Irs} observations (day/month/year).}
\begin{center}
\begin{tabular}{@{\excs}lc|ccc|ccc|ccc}
\hline \hline
 & & \multicolumn{3}{c|}{HD\,15407\,A} & \multicolumn{3}{c|}{HD\,113766\,A} & \multicolumn{3}{c}{HD\,169666}\\ 
 & & \multicolumn{3}{c|}{(09/10/2008)} & \multicolumn{3}{c|}{(01/03/2004)} & \multicolumn{3}{c}{(02/07/2005)}\\ 
 \hline
Facilities & $\lambda$ & Epoch & Flux & P$_{\mathrm{\textsc{irs}}}$ & Epoch & Flux & P$_{\mathrm{\textsc{irs}}}$ & Epoch & Flux & P$_{\mathrm{\textsc{irs}}}$ \\
                & [$\mu$m] & & [mJy] & [mJy] & & [mJy] & [mJy] & & [mJy] & [mJy] \\ \hline
MSX & 8.28 & 1996--1997 & 808\,$\pm$\,49  & 894\,$\pm$\,45 & - & - & - & - & - & - \\
AKARI/\textsc{Irc}  & 9.00 & 2006--2007 & 960\,$\pm$\,57 &  927\,$\pm$\,47 & 2006--2007 & 1308\,$\pm$\,92 & 1439\,$\pm$\,72 & 2006--2007 & 344\,$\pm$\,18  & 333\,$\pm$\,17 \\
WISE & 11.56 & Feb. 2010 & 641\,$\pm$\,30 &  673\,$\pm$\,34 & Jan.-Jul. 2010 & 1161\,$\pm$\,53 & 1372\,$\pm$\,71 & Feb.-Mar. 2010 & 172\,$\pm$\,8 & 182\,$\pm$\,9\\
IRAS & 12.00 & 1983 &   1050\,$\pm$\,63 &  891\,$\pm$\,45 & 1983 & 1590\,$\pm$\,79 & 1746\,$\pm$\,89 & 1983 & 232\,$\pm$\,16 & 242\,$\pm$\,12  \\
AKARI/\textsc{Irc} & 18.00 & 2006--2007 & 497\,$\pm$\,33 &  445\,$\pm$\,22  & 2006--2007 & 1428\,$\pm$\,87 & 1504\,$\pm$\,79 & 2006--2007 & 130\,$\pm$\,23  & 107\,$\pm$\,6\\
WISE & 22.09 & Feb. 2010 & 400\,$\pm$\,24 &  384\,$\pm$\,19 & Jan.-Jul. 2010 & 1656\,$\pm$\,96 & 1554\,$\pm$\,81 & Feb.-Mar. 2010 & 103\,$\pm$\,6 & 99\,$\pm$\,5\\
Spitzer/\textsc{Mips} & 23.67 & - & - & - & Feb. 2004 & 1459\,$\pm$\,30 & 1431\,$\pm$\,74 & Jun. 2005 & 90\,$\pm$\,4  & 94\,$\pm$\,5 \\
IRAS & 25.00 & 1983 & 432\,$\pm$\,39 &  402\,$\pm$\,20 & 1983 & 1760\,$\pm$\,123 & 1664\,$\pm$\,87 & 1983 & 101\,$\pm$\,13  & 112\,$\pm$\,6 \\
\hline
\end{tabular}
\end{center}
\end{table*}

As shown in Fig.\,\ref{fig:beta}, amorphous grains smaller than $\sim$\,3.5--4\,$\mu$m are short-lived around HD\,15407\,A. Interestingly, we find that some small $\beta$-cristobalite grains, mixed with 1\% carbon and with sizes in the range 0.1-0.25\,$\mu$m, can survive to radiation pressure. Larger grains or grains mixed with higher carbon fractions have $\beta_{\mathrm{rp}}$ greater than 0.5. According to our results (Sect.\,\ref{sec:HD15}) silica grains with an addition of 1\% carbon provide the best match to the data. This finding slightly relaxes the constraints on the production of silica grains. However, we could not obtain a satisfying fit to the data (especially to the emission features from $\beta$-cristobalite) without amorphous grains smaller than 3\,$\mu$m.

To assess the time variability of the debris disk around the primary, we gather photometric observations that were performed in the last decades for this object. Table\,\ref{tab:HD15} summarizes our findings and reports the instrument used and wavelengths of observations as well as measured fluxes. The column labeled ``P$_{\mathrm{IRS}}$" reports flux predictions based on the \textsc{Irs} spectrum and their uncertainties. To obtain these predictions and their uncertainties, we convolve the \textsc{Irs} fluxes to the filters widths of other instruments. The photometric observations cover a range of several decades. Unfortunately, in some cases determining an accurate epoch for the measurement is not trivial. All of the relevant facilities and instruments listed in Tab.\,\ref{tab:HD15} observed the vicinity of HD\,15407\,A several times during their mission and the quoted flux densities in the specific catalogues have been derived as a combination of the photometry extracted from individual scans or observations. However, we know that the MSX Galactic Plane survey observations were obtained between May 1996 and January 1997, while AKARI was active between May 2006 and August 2007. The IRAS mission was accomplished in 1983. Utilizing the SCANPI software we find that the relevant IRAS scans were performed between the 17$^{th}$ of August and the 15$^{th}$of September 1983. In the case of WISE, the observations were performed the 8$^{th}$ and 9$^{th}$ of February 2010.

The measured fluxes over almost 3 decades compare very well with the fluxes derived from the \textsc{Irs} spectrum (observed on the 9$^{th}$ of October 2008), especially at 9\,$\mu$m where the emission feature from small $\beta$-cristobalite grains peaks (960$\pm$57 versus 927$\pm$47\,mJy). At least one year passed between the AKARI and Spitzer observations, a timescale to be compared with a radiation pressure timescale of a couple months. We therefore conclude that the system has been collisionally active over several decades and that the population of grains smaller than 3\,$\mu$m is being efficiently replenished over time.

\subsubsection{Time variability of HD\,113766\,A}

In the disk around HD\,113766\,A, all grains (amorphous and crystalline) smaller than $\sim$\,3.5\,$\mu$m should in principle be evacuated from the system on a timescale of months. This result contrasts with the strong and sharp emission features detected in the spectra, arising from amorphous grains as probed by the 10\,$\mu$m feature and from crystalline olivine grains (e.g., at 11\,$\mu$m). As for HD\,15407\,A we gather photometric observations of HD\,113766\,A, and compare them with the \textsc{Irs} spectrum convolved with the corresponding photometric filters. Table\,\ref{tab:HD15} summarizes the measurements and predictions. Again, obtaining dates of observations is difficult, however the IRAS mission was accomplished in 1983, AKARI was active between May 2006 and August 2007, and \textsc{Mips} observations were performed the 23$^{rd}$ of February 2004. WISE observations were obtained in January (25-26$^{th}$) and July (23-24$^{th}$) 2010. The photometric observations compare very well with the convolved fluxes, meaning the emission in excess around HD\,113766\,A should have been stable for the last 25 years. In addition to their VLTI/\textsc{Midi} interferometric observations, \citet{Smith2012} also presented VLT/\textsc{Visir} spectroscopic observations of HD\,113766\,A, in the 8--13\,$\mu$m spectral range. Interestingly, the shapes of both \textsc{Visir} and \textsc{Irs} spectra compare very well with each other (even though the new observations return a flux level that is of about 1.3 times smaller compared to the 2004 Spitzer observations). The presence of emission features, and hence of warm $\mu$m-sized grains indicates that these dust grains must be replenished quickly, on a timescale of a few orbits. This strongly points towards on-going collisions in the disk.

\subsubsection{Time variability of HD\,169666}

Computing the efficiency ratio $\beta_{\mathrm{rp}}$ for grains of different compositions (MgFeSiO$_4$, MgFeSi$_2$O$_6$, MgSiO$_3$ and Mg$_2$SiO$_4$) we find that grains smaller than 4\,$\mu$m are expected to be short-lived in the disk around HD\,169666. And yet emission features arising from small crystalline olivine grains are detected at longer wavelengths. \cite{Mo'or2009} already discussed the time variability of this debris disk. First, they compared two \textsc{Irs} spectra, obtained on the 2$^{nd}$ of July 2005 and on the 6$^{th}$ of June 2008 (we model the first one in this study as it includes the 2$^{nd}$ order of module SL, between 5.2--8.7\,$\mu$m). \cite{Mo'or2009} concluded that, in a 3 year span, no significant differences are seen between both spectra, and emission features can still be detected at longer wavelengths. In a similar way as for HD\,15407\,A and HD\,113766\,A, we gather photometric observations of HD\,169666 and compare them with the \textsc{Irs} spectrum convolved with the photometric filters. Last two columns of Table\,\ref{tab:HD15} show both observed and predicted fluxes. The strongest constraint on the time variability comes from the IRAS observations of 1983, for which the observed and predicted fluxes compare extremely well ( $F_{\nu,\,\mathrm{obs}}$=232\,mJy against P$_{\mathrm{\textsc{irs}}}$=242\,mJy at 12\,$\mu$m, and $F_{\nu,\,\mathrm{obs}}$=101\,mJy against P$_{\mathrm{\textsc{irs}}}$=112\,mJy at 25\,$\mu$m). Spitzer/\textsc{Mips} observations were obtained on the 23$^{rd}$ of June 2005, and WISE photometric observations were performed between the 21$^{st}$ of February and 6$^{th}$ of March 2010. As for the previous two sources, there must be an on-going process in the disk that continuously replenishes the small dust grain population, as the transient phase has been stable for at last 25 years.

\subsection{Presence of outer, cold belts\label{sec:outer_belt}}

The dust detected around the debris disks we study is transient, except maybe in the cases of HD\,98800\,B (given its relatively young age) and HD\,169666 (given the uncertainties on the location of the dust belt because of the low SNR in its spectrum). Two main scenarios have been invoked in the literature to explain the origin of the transient dust. First, the dust may be produced directly close to the star. This scenario can be the consequence of several events. Transient dust may originate from a massive collision between two large bodies being disrupted close to the star. Even though the possibility of a steady-state asteroid belt is most likely ruled out by the high fractional luminosities of the disks, transient dust may be produced in an asteroid belt that has been recently excited and where a collisional cascade started. The second main scenario involves the presence of an outer belt dynamically unstable feeding the inner regions of the disk. Bodies from the outer belt are being scattered inward and release small grains as they collide or sublimate. As mentioned in \citet{Beichman2005} for HD\,69830, a ``super-comet" from the outermost regions may have been captured on a low eccentricity orbit in the inner regions and is being evaporated and broken into smaller bodies. However, detecting an outer belt further away from the star does not enable us to disentangle between the two scenarios, since the outer belt may be stable in time and not responsible for the production of the transient dust.

The most straightforward way to infer the presence of cold, outer dust belts is via far-IR, or mm observations. For one source in our sample (HD\,113766\,A), we can already mention that another dust belt is likely to be located further away from the star. Spitzer/\textsc{Mips} observations at 70\,$\mu$m reveal an excess of $\sim$120\,mJy compared to the predicted flux of our best model (270\,mJy). As \textsc{Irs} observations are unresolved, we cannot quantify by how much the outer belt contribute to the flux measured by the spectrograph. In this study, we assume the dust is in a single, continuous disk. Consequently, the flux contribution of the outer belt in the \textsc{Irs} range translates into rather large values for $r_{\mathrm{out}}$ (12\,AU). This can be explained as a need for a contribution of colder grains. The extension of disks, in the formalism of our model, may be a {\it possible} diagnostic for the presence of outer dust belts. According to our findings, HD\,113766\,A is not the only source for which we find an extended disk. The width of the disk around HD\,15407\,A is of about $\sim$19\,AU, which also suggests the need for cold dust grains in order to reproduce the unresolved spectroscopic data. Further observations are required in order to distinguish between the two possibilities of either an extended disk or two spatially separated dust belts. Even though the \textsc{Mips} observations at 70\,$\mu$m do not reveal emission in excess compared to our model, we discussed the possibility of a two belts scenario around BD+20\,307 (Sect.\,\ref{sec:BD}). The justification being that the disk is rather extended (18\,AU) and that we observe a gradient in the dust composition: the warm inner regions probed by the 10\,$\mu$m emission feature seem to contain pristine dust, while emission features from crystalline olivine grains are detected at longer wavelengths and may arise from another dust belt. With the assumption of one continuous, homogenous disk and even allowing olivine grains to have low temperatures (mixed with 1 or 5\% of carbon), we cannot achieve a satisfying fit to the 10\,$\mu$m feature. 

The four remaining objects in our sample (HD\,69830, HD\,169666, HD\,98800\,B, and ID\,8) could be modeled successfully with the assumption of a single dust belt. For the first three objects complementary observations at longer wavelengths do not show any emission in excess, except for the 160\,$\mu$m \textsc{Mips} photometric point for HD\,98800\,B. In this case, including larger grains in the model may compensate for the small difference. The absence of far-IR emission in excess for these sources points towards the scenario of a dust production mechanism directly close to the star. However, determining which declinations of this scenario (massive collision, grinding of an asteroid belt, super-comet) is the correct one is difficult. In the following Section, we will investigate if the results from spectral decomposition can provide additional information on the origin of the dust.

\section{The origin of crystalline grains\label{sec:origin}}

To have a better understanding of these transient debris disks, one can investigate the origin of crystalline grains. There are several possibilities to crystallize amorphous grains: {\it (i)} thermal annealing, where grains are continuously subjected to temperatures high enough (or timescales long enough at intermediate temperatures), for their internal structure to be re-arranged, {\it (ii)} shock-induced melting, where grains are subject to high temperatures for short amounts of time, due to collisions. Finally, {\it (iii)} crystalline grains detected in the disks may originate from the parent bodies. The parents bodies are either asteroids progressively grind down in a belt, cometary bodies that were scattered inwards in a LHB-like event, or large planetesimals that collided and were completely disrupted. In the latter case, either collisions produce enough heat for crystallization, or crystals were already formed through differentiation in parent bodies and are released through collisions. Scenarios {\it (ii)} and {\it (iii)} are intimately connected since they both involve collisions between planetesimals or smaller bodies, but in the first case, collisions release enough energy to melt the fragments, while in the second case, the fragments are not altered by the collisions. As debris disks are gas-poor, we do not considered ``present-day" gas-phase condensation as a possible significant source of crystalline grains. In the following paragraphs, we present and discuss which diagnostics on crystallization can help us better understand the origin of these grains.

\subsection{Where's the enstatite ?}

The spectra in our target list do not display any strong, recognizable emission features that can be attributed to enstatite crystalline grains (e.g., at 9.3\,$\mu$m). This translates into overall small abundances for this particular dust species, while enstatite has been commonly observed around a large variety of primordial disks around Class\,II objects, with different ages, stellar masses and star forming regions (\citealp{Bouwman2008}, \citealp{Juh'asz2010}, \citealp{Olofsson2010}, \citealp{Sargent2009}). One result of these studies being that enstatite seem to dominate the crystallinity fraction of the inner regions ($\sim$\,1\,AU for T\,Tauri stars), while forsterite dominates the outer regions of the disks ($\leq$\,10--15\,AU for T\,Tauri stars). Concerning debris disks, so far only one object seems to harbor significantly more enstatite grains than forsterite grains (HD\,165014, \citealp{Fujiwara2010}). This contrast between gas-rich, massive disks and gas-poor debris disk is intriguing. For the objects we study, if the observed dust was the product of a collisional cascade of planetesimals formed within the first AU of the star, we may expect to detect more enstatite dust grains. This is one possible scenario mentioned by \citet{Fujiwara2010}, who speculate that the debris disk around HD\,165014 may be the aftermath of a collision between small bodies and the surface of a Mercury-like planet (which shows evidence for enstatite, \citealp{Sprague1998}). The overall low abundances of enstatite, and the lack of strong emission features, pratically rule out such events for the sources in our sample.

The observed crystalline grains may therefore originate from bodies that were formed at larger distances from the star (where enstatite is scarcer), or the observed crystalline grains were recently formed inside the disks. Both possibilities fit rather well the picture of low enstatite abundances. If the debris disks modeled in this study are the results of collisions between bodies originally formed further away from the star, the lack of enstatite grains compares well with observations of younger disks. On the other hand, if the observed crystalline grains are formed at present days in the debris disk, laboratory experiments can help us understanding the low abundances of enstatite. For instance, with their annealing experiments of amorphous grains at 1000\,K, \citet{Thompson2002} have shown that enstatite is difficult to form at these temperatures, and forsterite grains are preferentially formed (as probed by X-ray powder diffraction). This result echoes the respective activation energies for annealing of enstatite and forsterite grains (see \citealp{Gail2010} and references therein). Several laboratory experiments have shown that activation energies for annealing are larger for enstatite grains compared to forsterite grains (e.g., \citealp{Fabian2000}). Consequently, if present day thermal annealing is the main production source of crystalline grains in debris disks, we can expect to have higher abundances of crystalline olivine grains compared to enstatite. Interestingly, our finding of low abundances of enstatite echoes the study of crystallization processes during the violent outburst around EX\,Lupi. \cite{Abraham2009} found evidences for production of forsterite, but not of enstatite, within the first AU of the disk, a result later on confirmed by VLTI/\textsc{Midi} observation (\citealp{Juhasz2011}).

\subsection{The iron content}

In order to distinguish between the different scenarios on the origin of crystalline grains, the petrology of the olivine dust grains, and more specifically the iron content of these grains, can provide new insights on the crystallization processes at stake in these disks. This means, we have to be confident in our results about the Fe content of the olivine grains. This study relies on the laboratory measurements for different olivine samples. We insure to use data with as little contamination as possible (aerosol data). Additional effects may come into play, such as the shape of the grains, that can possibly shift peak positions (\citealp{Koike2010}), but also porosity or crystalline grains being embedded in other materials (\citealp{Min2008}). Concerning the first point, many olivine grains are injected in the aerosol apparatus, with various sizes and shapes. Plus, we do not make use of any scattering theory, such as the DHS theory that may oversimplify some aspects of the problem (\citealp{Mutschke2009}). Concerning the last two additional effects, we unfortunately cannot take them into account as this will result in too many degeneracies for an already complex modeling. We would like to stress again the need for high angular resolution observations to constrain some crucial parameters in the modeling. To conclude, we are confident about the detection of Fe-rich olivine grains in several debris disks (see Sect.\,\ref{sec:ferich}), given the available data and observations and the assumptions made on the disk geometry (e.g., continuous dust belt). 

This finding of Fe-rich olivine grains contrasts with observations of Class\,II disks, for which most studies concluded on the absence of Fe-rich crystalline silicates. In their sample of Herbig Ae/Be stars, based on the 9.4\,$\mu$m emission feature, \citet{Juh'asz2010} reported a possible fraction of 10\% of Fe in crystalline pyroxene grains. However, the authors do not report any shift to longer wavelengths for emission features associated with crystalline olivine grains (e.g., 19 or 24\,$\mu$m), which indicates an overall small fraction of Fe in olivine grains. To shed light on these petrological differences of olivine grains, we  discuss the results of annealing experiments in laboratory. If amorphous dust grains are exposed to high temperatures for a significant amount of time, internal re-arrangement processes are activated in the material and a periodic order appears in the originally disordered network (see \citealp{Henning2010}). Crystallization is a process that depends on two main environmental parameters: temperature and time. Amorphous grains can crystallize either at intermediate temperatures if the exposure time is long enough, or on short timescales if the temperature is high enough. In the following, we summarize some of the relevant results from previous studies about thermal annealing, with respect to the influence of these two parameters on dust petrology. 

\cite{Nuth2006} investigated the production of Fe-bearing crystalline grains via thermal annealing. According to their results, both end-members of magnesium and iron crystalline grains (forsterite and fayalite for the olivine group, respectively) can be formed via thermal annealing, if the temperatures are high enough ($\sim$\,1100\,K and 1400\,K, for magnesium and iron, respectively). However, as the temperatures required to form crystalline iron silicates are higher than temperatures required to form crystalline magnesium silicates, the authors argue the crystalline iron grains will evaporate on timescales of several months, while magnesium grains can survive longer. According to \cite{Nuth2006}, this timescale issue is the reason why iron crystalline grains are not detected in young protoplanetary disks: the grains are destroyed before they can be transported outwards and survive. Since iron end-member crystalline grains (e.g., fayalite) were not required to fit our data, this temperature threshold (1100 vs. 1400\,K) is not directly applicable to our problem. Still, \cite{Nuth2006} discuss thermal annealing of Fe- and Mg-silicate aggregates and suggest as a first approximation that annealing conditions for such aggregates might fall in the 1100--1400\,K temperature range. Therefore temperatures of 1200--1300\,K might be sufficient to produce the Fe-rich (Fe / [Mg + Fe] $\sim$ 20\%) crystalline grains that we detect. One should still keep in mind that the exact conditions for annealing of such mixed aggregates are yet to be better studied, which renders the discussion challenging.

\citet{Murata2009} have conducted annealing experiments of Fe-bearing amorphous silicates during 600 min at temperatures of about $\sim$1000\,K, at a pressure of one atmosphere. The oxygen partial pressure in the furnace was controlled, to avoid possible oxidation of Fe$^{2+}$ cation. They conclude that the Mg fraction ($=$\,Mg\,/\,[Mg+ Fe]) of the crystalline phase decreases with time, as the amorphous phase is annealed. At the end of the experiment, the crystallinity fraction in the original material reached $\sim$\,80\% and the Mg fraction was close to 0.8. Such fraction is quite similar to the Fe-rich olivine grains we use in this study. We can speculate that annealing over long period of time will lower the Mg fraction to match the one of the inital amorphous material.

\subsection{Collisions and shock-induced melting}

Transient dust observed around debris disks may be the consequence of collisions, where either km-sized bodies are continuously being grind down, or larger planetesimals are completely disrupted after a collision at high relative velocity ($v_{\mathrm{rel}} \sim 10$\,km.s$^{-1}$). One can wonder how much energy these disruptive, or grinding collisions release, and by how much the temperature of the resulting fragments can increase. Using laboratory shock experiments as well as numerical simulations \citet{Keil1997} studied the efficiency of shock heating by collisions (with $v_{\mathrm{rel}}$ ranging from 3 to 7\,km.s$^{-1}$) for planetesimals with sizes between 10 and 1000\,km. Their conclusions were that impacts do not significantly heat up the fragments, mostly because of the low surface gravity of the planetesimals. For instance, they find a globally averaged increase of about 50\,K for a collision involving a 1000\,km-sized body. But the planetesimals considered by the simulations of \citet{Keil1997} were non-porous bodies. Since many asteroids in the Solar System have non negligible porosity fraction (\citealp{Britt2002}), more refined models should consider the case of porous bodies. As demonstrated by \citet{Davison2010}, including porosity in simulations can make a significant difference. The main conclusions from their study were that before disruption, shock-induced heating is relatively minor and localized, but if the collision is disruptive then heating can be significant, with more than $\sim$\,10\% in mass of the parent body is melted. Collisions between porous planetesimals of comparable sizes can melt almost the entire mass of the two bodies, even at collision velocities smaller than 7\,km.s$^{-1}$. Since many uncertainties remain, applying the results of these simulations to our results is difficult. Nonetheless, we cannot rule out disruptive collisions as a source of newly formed crystalline grains. As the maximum temperature reached during such events is unknown, we can hardly address the question of the petrology of olivine grains when produced by collisions.

\subsection{Crystallization in differentiated bodies\label{sec:differentiate}}

Interiors of non-porous planetesimals are known to be regions of high temperatures. The formation of planetary cores requires the melting of accreted bodies, and consequent differentiation. As discussed by \citet{Moskovitz2011}, the heating source can be the decay of short-lived radionuclides $^{26}$Al and $^{60}$Fe, which can result in temperatures high enough for the melting to take place. Differentiated bodies will then be the result of gravitational segregatation of molten silicates and metals within the planetesimal. According to studies of iron meteorites (e.g., \citealp{Schersten2006}, \citealp{Qin2008}), such meteorites are the consequence of accretion, differentiation as well as crystallization of their original parent bodies, suggesting that such event most likely happened in the early Solar System, when the decay of $^{26}$Al was significant enough to contribute to the heating. Even though the question of Fe-enrichment of olivine grains in differentiated bodies is at present day a poorly constrained question, we hypothesize that Fe-rich olivine grains may be produced inside young differentiated bodies. Disruptive collisions between these bodies can then release the crystalline grains (and not altering them in non-porous collisions as studied by \citealp{Keil1997}), an hypothesis that could account for the origin of the thermally metamorphosed LL chondrites, that contain Fe-rich olivine grains (\citealp{Nakamura2011}).

\subsection{Individual source discussions}

Here, we discuss for each source which scenario can best explain the inferred dust mineralogy, based on the several diagnostics presented previously.

\subsubsection{HD\,113766\,A} 

According to our best model, $r_{\mathrm{in}}$ is located close to the sublimation radius ($T_{\mathrm{d,\,max}} \sim 1200$\,K). Fe-rich olivine grains are found to be more abundant than Mg-rich olivine grains. Since olivine grains produced in Class\,II disk are preferentially Mg-rich, we are not observing the mineralogy of parent bodies inherited from the primordial disk (e.g., comets, see Sect.\,\ref{sec:comet}) as the origin of the dust around HD\,113766\,A. Additionally, from IRAS observations, we know that the transient phase has last for at least 25 years, suggesting a continuous replenishment of $\mu$m-sized grains. Since disruptive collisions of bodies inherited from the primordial disk do not support our findings, we first propose that we may be witnessing the continuous grinding of several differentiated bodies (see Sect.\,\ref{sec:differentiate} and \ref{sec:asteroid}). Fe-rich olivine grains could have been produced inside large planetesimals and only then released in the exo-zodiacal belt as a consequence of disruptive collisions. Second, given the relatively high dust temperatures we find for our best model, one can also hypothesize that crystalline grains are produced via thermal annealing, in a similar way as discussed by \citet{Nuth2006}. Crystalline Fe-rich silicates may be formed at temperatures in the range 1100--1400\,K from Fe-, Mg-silicate aggregates. The eventual timescale issue for the survival of these grains is then comparable to the timescale of evacuation by radiation pressure for the (sub-)\,$\mu$m-sized olivine grains. With this picture in mind, we do not have to assume that all the observed $\mu$m-sized {\it crystalline} grains are continuously replenished from the parent bodies. Instead, we can suggest that a fraction of the small amorphous grains is constantly being annealed into Fe-rich olivine grains. The tentative detection of enstatite (0.8$\pm$0.8\%) fits in both scenarios. Very few Solar System chondrites contain enstatite ($\sim$\,2\%, \citealp{Norton2008}), which points toward comparable composition of the parent bodies around HD\,113766\,A. In the case of present-day crystallization, the timescale for production of enstatite may be longer than the timescale for evacuation of the grains. 

\subsubsection{HD\,69830}

The dust belt is located slightly further away from the star compared to the sublimation radius of silicate grains ($T_{\mathrm{d,\,max}} \sim 700$\,K). Therefore, thermal annealing as described by \citet{Nuth2006} cannot account for the abundances of Fe-rich olivines we find. However, radiation pressure has very limited effect on sub-\,$\mu$m-sized grains, meaning the grains survive much longer in the disk. As explained by \citet{Murata2009}, and since survival timescales are no longer an issue, long exposure times to intermediate temperatures can produce Fe-rich crystalline grains, annealed from Fe-bearing amorphous grains that dominate the amorphous phase according to our results. We could possibly be witnessing the effect of Mg/Fe fractionation in olivine grains. However, one open question would be to compare the timescales for fractionation and collisions. According to our result the system can be described as a collisional cascade ($p=-3.54$, \citealp{Dohnanyi1969}), suggesting numerous collisions at all grain sizes. In a debris disk, relative velocities are too important for coagulation between grains to take place, and collisions most likely result in fragmentation of the colliding bodies. Then, how would fragmentation affects the efficiency of fractionation? One may hypothesize that the energy released in the collision may accelerate the fractionation process by increasing the grains temperatures, but we cannot conclude on such conjectures without further investigations on that matter. On the basis of the activation energy for Mg$_2$SiO$_4$ smoke in \citet{Fabian2000}, the characteristic crystallization timescale at 700K is of about 3000\,yr. However, at 600K the timescale increases up to 30\,Myr, to produce pure forsterite. Since the activation energy of iron-rich silicates should be higher than Mg-rich silicates this timescale can increase significantly for iron-bearing silicates. This renders the possibility of thermal annealing as the main production source for crystalline grains quite unlikely. 

As suggested by the absence of an outer belt, the origin of the transient dust around HD\,69830 is most likely the result of collisions close to the star. Since Fe-rich olivine grains are unlikely to be formed in Class\,II disks, we can potentially rule out that the parent bodies are inherited from the primordial disk. Therefore the possibility of an asteroid belt being slowly ground down may not fit very well in this picture. Comets like Hale-Bopp or Wild\,2 are thought to have formed in the outer regions of the Solar System, and the crystalline olivine grains observed in both comets are Mg-rich (see Sect.\,\ref{sec:comet}). In the scenario of a super-comet being broken in smaller bodies, as described in \citet{Beichman2005}, we would therefore expect to preferentially detect Mg-rich olivine grains, formed in the primordial disk and transported outside afterwards. The petrology of the olivine grains we infer in our best model does not support this scenario. Shock-induced melting could contribute as a source of crystalline grains, but it would imply that crystallization via melting strongly favors the production of Fe-rich crystalline grains, a question that is poorly constrained at the moment. However, we can hypothesize that the transient dust originate from differentiated bodies, inside which olivines grains were enriched in Fe (e.g., \citealp{Nakamura2011}). As mentioned by \citet{Keil1997}, if such planetesimals are non-porous (a reasonable assumption for differentiation to take place), their breakup into smaller fragments should not alter the mineralogy too much since shock-induced melting then becomes negligible. Interestingly, the system is the host of three Neptune-sized planets. The observed dust belt may be the result of a dynamical instability, possibly triggered by planet migration for instance. Such an instability may have resulted in the collision and disruption of large planetesimals.

The small abundance of enstatite may correlate with the abundance of silica (Mg$_2$SiO$_4$ + SiO$_2$ $\rightleftharpoons$ 2\,MgSiO$_3$). Production of enstatite via thermal annealing may not fit well in this picture, since the timescale for annealing at these intermediate temperatures (700\,K at maximum) is most likely longer than the collisional timescale. The finding of low abundance of enstatite compares well with the overall scarcity of enstatite chondrites in the Solar System. 

\subsubsection{ID\,8}

The origin of crystalline grains around ID\,8 is more difficult to discuss because of the low SNR at longer wavelengths. This renders the spectral decomposition more uncertain on the exact crystallinity fraction as well as on the iron fraction in olivine grains. The uncertainties from the MCMC run show that the detection of Fe-rich over Mg-rich olivine grains is less convincing compared to the previous two sources. We therefore prefer not to conclude on the exact petrology of the crystalline olivine grains in this system. Still, one of the interesting characteristic of the modeling results is the very steep grain size distribution ($p=-4.01$, for $s$ up to 1000\,$\mu$m). This suggests the system is at a peak of collisional activity, therefore producing large amounts of sub-$\mu$m-sized grains, a majority of them being amorphous. It indicates that whatever the parent bodies are (asteroids, comets or planetesimals), they should have had an overall low crystallinity fraction.

\subsubsection{HD\,15407\,A}

The spectrum of HD\,15407\,A does not show emission features associated either with olivine or enstatite grains. But the spectrum shows several emission features associated either with $\beta$-cristobalite or coesite (see Sect.\,\ref{sec:HD15}). The exact nature of the SiO$_2$ polymorph is difficult to constrain since they both show an emission feature around 16\,$\mu$m. Additionally, it is possible that other polymorphs (e.g., non-crystalline silica) are present in the debris disk around HD\,15407\,A, but inferring their presence is a degenerate problem given the strong similarities between the emissivity profiles, especially for the strongest emission feature, around 9\,$\mu$m. The $\beta$-cristobalite form of silica we use is close to the annealed silica studied by \citet{Fabian2000}, which is the result of an annealing experiment, for 5\,h at a temperature of 1220\,K. This indicates that the grains have experienced rather high temperatures, but as discussed in \citet{Sargent2009a}, once formed cristobalite grains must be cooled quickly enough to retain their crystalline structure. If not, the grains are expected to revert to the lower temperature polymorphs of silica ($\beta$- and $\alpha$-quartz). Formation of coesite, instead of $\beta$-cristobalite, is not a straightforward process either since it requires both high temperatures and pressures\footnote{Oral contribution by Koike Chiyoe et al., at the ``Grain Formation Workshop", Japan, 2011}. Such conditions may be fullfilled in a collisional environment. It is noteworthy to remark that in the case where pressures are not high enough to form coesite, $\beta$-cristobalite should be formed preferentially.

The mineralogy of the dust around this source is puzzling. No crystalline silicate grains are detected (olivine or enstatite), but a significant fraction of SiO$_2$ is required to model the spectrum. According to our best model, the maximum temperature is of about 1530\,K, a temperature high enough to anneal silicates grains and produce for example crystalline olivine grains. Even though small grains are short-lived around HD\,15407\,A because of radiation pressure, if the main source of silica was thermal annealing, we should expect to detect features associated with olivine grains (as in the case of HD\,113766\,A for instance). Their non-detection points towards a different source than thermal annealing, which in turn means that the maximum temperature we find is not appropriate and that grains should be colder than $\sim$\,1500\,K. Since the derived geometry of the disk is dominated by the fitting of emission features and given that there is only one strong emission feature around 9\,$\mu$m in the entire spectrum of HD\,15407\,A, this can explain why the temperature determination may not be accurate enough in our best model. Additionally, we discuss previously the possibility of a cold, outer dust belt further away from the central star. If confirmed, the geometry of the disk would have to be revised, and hence at the same time the maximum temperature of the grains. This temperature issue can be easily constrained with high-angular resolution observations. From the absence of crystalline olivine grains, ruling out thermal annealing as a source of crystallization process leaves us with little options to explain the peculiar dust composition of this dust belt. Since such crystalline grains are expected to be common in the outer regions of primordial disks where comets are formed, and since silica is not usually detected in cometary bodies, the scenario of an outer belt feeding the inner regions is rather unlikely. We may therefore be witnessing the aftermath of collisions between silica-rich parent bodies, or the production of coesite via collisions that provide temperatures and pressures high enough to form this particular polymorph of SiO$_2$. In the first case, collisions should be either non-disruptive, or the parent bodies should be non-porous, otherwise shock-induced heating may produce other crystalline grains, as discussed previously. In that scenario, the collision may be the consequence of dynamical stirring, triggered by the secondary of the binary system.

\subsubsection{Other sources}

For the three remaining sources, we can hardly discuss the origin of the crystalline grains or the petrology of olivine grains. For HD\,98800\,B, we conclude in Sect.\,\ref{sec:HD98} that the crystallinity fraction could only be considered as tentative because of the lack of strong emission features associated with crystalline grains. For HD\,169666, the overall low SNR of the data renders the analysis difficult, and we prefer not to over-interpret data. Concerning BD+20\,307, since the fit is not satisfying around 10\,$\mu$m, the confidence level of the crystallinity fraction and iron content is not sufficient for any interpretation of their origin. 

Overall, the conclusion of this Section\,\ref{sec:origin} is that many processes are active in these transient debris disks, from thermal annealing at high or intermediate temperatures, to (non-)\,disruptive collisions of (non-)\,porous parent bodies with various compositions. Spectroscopic observations contain a finite amount of informations, and we have a limited knowledge about dust emissivity properties (from near- to mid-IR). We are able to suggest possible scenarios to interpret our results within these unfortunate limitations but further investigations (observations, instrumentation and laboratory experiments) are mandatory to confirm, or infirm our findings. To summarize, it is not trivial to assume that the inferred dust composition originates from parent bodies inherited from the primoridal disk because of several other crystallization processes at stake. 

\section{Comparison with Solar System objects\label{sec:solar}}

In this Section, we discuss how our results compare with studies of bodies that were formed during the youth of our Sun. For instance, very few chondrites ($\sim$\,2\%, \citealp{Norton2008}) that fall on Earth seem to contain enstatite, a result consistent with our modeling results. The Fe content is also an interesting diagnostic to focus on, as it is quite a novel finding for dust in exo-zodiacal belts.

\subsection{Cometary bodies\label{sec:comet}}

As detailed in the reviews by \citet{Hanner2010} and \citet{Wooden2008} comets are formed in the outermost regions (Oort Cloud or the Kuiper Belt region) and trace the youth of the Solar System. Informations about their dust composition is a prime objective in order to better understand the formation of our Solar System. Using spectroscopic observations at different epochs, \cite{Wooden1999} studied the spectrum of the comet C/1995\,01 (Hale-Bopp). According to their modeling they found both Mg-rich olivine and crystalline pyroxene grains to be present in the coma, but no clear evidences for Fe-rich olivine grains. The Mg-rich pyroxene grains were required to reproduce the 9.3\,$\mu$m feature, which is not detected in any of our sources. The modeling results for the comet Hale-Bopp are very similar to the ones for the Oort Cloud comet C/2001 Q4 (\citealp{Wooden2004}). For both comets, the mineralogical findings echo the results from the Stardust mission, on comet P81/Wild\,2. Based on samples collected in situ, \citet{Zolensky2008} found that olivines in the comet mostly consist of Mg-rich olivine grains, with a Mg fraction (=Mg\,/\,[Mg + Fe]) of about 99. However, they also found the composition range of olivine to be extremely broad, ranging from Mg fractions between 4 and 100. From the peak position of the Mg distribution, the author conclude that composition changes due to capture heating must have been insignificant, otherwise the Mg distribution should instead peak around 40--60. Interestingly, from the peak position of the Mg distribution, the authors suggest that the petrology of the olivine grains can be explained by aqueous alteration; Fe-rich olivine grains may not survive as easily as Mg-rich grains in presence of water (\citealp{Wogelius1992}). The olivine grains may either trace the eventual presence of liquid water on P81/Wild\,2, or trace different regions of the initial disk, where Mg-rich olivine grains have preferentially survived aqueous alteration compared to Fe-rich olivine grains. The fact that crystalline grains detected in cometary bodies are Mg-rich rather than Fe-rich is consistent with observations of gas-rich Class\,II disks. Comets are expected to trace the mineralogy of the outermost regions of the primordial disk, therefore if Fe-rich olivine grains, as we detect them in debris disks, were already formed during the Class\,II phase, we should in principle observe significant fractions of them in comets as well. For the couple of objects for which we find more Fe-rich than Mg-rich olivine grains, this points towards a new generation of dust in the transient phase of the disks, which is subjected to alternative crystallization processes, more inclined to form Fe-rich crystalline grains. 

\subsection{Interplanetary Dust Particles}

The exact origin of Interplanetary dust Particles (IDPs) is difficult to track, they could have been trapped in asteroids or short-period comets (originating from the Kuiper Belt, see \citealp{Bradley2010} for a review). It is in principle possible to distinguish between an asteroid or comet origin, based on He release during their entry in the atmosphere (\citealp{Brownlee1995}), but the place of formation of these IDPs is thought to be presolar, based on isotopic considerations. An interesting point with respect to our study is that olivine grains contained in several IDPs are found to be Fe-rich (e.g., \citealp{Bradley1994}, \citealp{Brunetto2011}). According to \cite{Zolensky1994} the Mg distribution for hydrous IDPs ranges between 76 and 100, while it ranges between 52 and 100 for anhydrous IDPs, these differences being the possible consequence of aqueous alteration. \citet{Nguyen2007} discuss the origin of Fe in IDPs and suggest that the Fe enrichment may be the consequence of a secondary process taking place in the solar nebula or on the parent body, which echo our findings. Nevertheless, \citet{Nguyen2007} also discuss the possibility of condensation under non-equilibrium conditions in stellar outflows to explain the detection of Fe-rich crystalline grains. Overall, the existence of Fe-rich olivine grains in various astronomical environments is not to be questioned, the novelty of our study is to detect such grains in several exo-zodiacal belts. Such grains are not expected to be formed in gas-rich Class\,II disks, their detection in debris disks points toward alternative crystallization process during the transient phase of the disks.

\subsection{S-type asteroids\label{sec:asteroid}}

Investigating these alternative crystallization mechanisms can be supported by our current knowledge of other bodies from the Solar System, such as asteroids. \cite{Nakamura2011} study the mineralogy of samples collected by the Hayabusa spacecraft, directly on the near-Earth Itokawa S-type asteroid. The authors find the most abundant crystalline silicates to be of olivine composition, with an averaged Fe fraction of about 30\,\% (Fa$_{29.0 \pm 0.7}$). Based on estimations of equilibration temperatures, they conclude that these particles have experienced a peak temperature of about $\sim$\,1100\,K before cooling down slowly to $\sim$\,900\,K. To explain the slow cooling, the authors suggest that such particles shoud have been formed deep inside a large ($\geq$\,20\,km) asteroid, where intense thermal metamorphism can efficiently take place (supported by the decay of Al$^{26}$ for instance). \cite{Nakamura2011} conclude that results from the Hayabusa mission provide a link between the composition of S-type asteroids and ordinary chondrites. Together with their findings, we are able to connect the mineralogy and petrology of bodies from our Solar System with dust grains observed in exo-zodiacal belts around Solar analogs.

\section{Conclusion\label{sec:conclusion}}

In this study, we present a new and powerful radiative transfer code, dedicated to debris disks (\textsc{Debra}). Combining this code with recent laboratory experiments on dust optical properties we model the Spitzer/\textsc{Irs} spectra of 7 debris disks around solar-analogs, in order to derive the mineralogy of the dust. The entire SEDs, from optical to millimeter wavelengths, are successfully modeled within the optically thin regime. Based on the modeling of these rare objects, we obtain the following results:

\begin{enumerate}
\item  To contribute significantly in the mid-IR, grains must be ``dirty", especially grains with low absorption efficiencies in the near-IR, such as Fe-free amorphous grains, enstatite and crystalline olivine grains. If $Q_{\mathrm{abs}}$ values are too low at the wavelengths where the stellar radiation peaks, grains will be almost transparent and consequently too cold for their emission to be significant. Even though the crystalline olivine grains we use in this study contain some Fe (Mg$_{\mathrm{2x}}$Fe$_{\mathrm{2(1-x)}}$SiO$_4$, with $x=$\,0.925 or 0.8), \citet{Zeidler2011} have demonstrated their absorption efficiencies to be very low in the near-IR. Adding impurities, such as carbon, helps circumventing this issue. One should note that we do not assume the dust species to be in thermal contact. Each grain, for all sizes and all compositions has its temperature determined as a function of its distance $r$ and absorption efficiencies.
\item For several sources, especially HD\,113766\,A and HD\,69830, we find that crystalline olivine grains contain Fe inclusions, up to 20\% compared to Mg. Fe-rich crystalline grains are usually not observed in Class\,II protoplanetary disks, which can be explained by several studies of crystallization processes (e.g., \citealp{Gail2004}, \citealp{Nuth2006}, \citealp{Murata2009}). Therefore our findings of more Fe-rich compared to Mg-rich olivine grains points towards a new generation of crystalline dust grains, for which the crystallization processes are slightly different in debris disks than in Class\,II disks (e.g., inside differentiated bodies), leading to a higher fraction of Fe.
\item  In the formalism of \cite{Wyatt2007}, using the ratio between observed and maximal fractional luminosities, we find that most, if not all, of the debris disks we study are in a transient phase, that cannot be explained by a steady-state evolution from the Class\,II phase, meaning that $\mu$m-sized dust grains were produced relatively recently compared to the stellar ages. Computing the radiation pressure efficiency $\beta_{\mathrm{rp}}$, we find that $\mu$m-sized grains are short-lived around at least three objects (HD\,15407\,A, HD\,113766\,A and HD\,169666). Investigating their variability in time via archive mining, we conclude there must be an on-going process that replenishes efficiently the population of smallest grains on short dynamical timescales. At least 5 out of 7 sources are either in binary systems or have planets orbiting the central star, suggesting the transient phase might have been triggered by dynamical stirring from these massive bodies.
\item We find that relative abundances of enstatite grains are overall small for all the sources we study. This is the consequence of non-detections of recognizable emission features associated with pyroxene crystalline grains (especially at 9.3\,$\mu$m). If the observed crystallinity is the consequence of present day crystallization processes, this may be explained by higher activation energies for annealing compared to crystalline olivine grains. If the observed crystallinity fraction is inherited from parent bodies that collided (with no shock-induced heating), this echoes the scarcity of enstatite chondrites in our own Solar System.
\item The quality of the fits we present in this study is intimately connected to the $Q_{\mathrm{abs}}$ values we use in the modeling, especially for olivine grains (which are responsible for the strongest emission features in the spectra we model). This underlines the extreme importance of laboratory experiments to improve our understanding of cosmic dust properties. The aerosol technique is a promising technique that provide direct measurements of optical properties, with as less environmental contamination as possible (e.g., \citealp{Tamanai2009}).
\item According to far-IR observations, only one source in our sample (HD\,113766\,A) harbors a cold, outer belt. Based either on disk properties from our modeling results ($r_{\mathrm{out}}$), or on mineralogical considerations (fit to emission features), we discuss the possibility of outer belts around two other sources (HD\,15407\,A and BD+20\,307). 
\end{enumerate}

As a conclusion, we would like to stress the fact that SED modeling is a challenging problem, for proto-planetary and debris disks. Since debris disks are optically thin at all wavelengths, the problem becomes even more difficult as several assumptions no longer hold, such as thermal contact between dust grains (which renders the temperature determination critical) or the continuum emission as commonly used when modeling spectra of Class\,II disks. Spectroscopic observations do not contain any spatial informations, which introduces uncertainties in the modeling as demonstrated by the bayesian inference performed on the disk parameters. Therefore more refined modeling would require high angular resolution observations (e.g., near-IR interferometry or scattered light imaging, \citealp{Lebreton2011}) to lower the number of degenerate parameters ($r_{\mathrm{in}}$, $r_{\mathrm{out}}$ and possibly $\alpha$). Further work is also required on the dust properties to improve the modeling results, not only with laboratory experiments (e.g., aerosol measurements for other dust species), but also to investigate additional effects such as grains porosity, or inhomogeneous aggregates of different species (as described in \citealp{Min2008} and \citealp{Voshchinnikov2008}). This latter point may eventually provide a suitable solution for disks where $\mu$m-sized grains are subjected to efficient radiation pressure. If small grains are embedded in larger grains, they may survive longer in the dust belt. Unfortunately, this increases the complexity of the modeling, and again underlines the need for additional constraints on the disk geometry. With observations and laboratory data available at present days, we are able to successfully model the \textsc{Irs} spectra of several warm debris disks, with a limited number of dust species, and provide new insights into the origin of this transient dust. As exemplified by Figure\,\ref{fig:fe_proba}, we are confident in our mineralogical results, especially concerning to the relative abundances of Fe-rich and Mg-rich olivine grains.

\begin{acknowledgements}
The authors thank the anonymous referee for the constructive and positive comments, as well as the A\&A Editor, Malcolm Walmsley, for his careful reading and for providing additional suggestions to improve the paper. J.\,O. acknowledges the Alexander von Humboldt foundation for critical financial support. A.\,M. and P.\,\'A. appreciate financial support from the Hungarian Research Fund OTKA 101393.
\end{acknowledgements}
\bibliography{biblio}

\end{document}